\documentclass[a4paper,11pt]{article}
\pdfoutput=1
\usepackage{jcappub}
\usepackage[T1]{fontenc}
\usepackage[dvipsnames]{xcolor}
\usepackage{tikz}
\usepackage{amssymb, amsmath, epsfig, natbib}
\usepackage{graphicx}
\usepackage{tabularx}
\usepackage{float}
\usepackage{soul}
\usepackage{comment}
\usepackage{layouts}
\usepackage{ifthen}
\usepackage{multicol}
\usepackage{footnote}
\usepackage{tablefootnote}
\usepackage{caption}
\usepackage{subcaption}

\makesavenoteenv{tabular}

\def\fnl{$f_{\rm{NL}}$}

\newcolumntype{"}{@{\hskip\tabcolsep\vrule width 1pt\hskip\tabcolsep}}

\newcommand{\lt}{\ensuremath <}
\newcommand{\gt}{\ensuremath >}

\newcommand*{\hompc}  {h \rm{Mpc}^{-1}}
\newcommand*{\mpcoh}  {h^{-1}\rm{Mpc}}
\newcommand*{\gpcoh}  {h^{-1}\rm{Gpc}}
\newcommand{\specsim}{{\tt SpecSim}}

\newcommand{\warn}[1]{{\color{black}{#1}}}
\newcommand{\rfree}[1]{{\color{black}{#1}}}
\author[a,b]{M. J. Wilson}
\author[a,b,c]{and Martin White}
\affiliation[a]{Lawrence Berkeley National Laboratory, One Cyclotron Road,
Berkeley, CA 94720, USA}
\affiliation[b]{Berkeley Center for Cosmological Physics, UC Berkeley, CA 94720, USA}
\affiliation[c]{Department of Physics, University of California, Berkeley, CA 94720, USA}
\emailAdd{mjwilson@lbl.gov}
\title{Cosmology with dropout selection: Straw-man surveys \& CMB lensing} 
\abstract{
We seek to prove the means, motive and opportunity of `dropout' selected $2~\leq~z~\leq~5$ \ galaxies for large-scale structure.  Together with acquired low-$z$ tracers, these samples would map practically every linear mode and facilitate a tomographic decomposition of the Cosmic Microwave Background (CMB) lensing kernel over an unprecedented volume\rfree{.  With this, one may infer} (the time evolution of) matter density fluctuations and perform compelling tests of horizon-scale General Relativity, neutrino masses and Inflation-- viz., curvature, running of the spectral index and a scale-dependent halo bias induced by (local) primordial non-Gaussianity.  This is facilitated by the order-of-magnitude increase in sensitivity achieved by planned CMB, \rfree{optical-to-near-infrared} imaging and spectroscopy.    

Focusing on traditional color-color -- rather than photometric redshift -- selection, we estimate the expected completeness, contamination, and spectroscopic survey speed of tailored Lyman-break galaxy (LBG) samples.  With these in hand, we forecast the potential of CMB lensing cross-correlation, `clustering redshifts' and Redshift-Space Distortions (RSD) analyses.  In particular, we estimate: the depth dependence of interlopers based on CFHTLS-Archive-Research Survey (CARS) data and propagate this to biases in cosmology; a simple relation for the dependence of the linear galaxy bias on redshift and depth; new inferences of non-linear halo bias at these redshifts using legacy data; detailed forecasts of LBG spectra as would be observed by the Dark Energy Spectroscopic Instrument, Prime Focus Spectrograph, and their successors.  We further assess the relative competitiveness of these spectroscopic facilities based on an intuitive figure-of-merit and \rfree{define modern equivalents to traditional color selection criteria for the Large Synoptic Survey Telescope, where necessary}.  

We confirm these science cases to be compelling for achievable facilities in the next decade by defining a LBG sample of increasing Lyman-$\alpha$ equivalent width with redshift, which delivers both percent-level RSD constraints on the growth rate at high-$z$ and measurements of CMB lensing cross-correlation at $z=3$ and 4 with a significance measured in the hundreds, given sufficient area overlap.  \rfree{Finally, we discuss the limitations of this initial exploration and provide avenues for future investigation}.
}  
\arxivnumber{1904.13378}
\begin{document}
\maketitle
\flushbottom
\section{Means, motive and opportunity for $z \geq 2$ large-scale structure}
\label{sec:sciencecase}
The study of the large-scale structure in the Universe, as indirectly traced by both galaxies and gravitational lensing, \rfree{promises great insight into the many fundamental physics and cosmology \cite{Wei13, PDG18} questions that remain}, not least gravitational physics \cite{JaiKho10, Joyce15, Joyce16, Amendola18}, the neutrino mass hierarchy \cite{Les13, Pat15, Arc17, LatGer17} and inflation \cite{Liddle00}.  Completed and forthcoming imaging surveys, including the Legacy Survey\footnote{\url{http://legacysurvey.org}}, Dark Energy Survey\footnote{\url{https://www.darkenergysurvey.org/}}, Hyper-Suprime Camera\footnote{\url{http://hsc.mtk.nao.ac.jp/ssp/}}, Clauds\footnote{\url{http://www.ap.smu.ca/~agolob/clauds/survey/}}, CFIS/UNIONS\footnote{\url{http://www.cfht.hawaii.edu/Science/CFIS/}}, LSST\footnote{\url{https://www.lsst.org}} and Euclid\footnote{\url{https://www.euclid-ec.org}}, will deliver data of unprecedented depth and area for this purpose. 
\begin{table}[]
    \centering
    \begin{tabular}{|c|c|c|c|c|c|c|c|c|}
      \hline
      Survey              & $u$ & $g$ & $r$ & $i$ & $z$ ($Z$) & $y$ ($Y$) & Area & Ref. \\
      & & & & & & & [$10^4$ deg$^2$] & \\ 
      \hline
      \hline
      HSC            & \ \rfree{--}\tablefootnote{The Clauds survey will observe 19 deg$^2$ to a $u$-band depth of 27.1 in the HSC Deep fields.} & 26.4     & 25.9 & 25.7     &    25.0  & 24.3 & \rfree{0.14} & \cite{HSC18} \\  
      \hline
      KiDS - Viking       & 24.2   & 25.1   & 25.0   & 23.6   & (22.7) & 22.0 & 0.13 & \cite{Hildebrandt17}\\
      \hline
      \hline
      Dark Energy Survey  & --   & 25.4 & 24.9 & 25.0 & 24.7  & (21.69) & 0.50 & \cite{DES} \\
      \hline
      CFIS / UNIONS              & 24.4 & 25.4     & 24.8 &  24.5    & 24.3    & --  &  1.00 & \cite{Ibata17} \\
      \hline
      Legacy Survey       & --   & 24.0 & 23.4 & --   & 22.5  & -- & 1.40 & \cite{Dey18}\\
      \hline
      \hline
      LSST-Y1                & 24.07 & 25.6 & 25.81 & 25.13 & 24.13  & 23.39 & 1.23 & \cite{LSST18}\\
      \hline
      LSST-Y10                & 25.3 & 26.84 & 27.04 & 26.35 & 25.22  & 24.47 & 1.43 & --\\
      \hline  
    \end{tabular}
    \caption{A comparison of the available depths of current and forthcoming imaging surveys;  We quote the publicly available (inhomogeneous) metrics for each survey, $5\sigma$ depths for galaxy  sources for KiDS, \rfree{the Legacy survey and HSC, with point-source depths for the remainder}.  The difference can be $0.5-1.5$ magnitudes, with the latter likely the more appropriate for our science case.
    }
    \label{tab:depths}
\end{table}

While there are compelling tests that can be uniquely performed without mode-sampling variance in principle \cite{McDSel09}, many continue to require a large volume (as is the case for multi-tracer analyses, in practice).  For our fixed $4\pi$ steradians, we are then forced to great distances or small scales to achieve greater precision.  The number of uncertainties on megaparsec scales, and proven successes of large-scale studies, e.g. Baryon Acoustic Oscillations \cite{Eisenstein05, DESI16}, motivate consideration of the former as the program of choice.  The open questions are thus to what distances can we reach, and how effectively, using traditional selection of high-$z$ galaxies and this unprecedented data.  

While we explore the multiple science returns in \S \ref{sec:cosmology}, we first provide initial motivation with  Fig.~\ref{fig:ang2spatial}, the left panel of which shows that $\Lambda$CDM, when constrained by current observations \cite{Planck18Leg, Planck18Pars, Alam17}, predicts $\sigma_8(z)$ with 1.1\% precision up to $z=5$.  \rfree{Such a definitive prediction deserves a challenge by facilities capable of an equally precise constraint, rather than simply assuming this to be the vanilla Einstein-de Sitter epoch that $z<1$ was also once assumed to be.  We will argue that samples of dropout galaxies, in combination with upcoming sub-millimetre surveys, can meet such a challenge.}

Further unique opportunities to test General Relativity (GR) are provided by the large scales made available.  Within GR, gravitational lensing is sourced by the Weyl potential, which is uniquely predicted by the Newtonian potential inferred from the non-relativistic motion of galaxies -- in the absence of anisotropic stress.  This constraint fails to hold for a number of alternative theories posited.  Direct tests of this relationship would augment the stringent diagnostics of gravity placed by galaxy-galaxy lensing, cosmic shear and magnification with Euclid and LSST \cite{Hildebrandt17, Troxel017} to a greater volume, $\simeq 850 \ (\gpcoh)^3$ to $z=6$, and to larger scales.  The $2 \leq z \leq 5$ volume is entirely unreachable with galaxy lensing, while the systematic biases due to e.g.\ stellar contamination, varying depth, etc., propagate dissimilarly.  See ref.~\cite{Chen18,CVDE-21cm} for the ultimate culmination of testing GR to $z\simeq 7$ with 21cm intensity mapping.  In the following subsection, we briefly review CMB lensing as a much more promising test of these effects.  This reflects our initial focus on \rfree{this first synergy, with detailed analyses of the others, e.g.\  redshift-space distortions, left to future work}.

\subsection{Cosmic Microwave Background lensing tomography}
\begin{table}
\centering
 \begin{tabular}{|| l | c | c | c | l ||} 
 \hline
 Survey & Map RMS & Resolution & Area & Ref. \\ [0.5ex] 
 & [$\mu K$-arcmin] & [$\prime$] & [$10^4$ deg$^2$] & \\ 
 \hline\hline
 Planck & \rfree{33.0} & 7.2  & 2.37 & \cite{Planck18Pars} \\
 \hline 
 Advanced ACT & 12.0 & 1.5 & 0.825 & \cite{AdvACT} \\ 
 \hline
 SPT-3G & \rfree{3.3} & 1.4 & \rfree{0.25} & \cite{Anderson18} \\ 
 \hline
 Simons Observatory & 6.0 & 1.0 & 1.65 & \cite{SO18} \\
 \hline
 CMB-S4 & 1.0 & 1.4 & 1.65 & \cite{CMBS4} \\
 \hline
 LiteBIRD & \rfree{4.1} & 30.0 & 2.89 & \cite{LiteBIRD}  \\ 
 \hline
 COrE & 3.6 & \rfree{7.7} & 3.30 & \cite{CORE} \\
 \hline
\end{tabular}
\caption{A census of the basic instrumental properties of current, planned and proposed CMB experiments; see also Table 5 of ref.~\cite{Errard16}.  \warn{These should be interpreted as indicative values close to \rfree{150 GHz}.}}
\label{Table:CMBxps}
\end{table}

With recent experiments yielding up to $40 \sigma$  \warn{detections}
\cite{Sherwin17, Omori17, PlanckLens15,Planck18Lens} of arcminute-scale lensing distortions to the 
primary Cosmic Microwave Background (CMB) by intervening matter, CMB lensing has taken on a greater significance as a looking glass onto the entirety of matter fluctuations up to (the single source plane at) last scattering, $z\simeq 1100$, with significant weighting beyond $z=2$ -- see Fig.~\ref{fig:pz} and Fig.~3 of ref.~\cite{LewCha06}.  This gravitational lensing will be a driving force of cosmological research, with future experiments (Table \ref{Table:CMBxps}) 
\warn{capable of achieving ever greater significance} \cite{SO18, S416}.  For example, the CMB-S4 $\kappa$ auto-power spectrum will reach a S/N of 406 for $L \simeq 1000$, with further gains to $L \simeq 2000$ \rfree{by which point} the statistical error and resolution will be fully \warn{saturated} \cite{Schmittfull18}.

However, the full potential of this lensing map may only be achieved in combination with a wide-area, \rfree{dense} galaxy population with sufficiently precise redshifts to \rfree{trace} the radial structure and unlock the redshift evolution of density fluctuations, $\sigma_8(z)$; \warn{such a measurement would have} ramifications for many fundamental questions \cite{Schmittfull18, Yu18}.  \rfree{Not least, estimates of the (sum of) neutrino mass from the linear growth rate, $\delta_m\propto a^{1-3f_\nu/5}$, where $f_\nu$ is the neutrino fraction, which typically rely on challenging large-scale CMB polarisation measurements.  This may potentially be avoided with a sufficiently long redshift baseline \cite{Yu18} or by exploiting subtle differences between the lensing and halo power spectrum \cite{LoVerde14, Schmittfull18}.}  

With the lensing map itself acting as an integral constraint on the total variance, high~-~$z$ tracers \rfree{would} augment low~-~$z$ measurements via delensing and vice versa \cite{Smith12, Sherwin15}, as we explore in \S \ref{subsec:fgdelensing}.  This allows for the CMB lensing measurements to be \rfree{optimally tailored} to the low-$z$, Dark Energy dominated, Universe by delensing with the high-$z$ samples.   

\begin{figure}[t]
  \begin{subfigure}{.5\textwidth}
    \centering
    \includegraphics[width=\linewidth]{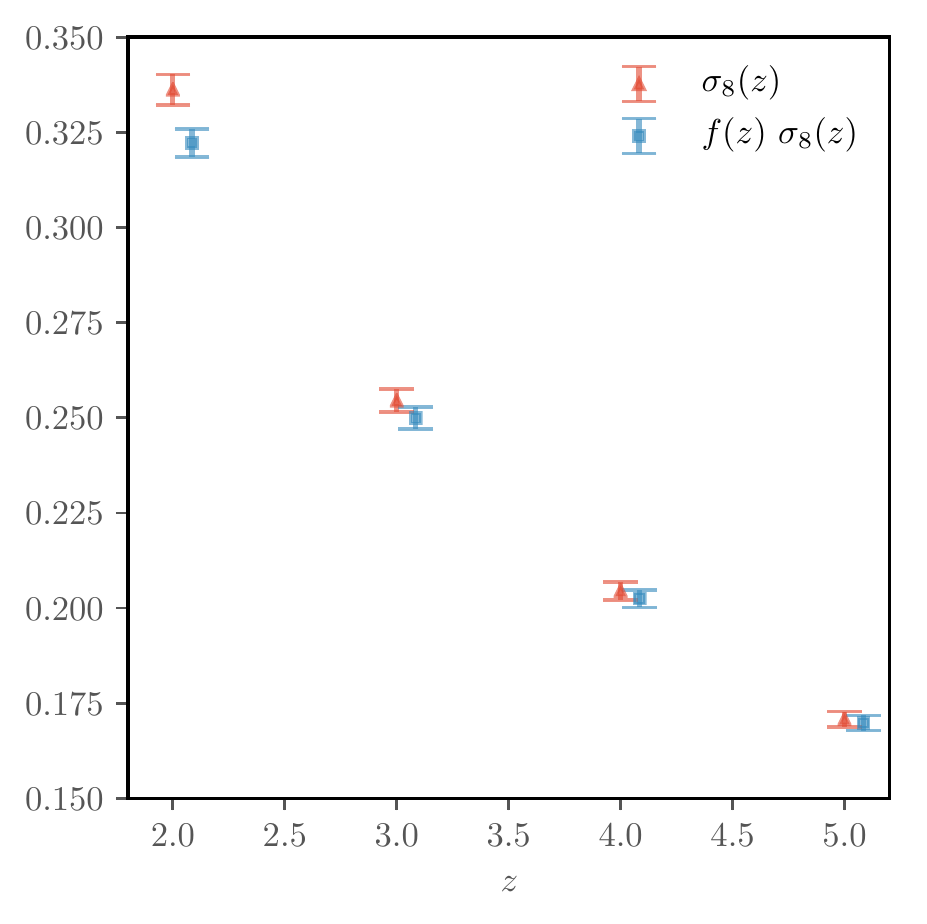}
  \end{subfigure}
  \begin{subfigure}{.5\textwidth}
    \centering
    \includegraphics[width=\linewidth]{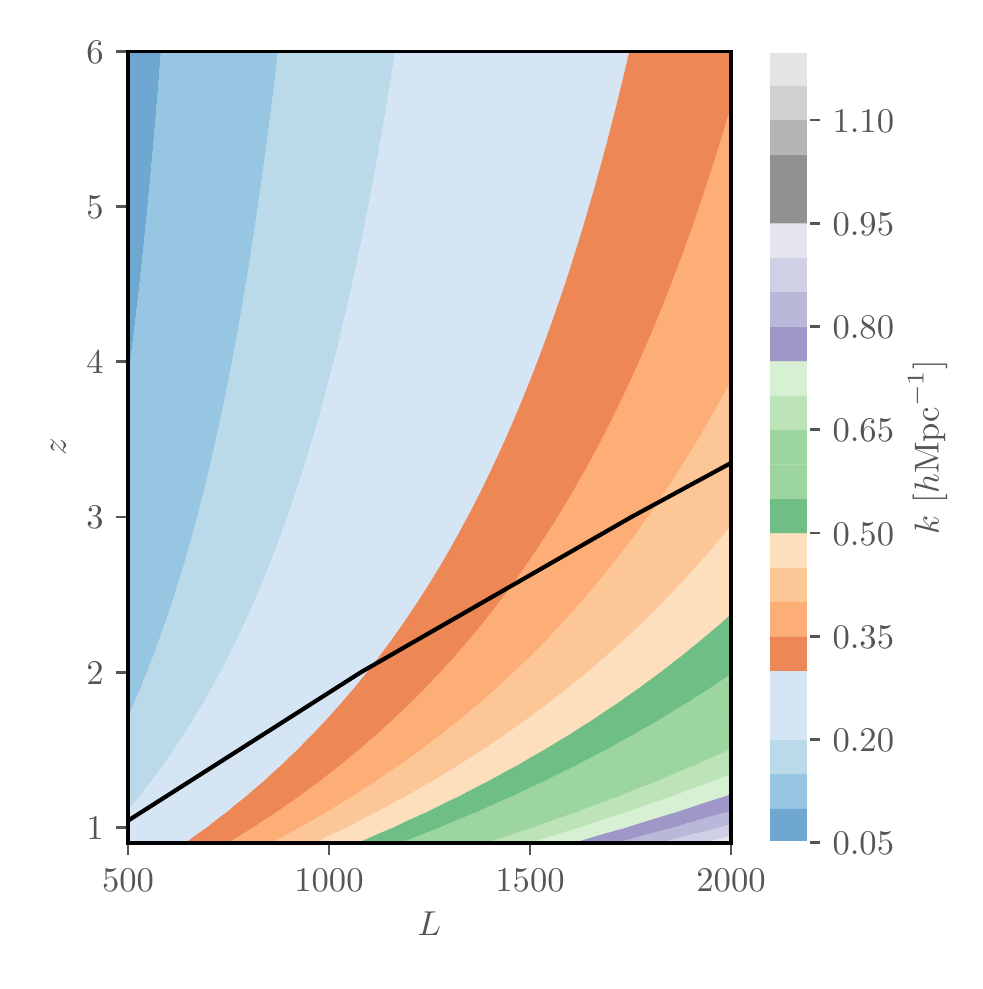}
  \end{subfigure}
\caption{\textbf{Left:}  Combined Planck (2018) and BAO \cite{Planck18Lens, Planck18Pars, Alam17} constraints on $\sigma_8(z)$ and $f\sigma_8(z)$ within $\Lambda$CDM with varying $m_\nu$.  Constrained by this combination, the $\Lambda$CDM prediction has a 1.1\% precision up to $z=5$ for both the amplitude and growth rate.  With fixed neutrino mass ($m_\nu=0.06\,$eV), this increases significantly to $0.4\%$.  Confirming this prediction requires challenging sensitivity, but provides stringent tests of the Standard Model \cite{Schmittfull18}.  \textbf{Right:}  The spatial, $k$, to angular, $L$, wavenumber mapping with redshift in the (extended) Limber approximation.  At these redshifts, $L<2000$ corresponds to mildly non-linear scales with significant S/N that may be fully utilised without imposing empirical modelling limits \cite{ZA70, Whi14, CLPT}. In particular, the expected Zeldovich displacement (solid) is shown as a likely limit 
beyond which nuisance parameter marginalisation limits potential gains.}
\label{fig:ang2spatial}
\end{figure}

Despite recent successes for \warn{CMB lensing} cross-correlation with high-$z$ galaxies and QSOs \cite{Sherwin12, Geach13, Bianchini15, Omori17, Peacock18, Geach19}, there is significant potential for improvement as QSOs do not provide sufficiently dense samples to fully exploit the \warn{CMB limited} signal-to-noise at $z=2$ and $z=3$.  Fig.~\ref{fig:ang2spatial} (right) further shows that the modes that are likely detected with significance, $L < 2000$, correspond to spatial wavenumbers at which current model predictions are accurate at the redshifts of interest.  Hence, it will likely be the galaxy sample or CMB experiment that will be the limiting factor, rather than theory.  In particular, the \rfree{lesser impact of halo-bias} on the cross-spectra, relative to the auto-spectra, may be exploited \cite{Modi17b} and motivates a theoretical treatment based on the complete set of available symmetries \cite{McDRoy09}; see ref.~\cite{Desjacques16} for a recent review.  We find in \S\ref{sec:clustering} that a linear galaxy bias approaching $b\simeq 10$ is certainly possible \warn{at relevant magnitudes} and explore the resulting scale-dependence within a halo occupation distribution model.  In addition, the modelling of projected statistics is sensitive to the source redshift distribution (including interlopers), $dN/dz$.  To date, the uncertainty on this distribution is commonly neglected or crudely approximated by a trade-off with the effective redshift, e.g.\ ref.~\cite{DESY1}, but this may not be sufficiently accurate for future surveys;  we investigate this in \S\ref{sec:interlopers}.

\subsection{Distinctive signatures of Inflation}
Standard Inflationary theories predict distinct signatures solely on large scales, which suggests that volume is typically the limiting factor for proposed surveys.  For instance, if we consider typical models with their current best bounds\footnote{This discussion closely matches that within Table 8 of ref.~\cite{Planck18Leg}.} (in brackets), Inflation makes testable predictions for: a spatially flat Universe ($\Omega_k = 0.0007 \pm 0.0019$), dominated by Gaussian and adiabatic scalar perturbations ($r_{0.002} < 0.07$), with a spectrum of density perturbations that is expected to be nearly scale-invariant, but marginally red ($n_s = 0.967 \pm 0.004$) and otherwise a featureless power law ($dn/d\ln k = −0.0042 \pm 0.0067$); non-minimal models predict a unique set of alternatives, see e.g.\ ref.~\cite{Slosar19}.  

Multi-field models make an additional prediction of (local) primordial non-Gaussianity of the density perturbations, with an amplitude \rfree{often} denoted \fnl, which induces a scale-dependent halo bias $\propto k^{-2}$ of primordial origin \cite{Dalal08}.  The largest amplitude predicted is of order unity, with constraints below this able to exclude many of the available models;  Current upper bounds are \fnl=$0.8 \pm 5.0$ \cite{Planck16XIII} and \warn{$-51< f_{\rm{NL}} <21$ from a spectroscopic survey \cite{Castorina19b}}.  Together with volume and access to large spatial scales, tracers with large linear galaxy bias help to maximise the potential signal.  A detection of such a signature would be of great significance as all single-field models (with standard Bunch-Davies initial conditions) would be excluded given the single-field consistency relations \cite{Maldacena03, Creminelli04}.

Considering a cross-correlation of the CMB-S4 lensing convergence map with $i < 27$ LSST tomographic redshift slices spanning $0 < z < 7$, 
ref.~\cite{Schmittfull18} forecasts an upper limit below unity, viz.\ $\sigma$(\fnl) = 0.7; without a high-$z$ Lyman-break galaxy sample these constraints would be roughly half as competitive.  Moreover, refs.\ \cite{Ferraro19, Percival19} strongly advocate complete spectroscopic followup of high-$z$ Lyman-break galaxies to constrain \fnl\ via the scale-dependent halo biasing present in the auto spectrum;  \warn{noting, in particular, that this removes stellar contamination as a significant systematic concern}.  Again, preliminary forecasts suggest that the long sought limit below unity is feasible.  For this, the required comoving number density is typically $\simeq 10^{-4} \ [(h^{-1} \rm{Mpc})^{-3}]$, while the high-$z$ bispectrum is additionally constraining if the sample density samples quasi-linear modes with high fidelity \cite{Karagiannis18}.

\subsection{Breaking the \{$\Omega_\Lambda$, $\Omega_\nu$, $\Omega_k$\} degeneracy}
Since current curvature constraints are primarily derived from the angular-diameter distance to last scattering, they are limited by our understanding of the late-time evolution of dark energy \cite{Hannestad05, Ichikawa06, dePutter09, Allison15, Lorenz17, MishraSharma18}, modified gravity and the sum of neutrino masses.  Constraints on the luminosity and angular-diameter distances at redshifts much less than last scattering greatly reduce this uncertainty;  e.g.\ Planck $TT$ + low $P$ + lensing + BAO suggests $\Omega_k = 0.000 \pm 0.005$ at 95\% confidence \emph{for fixed} $w$ \cite{Planck16XIII}.  In comparison, the Dark Energy Spectroscopic Instrument (DESI) will deliver an error on $\Omega_k$ of $\simeq 10^{-3}$ \cite{Font-Ribera14}.  An idealised full-sky, cosmic variance limited $z \leq 4$ BAO survey would achieve $10^{-4}$, but be degraded by a factor of order unity by either marginalisation over evolving $(w_0, w_a)$ models or tracers that are no more dense than $10^{-3} (\mpcoh)^{-3}$, assuming a linear galaxy bias of unity \cite{Takada15}.  As such, the gauntlet of $\sigma(\Omega_k) \simeq 10^{-4}$ set by eternal inflation models \cite{Kleban12} is within reach on an achieveable timescale, while remaining reassuringly distant from the fundamental uncertainty of $1.6 \times 10^{-5}$ due to horizon-scale density perturbations \cite{Waterhouse08}.

\subsection{Lyman-break galaxies as a tracer}
\label{sec:highztracers}
Currently ground-based surveys do a highly efficient job of sampling density fluctuations to $z \simeq 1.6$, including the Luminous Red Galaxies \rfree{(LRGs)}, [OII] Emission Line Galaxies \rfree{(ELGs)} and Quasars \rfree{(QSOs)} targeted by BOSS, eBOSS and DESI \cite{Paris17, DESI16}.  The Euclid and WFIRST satellites will be highly competitive at producing dense H$\alpha$ emitter samples at $z \simeq 2$, achieving $10^{-4} (h^{-1} \rm{Mpc})^{-3}$ for $b(z) \propto \sqrt{1+z} \simeq 1.7$ tracers -- see Table 3 and 4 of ref. \cite{Amendola18}.  At higher redshifts, fainter QSOs provide potential targets, but their number density rapidly declines to higher redshift and they become difficult to select near $z \simeq 2-3$ as the optical colors of unobscured QSOs around $z\sim 2.7$ resemble the more abundant metal-poor A and F halo stars and compact galaxies dominated by A and F stellar populations.  \warn{At lower redshift, $2.2<z<2.6$ QSOs can have similar colors to those contaminated by host light $z\simeq 0.5$}.  This makes highly complete but efficient detection with minimal contamination difficult \cite{Fan99, Richards01, Myers15}.  

As a result, the focus of this work is \rfree{arguably the next target of opportunity} for large-scale structure surveys: Lyman-break `dropouts', so named for the flux decrement blue-wards of the Lyman limit due to absorption by (intrinsic) neutral Hydrogen.  For example, a `$u$-dropout' sample is isolated by the lack of a detection in the $u$-band, but significant flux in one or more redder bands.  The detection band is often chosen to match the rest-frame UV (1500\AA) \ for the given redshift, to simplify the SED dependence of the $k$-correction.  In reality, a large magnitude difference or `drop' is required between the dropout and detection bands, as opposed to an actual non-detection, and selection is achieved on the basis of color-color cuts.  This results in a $u$-dropout sample localised near $z\sim 3$.  Similarly, $g$ and $r$-dropouts deliver samples at $z\sim 4$ and $5$ for detection bands of $i$ and $z$ respectively \cite{Ono18}.  We will refer to the idealised SED on which the selection is based, thought to be physically similar across these dropout classes, as Lyman-Break Galaxies (LBGs).

Color-selected samples \rfree{isolate} a subset of galaxies from a purely flux-limited survey, which may have photometric redshifts available for a much larger superset of objects.  We consider tailored color-color selections of physically motivated populations, rather than photometric redshift selection, as it better informs survey design considerations given the transparent dependence on given (detection) bands and their depth.  For given survey characteristics, an improvement might be expected from photometric redshift selection given the additional bands included.  However, we find a number of further advantages to color selection that warrant further investigation and quantification.  There is evidence that (at sufficient depth), the Lyman-break star-forming galaxies on which dropout selection is based comprise the majority of galaxies available at high redshift \cite{Giavalisco02, Shapley11, Ly09}.  Thus, selecting on perfect photometric redshifts would select essentially the same sample, but with often opaque assumptions of the template set, presence of emission lines and dust extinction (with their associated priors \cite{Speagle17}), the data space -- (positive definite) flux or magnitudes, inference of zero-point corrections, together with a varying redshift boundary due to the typical underestimate of the confidence interval for point estimates derived from the posterior \cite{Dahlen13}.

The basis of color-color selection is the well founded physical understanding of the Lyman-break subset, with the prominent SED features exploited derived from the atomic physics of neutral hydrogen.  \rfree{This} allows dropout selection to robustly identify physically very similar galaxies over an extended redshift range \cite{Giavalisco02, Shapley11}.
The likelihood is that a great number of observed (LSST) galaxies will not meet the `gold' level requirements necessary for cosmological studies.  As a result, we posit that the physically well understood samples described above will prove to be critical populations that remain.  Finally, given the minimal imaging requirements for dropout selection (e.g.\ a dropout and one to two detection bands), this allows for the largest area and homogeneous selection, as favoured by our chosen science cases.  \rfree{Additionally, this facilitates a highly efficient observing strategy focused on the dropout bands, with modest followup requirements for redder bands.  Practically, $g$ and $r$ dropouts are appealing due to the reduced sky background and associated ease to acquiring depth}, while $u$-dropouts \rfree{can be} more challenging due to the non-stationary atmosphere -- absorption by the ozone and scattering can introduce a spatially varying selection and hence artificial inhomogeneity.

As we will review, these samples are particularly well established, with numerous classic analyses providing the necessary sample properties such as completeness and interloper rates -- based on narrow band and spectroscopic followup for a range of depths, areas and configurations (ground and space based, with UV, optical and near-IR).  In particular, we make extensive use of the CFHTLS-Archive-Research Survey (CARS, \cite{Hildebrandt09}) catalogue of 80,000 LBGs for studies of the color selection, redshift distribution and clustering of $u$-dropouts.  These sources were identified in the $ugriz$ CFHTLS Deep fields with additional COSMOS $u$-band imaging, at depths approaching $29^{\rm th}$ magnitude in all but the (approximately one magnitude shallower) $z$ band.  The derived BPZ \cite{Benitez09} photometric redshifts are found to have a 3\% precision, as calculated from 3.1 deg$^2$ of overlapping spectra.  Similarly, given the much greater area, we utilise the Great Optically Luminous Dropout Research Using Subaru HSC (GOLDRUSH) survey, that targets $g$, $r$, $i$ and $z$-band dropouts in wide and deep optical Hyper Suprime-Cam images obtained as part of the Subaru Strategic Program \cite{Aihara18}.  The wide layer will eventually cover $1400$ deg$^2$ to approximately $26^{\rm th}$ magnitude with the deep and ultra-deep layers going one or two magnitudes deeper.  The preliminary release covers $100$ sq.~deg.\ and over 0.5 million dropout galaxies in the range $4<z<7$ \cite{Ono18}.

\warn{Such studies provide important precursors for survey design, in particular of the necessary spectroscopic followup for our science cases.  The stringent redshift precision required, $\sigma_z / (1+z) = 10^{-3}$, may potentially be achieved with absorption line redshifts of bright targets or prevalent emission lines for fainter objects, e.g.\ the Lyman-$\alpha$ line.  The following outline describes how such properties leads to the definition of straw-man surveys for which we may derive first estimates of the cosmology returns}.  

Where necessary, we assume a Planck (2018) + BAO $\Lambda$CDM cosmology \cite{Planck18Leg} with 
$(\Omega_m, \Omega_b, \Omega_\Lambda, h, \sigma_8)$ = 
$(0.311, 0.04898, 0.6894, 0.6770, 0.811)$\footnote{Rightmost column of Table 6.}.
Throughout we 
follow standard practice in assuming the Born and Limber \cite{Limber53} approximations for lensing.  As the precision improves, it will be necessary to reconsider such approximations for both auto and cross-correlations \cite{Bohm16, MarzFanz16a, MarzFanz16b, MarzFanz16c, LewPra16, Fab17, Simon07}.  In particular, with respect to sufficient cleaning of contaminants \cite{Amb04, vanE12, FerHil17}.

\subsection{Outline}
\warn{This work proceeds as follows:} in \S\ref{sec:LyBreakSelection}, we review physical properties of $2 \leq z \leq 5$ Lyman-break galaxies and their color-color selection.  In particular, we establish the expected redshift distribution, depth dependence of the areal density, the (linear) clustering redshift evolution, non-linear clustering and expected interloper rates assuming LSST and Euclid-like imaging.  \warn{We propagate these results to cosmological parameters in the context of CMB cross-correlations to confirm interlopers introduce a significant bias}.  

As a result, in \S\ref{sec:specz}, we explore the necessity and potential for dedicated spectroscopy.  We define fiducial photometric and spectroscopic samples that are likely achieveable for a range of future facilities and discuss the complications that will likely be faced.  In particular, we forecast spectra for a range of LBG candidates with both \rfree{M-DESI} and the Prime Focus Spectrograph (PFS) and judge the redshift efficiency with {\tt redrock}.  We consider various simple proxies and available observations to ensure these estimates are reasonable.  In Appendix~\ref{app:FOM}, we \rfree{define a figure-of-merit that quantitatively ranks contending spectroscopic facilities}.  

In \S\ref{sec:cosmology}, we establish the synergy of LBGs with CMB lensing -- calculating the achieveable signal-to-noise for our fiducial photometric sample, both with and without delensing with low-$z$ spectroscopic tracers.  We further provide preliminary estimates of the degree to which the fiducial spectroscopic sample constrains the LBG redshift distribution with clustering redshifts and potential for redshift-space distortions studies.  Finally, we suggest avenues for further work and conclude in \S\ref{sec:discussion}.  We make the associated package publicly available\footnote{\url{github.com/michaelJwilson/LBGCMB}}.

\rfree{Having outlined LBGs as one of the most competitive $z \geq 2$ targets, we first review and reference the current best knowledge of their physical properties}.

\section{Properties of Lyman-break galaxies and their selection}
\label{sec:LyBreakSelection}
\begin{figure}[t]
  \centering
    \begin{subfigure}{0.5\textwidth}
    \includegraphics[width=\linewidth]{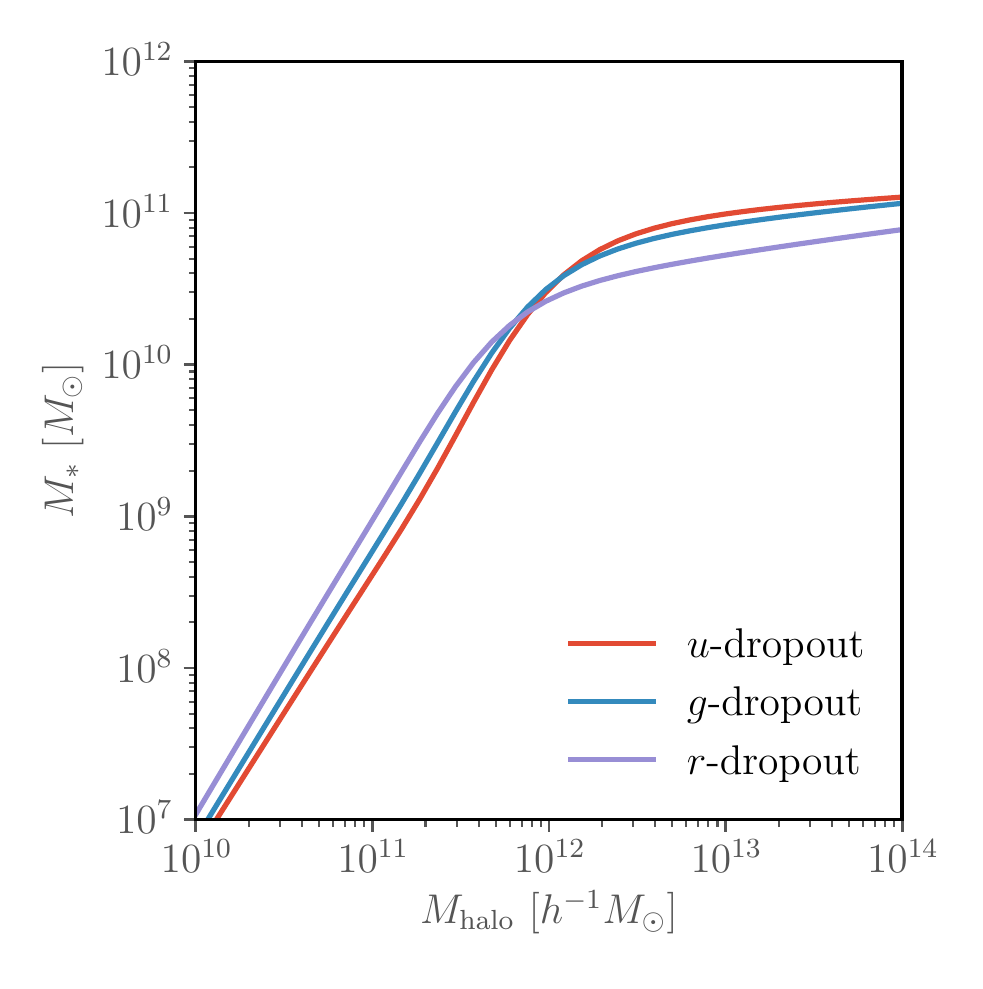}
    \end{subfigure}%
    \begin{subfigure}{0.5\textwidth}
    \includegraphics[width=\linewidth]{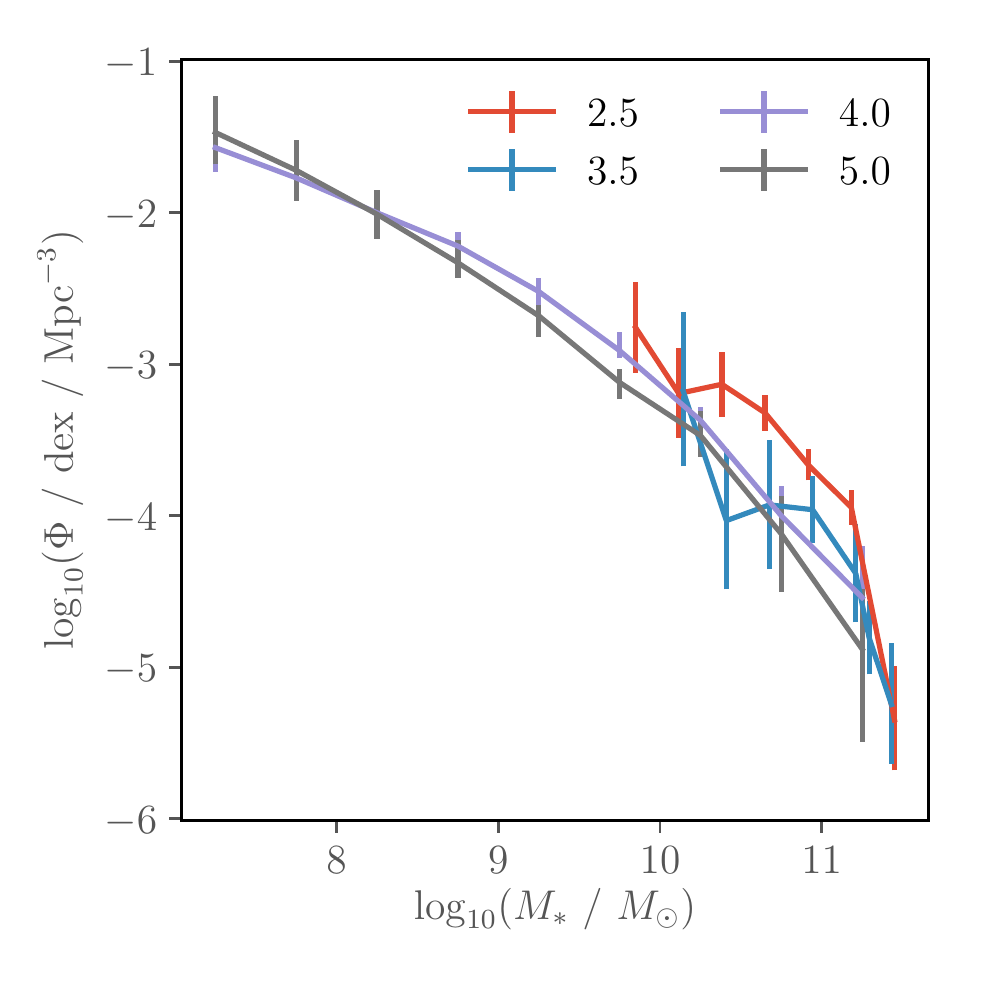}
    \end{subfigure}%
  \caption{\textbf{Left}:  The stellar-mass--halo-mass relation for $u$, $g$ and $r$-dropouts from ref.~\cite{Ishikawa17} (using the functional form of ref.~\cite{Behroozi18}).
  \textbf{Right}:  The stellar mass function for $z\simeq 2.5-5.0$, from the compilation by ref.~\cite{Behroozi18}.
  }
  \label{fig:phys_properties}
\end{figure}
\subsection{Astrophysical properties}
Dropout selection on magnitude limited surveys naturally selects massive, actively star-forming galaxies that comprise the majority population over the redshift range of interest.  Except at the bright end, the fraction of obscured or reddened galaxies which are missed by the selection is small, though the completeness does vary with redshift \cite{Park16, Jose17, Malkan17, Harikane17}.  Their rest-frame UV spectra are typically dominated by O \& B star emission with stellar masses $>10\,M_{\odot}$ and temperature $>2.5\times10^4$K.  LBGs lie on the main sequence of specific star formation\footnote{For halos defined with mean density $\mathcal{O}(10^2)$ times the background (or critical) density and $t_{\rm dyn}~\simeq~0.1/H(z)$, this sequence corresponds to $\dot{M}_\star\propto M_\star/t_{\rm dyn}$.} with $( \dot{M}_\star/M_\star ) \simeq 0.2(1+z)^{1.5}\,{\rm Gyr}^{-1}$ \cite{Ishikawa17} and UV luminosity approximately proportional to stellar mass \cite{Song16,Qiu18}.  Refs.~\cite{RodriguezPuebla17, Behroozi18} provide compilations of stellar mass functions and star formation rates, respectively, with the latter also containing model fits for quenched and satellite fractions.  Fig.~\ref{fig:phys_properties} shows the stellar mass functions at $3\le z\le 5$.  Note the steady build up of stellar mass with time.  For the samples of interest, the galaxies will have $M_\star\sim 10^{10-11}\,M_\odot$, close to the knee of the stellar mass function.

There has been an extensive effort to model\footnote{Care should be exercised in interpreting  
these models given the heterogeneity of assumed cosmologies, which can be significant for distances, volumes and halo mass functions at high redshift.}  LBGs within the context of large-scale structure since the first samples were identified.  Most recently, these include empirical \cite{Finkelstein15, Mashian16, Jose17, Behroozi18}, semi-analytic \cite{Weinmann11, Somerville15, Garel15, Park16, Yung18, Qiu18, Yung19, Samui19} and hydrodynamic models \cite{Nagamine10, Katsianis15, Dave17}.  These commonly suggest that the majority of LBGs will grow into `typical' galaxies by the present day, living in a wide range of environments (e.g.~refs.~\cite{Conroy08, Cai17}).  For instance, Fig.~\ref{fig:phys_properties} shows the stellar-to-halo mass relations for $u$, $g$ and $r$ dropouts, which we see evolves only slowly.  The galaxies we will highlight later occupy halos near the peak of the stellar-to-halo mass ratio ($M_h\sim 10^{12}\,h^{-1}M_\odot$).  See refs.~\cite{Giavalisco02, Shapley11} for comprehensive reviews on this area (and more).  We defer further detailed discussion of physical modelling, and the implications for potential surveys, to future work.

\subsection{Color tracks and selection}
\label{sec:sed}
\begin{figure}
  \begin{subfigure}{0.5\textwidth}
    \centering
    \includegraphics[width=1.\linewidth]{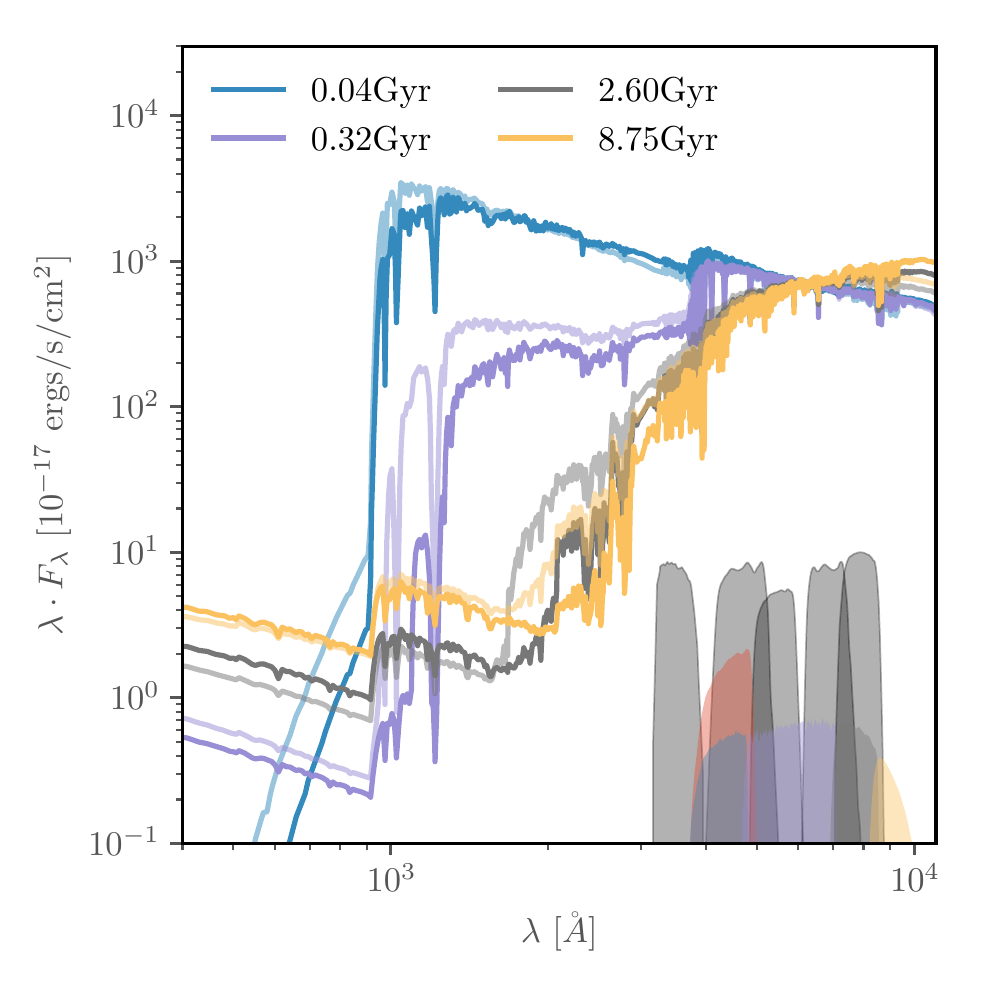}
  \end{subfigure}
  \begin{subfigure}{0.5\textwidth}
    \centering
    \includegraphics[width=1.\linewidth]{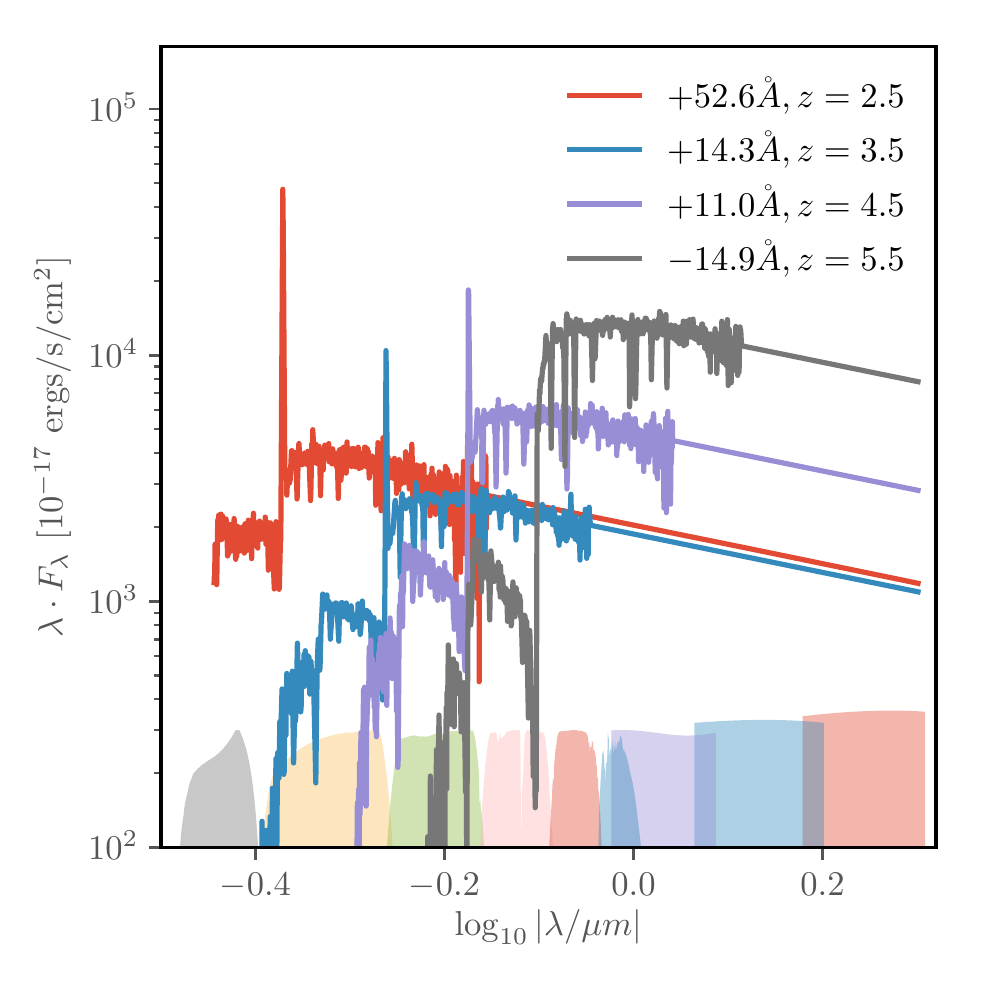}
  \end{subfigure}
\caption{\textbf{Left}:  Bruzual and Charlot \cite{Bruzual03} single-stellar population rest-frame spectra for a Salpeter initial mass function, a range of age (0.04 to 9 Gyr) and metallicities of $1.0$ (solid) and $0.2\,Z_\odot$ (translucent) respectively.  Note the characteristic break due to intrinsic neutral hydrogen absorption approaching rest-frame 912\,\AA\ for young galaxies, which gradually reddens to a dominant 4000\,\AA\ break with age.  \textbf{Right}:  Shapley composites \cite{Shapley03} of given Lyman-$\alpha$ equivalent widths, to which a Madau IGM extinction \cite{Madau95} has been applied. 
Background filters show the LSST $ugrizy$ and \rfree{Euclid $YJH$} pass-bands.  The practically flat $F_{\nu}$ spectrum for $\simeq 50\,$Myr LBGs show zero colors for the LSST filters approaching $1\,\mu$m, with a magnitude or greater break at that corresponding to rest-frame UV ($\approx 1500$\AA).}
\label{fig:lbgsed}
\end{figure}
`Dropout' color-color selection of Lyman-break galaxies targets the steep break -- in an otherwise shallow $F_{\nu}$ spectra -- that occurs bluewards of the 912\AA \ Lyman limit due to absorption by the neutral hydrogen rich stellar atmospheres and interstellar photoelectric absorption.  Lyman-series blanketing along the line-of-sight further suppresses flux short-ward of 1216\AA \ for $z>2$ sources \cite{Madau95}.  We show this explicitly in Fig.~\ref{fig:lbgsed} for  commonly assumed LBG SED templates -- the rest-frame single-stellar populations of ref.~\cite{Bruzual03} (left) and the Shapley composites of given Lyman-$\alpha$ equivalent widths (right); the latter have had the appropriate mean Lyman-forest for that redshift applied and been extended redwards assuming a flat $F_\nu$.  The left figure includes background filters showing the classic Steidel $UGVRI$ (grey) traditionally used for LBG searches, while the foreground shows the Subaru-$B$ filter and HST ACS (F435, F606, F775, F8501) used for more modern, ultra-deep, studies, e.g.\ ref.~\cite{Stark10}.  Similarly, the right figure shows the optical $ugrizy$ LSST filters and near-IR $YJH$ Euclid filters.    

In implementation, approximately zero colors (in AB magnitudes) are expected towards the red end of the optical window, with $\simeq 0.7$ magnitudes or greater difference between the two filters bracketing the Lyman break \cite{Meier76, Guhathakurta90, Steidel92}.  Beyond this, the detectability of a rest-frame UV continuum preferentially selects modest dust extinction, with observations typically consistent with a Calzetti-like \cite{Calzetti00} extinction and a mean color excess of 0.3.  \rfree{Although bounded by $0.0 < E(B-V) < 0.5$, typically, the variance is high due to the distribution of dust geometries and unaccounted for emission lines leading to negative $E(B-V)$ color excesses for some best-fit models \cite{Steidel96}}.
In these respects, spectra of $z>3$ galaxies have a close similarity to local starbursts, with highly variable degrees of Ly-$\alpha$ emission or absorption, distinctive high-ionisation stellar emission lines of He{\sc ii}, C{\sc iv}, Si{\sc iv}, and N{\sc v} and strong interstellar absorption lines due to low-ionisation states of C, O, Si and Al (with rest-frame equivalent widths of order 2 to 3.5\AA) -- see Fig.~4 of ref.~\cite{Giavalisco02} and ref.~\cite{Shapley11}.
\begin{figure}[t]
  \begin{subfigure}{0.5\textwidth}
    \centering
    \includegraphics[width=\linewidth]{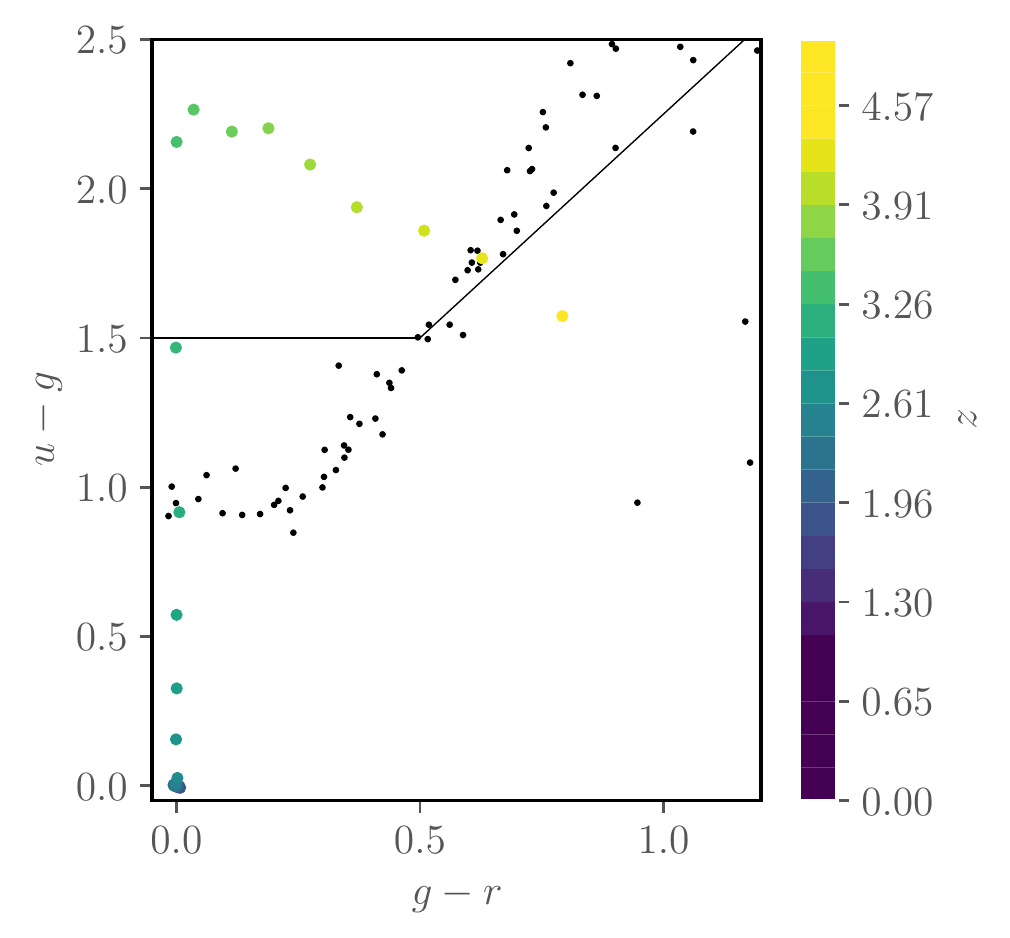}
  \end{subfigure}%
  \begin{subfigure}{0.5\textwidth}
    \centering
    \includegraphics[width=\linewidth]{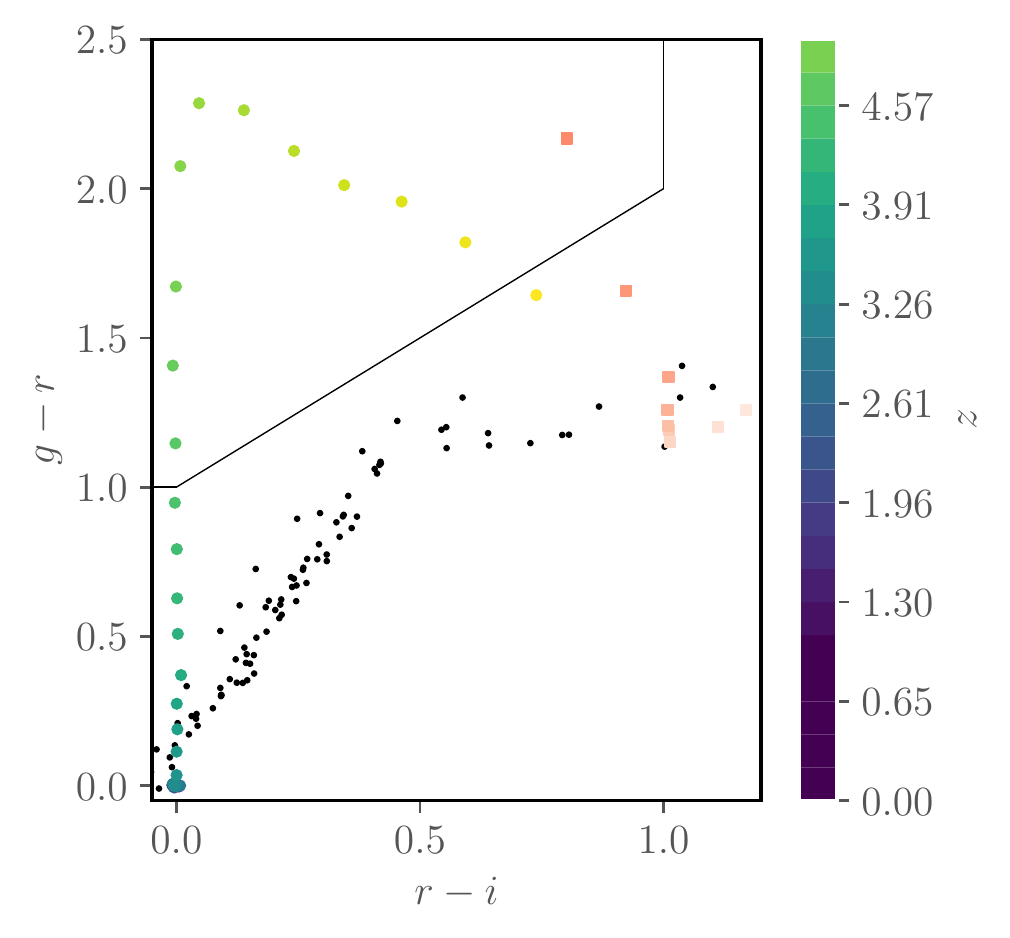}
  \end{subfigure}
  \caption{Color tracks for a $25^{\rm th}$ AB magnitude source with a $F_{\nu}$ spectrum assumed to be flat up to a total break at the rest-frame Lyman limit, $912$\AA.  \warn{A 26.5 imaging depth is assumed to determine the LBG colors}.  The CARS \cite{Hildebrandt09} $u$-dropout selection is shown (left), together with the Goldrush $g$-dropout selection\cite{Ono18} (right), with both the stellar locus (black dots) and low-$z$ ellipticals (red squares) also shown.  The colorbar indicates the corresponding LBG redshift.
}
\label{fig:ctrack}
\end{figure}

Figure~\ref{fig:ctrack} shows the realisation of this in the color-color spaces typically used for selection, with tracks for an idealised $25^{\rm th}$ AB magnitude source with an assumed flat $F_{\nu}$ spectrum up to a total Lyman break at 912\AA.  The loci show a $u$-dropout selection, as used by the CARS survey \cite{Hildebrandt09} (left black polygon):
\begin{equation}
    (u-g) > 1.5; 
    \quad -1.0 < (g-r) < 1.2; 
    \quad 1.5 \ (g-r) < (u-g) -0.75, 
\end{equation}
and the $g$-dropout selection of the Goldrush survey \cite{Ono18} (right black polygon):  
\begin{equation}
(g-r) > 1.0; 
\quad (r-i) < 1.0; 
\quad (g-r) > 1.5 \ (r-i) + 0.8,
\end{equation}
In both cases, a given LBG systematically tracks across color-color space in a manner that is effectively isolated with two colors given the distinctive SED.  The craft in designing such selection is ensuring limited encroachment by non-LBG sources, particularly at the reduced depths where the photometric scatter is larger.  The primary choice in this respect is the magnitude of the \rfree{drop} required, as reflected by the intercept with the ordinate.  Typical values range from 0.8 to 1.5 magnitudes, depending on the depth of the dropout bands available.  

From the figure, we can see the magnitude of the drop primarily affects the lower redshift tail and those with less intrinsic extinction.  The stellar library of ref.~\cite{Pickles98} shows increased contamination from the stellar locus (black dots) is a concern for a reduced break, particularly for $u$ dropouts at the high redshift end, but allows for increased number density given the restricted $u$-band depth of future datasets.  Both the redshift tails and completeness may be reduced in order to minimise this contamination, but this requires accurate colors and therefore sufficient depth.  For instance, $u$-dropout selection could be steepened for redder $g-r$ to reduce stellar contamination in this case.  Early-type (luminous red) galaxies at low redshifts (red dots) are a greater concern for $g$-dropouts, which seemingly receive very little contamination from stars; see also Fig.~1a of ref.~\cite{Steidel99}.  

Such contamination would be a concern for accurate cross-correlation studies, as low-redshift galaxies will correlate with the lensing map and this contamination would need to be accurately cleaned or modelled (see \S\ref{sec:interlopers}).  With sufficient frequency coverage for foreground dust removal, stellar contamination is less of a concern.  Particularly as it is the larger $L$ measurements which typically provide the most signal-to-noise, \fnl \ being the exception.  We explore the suitability of these selection boxes for our chosen science cases in \S\ref{sec:tailor}.

`True' dropout selection is limited to $z>3$, where the Lyman break redshifts into the optical.  At lower redshift, the break moves to the blue or UV (0.25 to $0.3\,\mu$m -- see Fig.~1 of ref.~\cite{Ly09}) and space-based missions are necessary given the strong atmospheric cut-off.  As a result, a further $UGR$ optical selection has been designed to simply assure the  absence of any break (Lyman, Balmer or 4000\AA) and a shallow $F_\nu$ spectrum.  This $z\simeq 2.3$ BX selection \cite{Adelberger04, Adelberger05}, 
\begin{alignat*}{2}
G - R &\geq −0.2, \qquad  &&U_n - G \geq G - R + 0.2 \nonumber \\
G - R &\leq 0.2(U_n - G) + 0.4, \qquad &&U_n - G \lt G - R + 1.0,
\end{alignat*}
has been designed to limit the potential contamination.  \rfree{Many successful studies have resulted}, e.g.\ refs.\ \cite{Sawicki06, Ly09, Reddy08, Hathi10}, but clearly may be improved by utilising all filters to ensure the absence of a break.  The lower redshift BM selection, $z \simeq 1.7$,  is distinguished by a bluer $U-G$ color due to the lesser IGM extinction.  In combination, BM/BX selection yields a $z\simeq 2$ sample with an angular density similar to QSOs for $g \simeq 22.5$, which rises exponentially for LBGs at fainter magnitudes and yields an angular density $30\times$  greater than QSOs alone -- see Fig.~1 of ref.~\cite{Lee14}.  \warn{Given the apparent lack of previous conversion to modern CCD photometric systems, we provide an approximate conversion of these colors to LSST-like filters in Appendix~\ref{app:filter_conversion}}.  Ref.~\cite{Ly09} performed followup spectroscopy based on GALEX/NUV imaging at $\simeq 27$ mag depth with LRIS on Keck, finding interloper contamination of 17\% with ground-based imaging -- where star removal based on a PSF-like profile is much less effective.  An additional $R > 23.5$ brightness cut has been shown to be effective in limiting this interloper contribution.

Alternatively, the availability of deep yet wide near-IR data would facilitate $BzK$ selection \cite{Daddi04}, designed to look for a flat spectrum redwards of the Lyman-break and a magnitude difference across the Balmer break, 
\begin{equation}
    BzK \equiv (z - K) - (B - z) > -0.2
\end{equation}
for star-forming (significant [OII] emission) galaxies at $z>1.4$, and 
\begin{equation}
    BzK < -0.2 \quad \& \quad (z - K) > 2.5,
\end{equation}
for passively evolving galaxies in the same range.  For surveys deeper than $\simeq 20^{\rm th}$ magnitude, the areal density can be several thousand per sq.\ deg., which make them an attractive population.  Unfortunately, spectroscopic followup would be challenging in this case.  Generally, $1 < z < 3$ galaxies can be identified in this manner due to the Balmer break at 3646\AA \ and the $4000\,$\AA \ break -- with an onset defined by Ca{\sc ii}, H and K lines and the bulk photospheric opacity derived from ionized metals \cite{Hamilton85, vanDokkum04, Hayashi07} -- while the $1.65\,\mu$m bump, at the minimum of the $H^{-}$ opacity, yields accurate identification of $0<z<1$ interlopers \cite{Sawicki02}; see ref.\ \cite{Buchs19}.  

Nearer term datasets are limited to the VISTA Hemisphere Survey, which covers the Southern Hemisphere to $K=18.5$, 
the Viking survey, 1500 deg$^2$ to $K \simeq 21.2$ \cite{Viking13} and VIDEO, 15 deg$^2$ to $K \simeq 23.5$\footnote{\url{https://www.eso.org/public/teles-instr/paranal-observatory/surveytelescopes/vista/}}.
  Euclid\footnote{\url{https://www.euclid-ec.org/?page_id=2581}} will also acquire $Y$, $J$ and $H$ $\simeq 24$ over 15K sq.~deg, with a more red sensitive $H$ filter than the norm; viz.\ central wavelengths of 1.085, 1.375 and 1.7725$\mu$m, see Fig.~2 of ref.~\cite{Inserra18}.  The corresponding deep fields total 40 deg$^2$ at up to two magnitudes deeper.  We shall return to the particularly promising combination of LSST and Euclid later.

\subsection{Expected redshift distributions}
\begin{figure}[t]
\centering
\includegraphics[width=\textwidth]{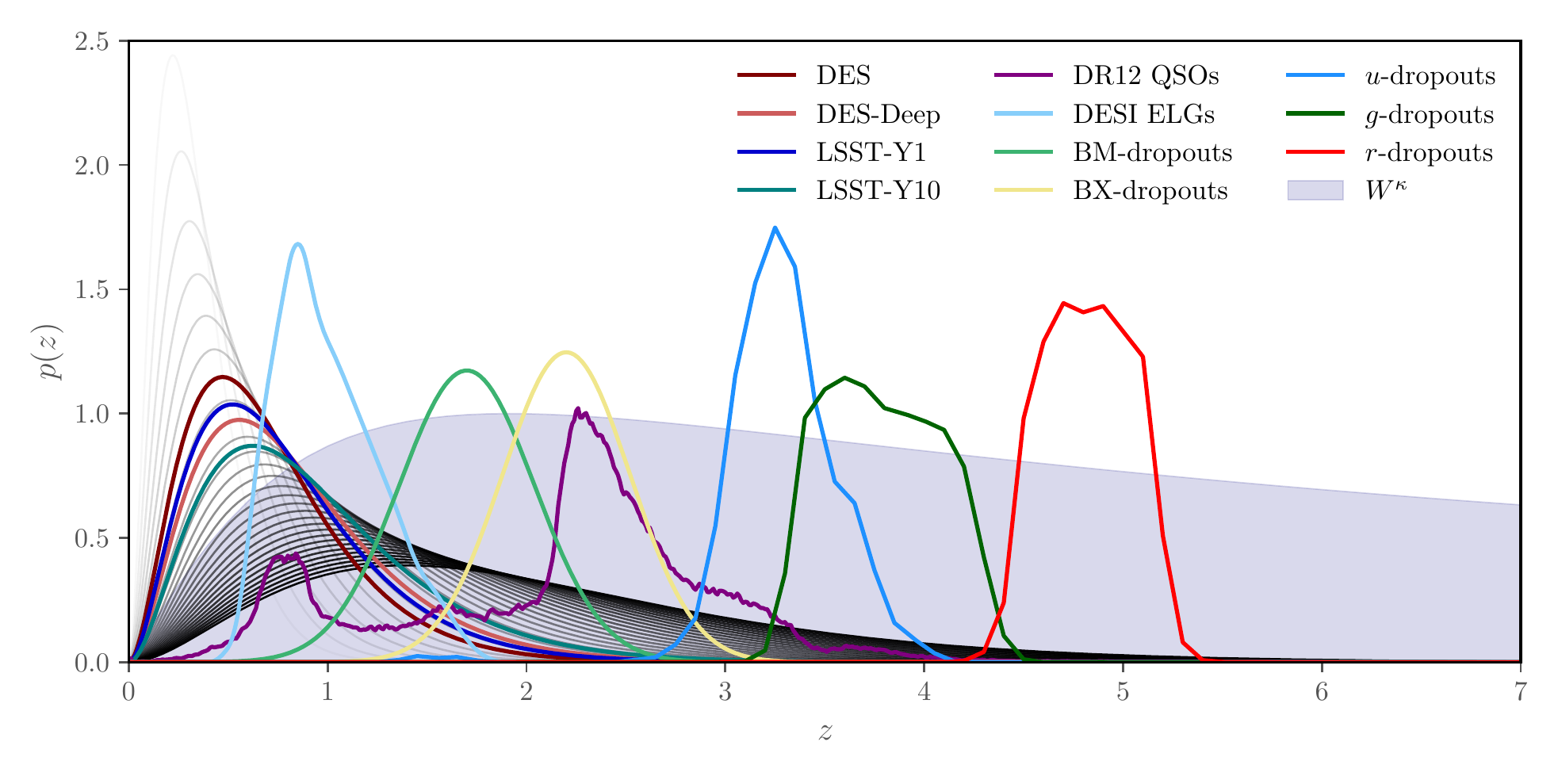}
\caption{The CMB lensing kernel ($W^{\kappa}$, shaded) and $p(z)$ distributions for established forms of photometric selection for high-$z$ galaxies.  Lyman-break dropout selection naturally yields samples centered on $z=1.70$, 2.20, 2.96, 3.8 and $4.9$ for BM, BX, $u$, $g$ and $r$-dropouts with FWHM of $0.75$, $0.80$ \cite{Steidel04, Reddy08}, $0.61$ \cite{Hildebrandt07}, $0.68$ and $0.80$ \cite{Ono18} respectively. 
This places the samples at significantly higher redshift than those of which large-scale structure studies are currently based, e.g. ELGs and QSOs \cite{DESI16}.  A common approximate form for magnitude limited samples \cite{LSST09,LSST18} is also shown, for depths between 20 and 35 in half magnitude steps.
}
\label{fig:pz}
\end{figure}
Fig.~\ref{fig:pz} shows that a tomographic decomposition of the galaxy populations across much of the CMB lensing kernel (shaded) is entirely possible with these known color-color selections.  Ideally, such samples would be non-overlapping step functions, but it is clear that the distributions achieved are each closer to a Gaussian due to the highly varying completeness in the tails.
In detail, this is sensitive to many determining factors: scattering out of the selection due to photometric errors, the intrinsic distribution of spectra due to the stochasticity of the Ly-$\alpha$ forest, intrinsic reddening by dust, line emission, etc.  Nevertheless, dropout selection represents a relatively clean form of redshift selection given the dominant central peaks at relevant depths.  For instance, a FWHM of $\simeq 0.6$ is achieved for both $u$ and $g$ dropouts, with a peak at $z\simeq 3$ and $4$ respectively.  This is to be contrasted with apparent magnitude selection (black lines), which yield samples heavily dominated by the $z<1$ Universe.  We explore how this picture can change due to interlopers in \S\ref{sec:interlopers}. 
  
\subsection{Schechter function estimates of the volume and angular number density}
Existing surveys have constrained the UV luminosity function over a broad range of redshifts.  We follow standard practice in the definition of the Schechter function \cite{Schechter76, Ono18} with absolute magnitude:
\begin{equation}
    \Phi(M_{\rm{UV}}) \ dM_{\rm{UV}} = \left ( \frac{\ln 10}{2.5} \right ) \phi^\star \ 10^{-0.4 (1 + \alpha) \left ( M_{\rm{UV}} - M_{\rm{UV}}^\star \right ) } \exp\left( -10^{-0.4 \left ( M_{\rm{UV}} - M_{\rm{UV}}^\star \right ) } \right) \ dM_{\rm{UV}},
\label{eqn:SchechterForm}
\end{equation}
where $(L / L_\star) = 10^{-0.4(M_{\rm UV} - M^\star_{\rm UV})}$ and convert to this form where necessary.  A collection of current best-fit parameters with redshift is provided in Table \ref{Table:Schechter}.  In converting from the published numbers to those in Table \ref{Table:Schechter} we have used the authors' value of $h$ to change units to $h^{-1}$Mpc, but otherwise made no corrections for the differences in fiducial cosmology.  We find the change in the distance modulus and differential volume element between the cosmologies listed and our fiducial cosmology are smaller than, or comparable to, the quoted errors on $M_\star$ and $\Phi_\star$.
We further assume a $k$-correction appropriate for the shallow $F_\nu$ spectrum of LBGs, i.e.
\begin{equation}
  M_{UV} = m  -5\log_{10}\left(\frac{D_L(z)}{10{\rm  pc}}\right) + 2.5\log_{10}(1+z)
  + \underbrace{m_{UV}-m}_{\approx 0},
\end{equation}
with $D_L(z)$ the luminosity distance.  The final SED $k$-correction is practically zero if the detection band is chosen to match the redshifted rest-frame $UV \approx 1500$\AA \, as is commonly the case; see e.g.~Fig.~3 of ref.~\cite{Sawicki06}.  As a result, the detection-band magnitude is a good proxy for the rest-frame UV luminosity.  Conveniently, $m_{UV}$ at $1500\,$\AA\ matches the peak of the $g$, $r$, $i$ and $z$ pass bands of LSST at $z\simeq 2$, 3, 4 and 5 respectively.  \warn{Note that a Schechter function typically under-predicts the number of bright galaxies in these samples \cite{Ono18}, but the difference is insignificant for our science cases}.

\begin{figure}[t]
\begin{center}
\resizebox{\columnwidth}{!}{\includegraphics{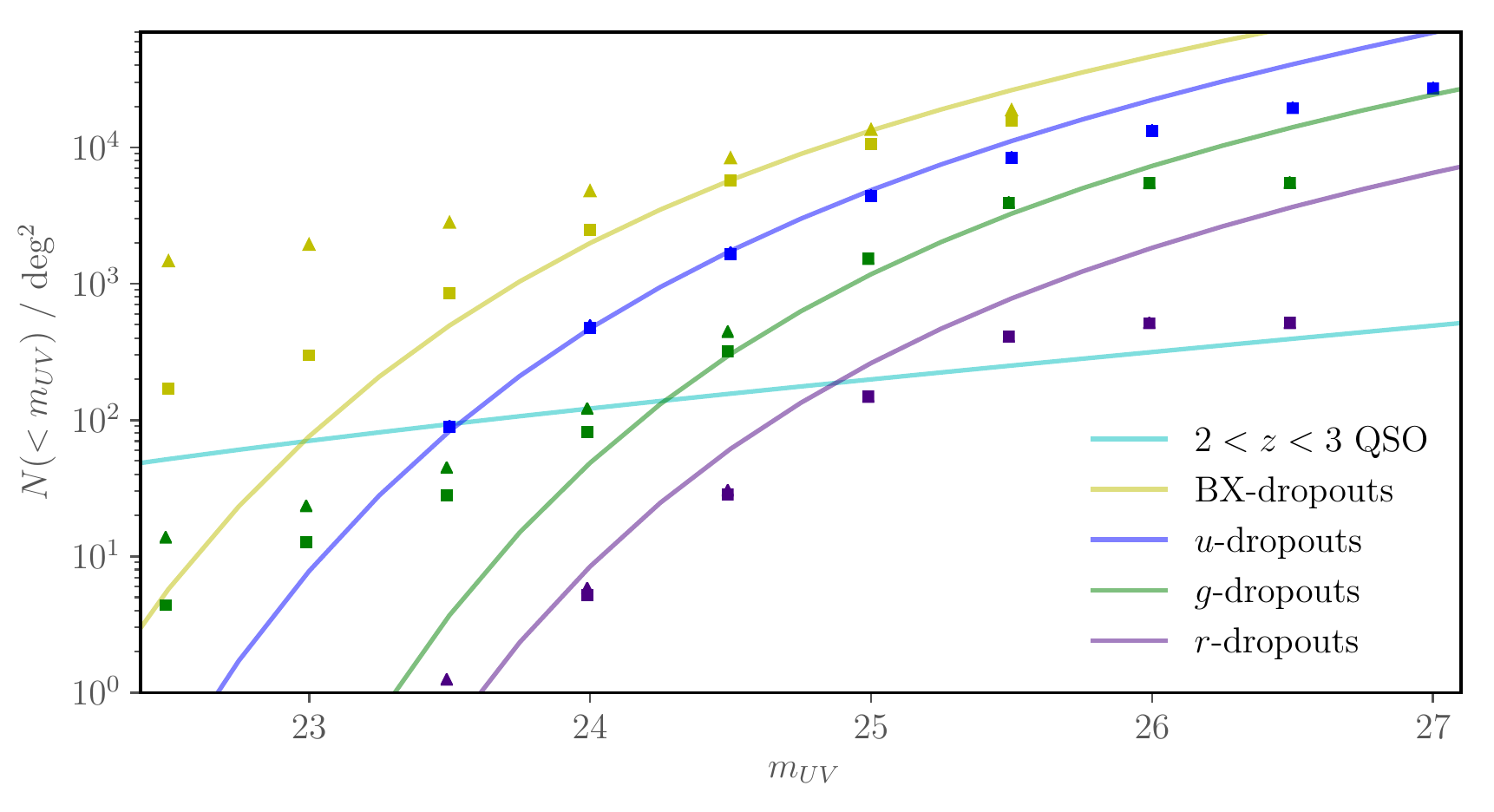}}
\end{center}
\caption{The cumulative angular number density of LBGs and quasars with detection band magnitude, $m_{UV}$ -- typically taken to be that closest to the redshifted rest-frame UV ($\simeq 1500$~\AA).  We take the completeness curves of refs. \cite{Malkan17, Ono18} and use Fig.~1 of ref. \cite{Red08} to derive $p(z)$ in each case.  Triangles indicate the raw counts at the quoted magnitude limits from refs.\ \cite{Reddy08, Malkan17, Ono18}. From the contamination corrected counts (squares), it is clear that the fractional contamination largely occurs at the bright end.  The suppression at the faint end is due to the known incompleteness, as corrected for in the luminosity function estimates.
}
\label{fig:SchCounts}
\end{figure}

Fig.\ \ref{fig:SchCounts} shows estimates of the expected angular number density for BX, $u$, $g$ and $r$-dropout samples \cite{Malkan17, Ono18}.  With respect to limiting magnitudes, of particular interest for their combination of area and depth are DES, HSC and LSST, as shown in Table~\ref{tab:depths}.  Given the ~25.6 and 26.84 $g$ band depth of LSST-Y1 and LSST-Y10 \cite{LSST18}, we expect $\simeq 3$K and $30$K $g$-dropouts per sq.\ deg.\ respectively.  \warn{This Schecter estimate assumes the completeness is simply derived from the apparent magnitude, rather than having a significant dependence on intrinsic extinction etc., as is the case for dropout selection.  The observed counts of realised surveys (markers) include this effect and thereby represent the actual density on the sky, including interlopers at the bright end.  See refs.~\cite{Red08, Malkan17, Ono18} for detailed blueprints of how these contamination rates are derived, e.g.\ fig.~6 of ref.~\cite{Yoshida08} which suggests the interloper contamination of $u$-dropouts is at most $\simeq 6$\% below 24.2 magnitude and 1\% at greater depth -- a difference smaller than the marker size}.

\begin{table}
\begin{center}
\begin{tabular}{||c | c | c | c | c | c ||} 
 \hline
  $z_{\rm{eff}}$ & $M^*_{\rm{UV}}$ & $\phi^* / \left ( 10\,h^{-1}{\rm Mpc} \right)^{-3}$ & $\alpha$ & $m_{UV}^\star$ & Ref. \\
 \hline
  \hline
  2.0 & -20.60 & 9.70 & -1.60 & 24.2 & \cite{Red08} \\
  \hline
  3.0 & -20.86 & 5.04 & -1.78  & 24.7 & \cite{Malkan17} \\ 
  \hline
  3.8 & -20.63 & 9.25 & -1.57 & 25.4 & \cite{Ono18} \\ 
  \hline
  4.9 & -20.96 & 3.22 & -1.60 & 25.5 & \cite{Ono18} \\     
  \hline
  5.9 & -20.91 & 1.64 & -1.87 & 25.8 & \cite{Ono18} \\
 \hline
\end{tabular}
\end{center}
\caption{Best-fit parameters for a Schechter function form (Eqn.~\ref{eqn:SchechterForm}) to the UV luminosity function, and the apparent magnitude, $m_{UV}^\star$, of an $M_\star$ galaxy, as a function of redshift.  Note the expected trend to brighter magnitudes at higher $z$, but largely the selection picks out similar galaxies at each epoch.  The spatial density shows a stronger trend with $z$ -- see refs.~\cite{Bouwens15, Bowler17, Schenker13a}.  At any redshift, the counts increase rapidly until $m_{UV}^\star$ and then flatten.
}
\label{Table:Schechter}
\end{table}

\subsection{Inferred clustering}
\label{sec:clustering}
Dropout selection in magnitude limited surveys naturally selects massive, star-forming, galaxies whose correlation function remains consistent with a power law on megaparsec scales at current precision, with an amplitude and slope comparable to low-$z$ bright spiral galaxies.  \rfree{The large bias implied by the measured clustering and abundance has important implications for the use of these samples as tracers of the matter field.  Highly biased objects exhibit a greater scale-dependent bias and their density field decorrelates with the matter field on larger scales than less biased objects; modelling such populations thus requires some care.  This need not be a blocking factor, given sufficient preparatory work with simulations etc., but may impact the precision of the inferred cosmological constraints.  Certainly, non-linear effects should not be ignored on the basis of the linearity of the density field, see \S \ref{sec:nonlinear-bias}.}  \warn{This high bias further suppresses the anisotropy of the redshift-space clustering and therefore provides a less sensitive test of gravity \cite{Weinberg13}}; we briefly explore the ramifications of this in \S \ref{sec:rsd}.  In the following, we first consider the dependence of the linear bias on redshift and apparent magnitude limit, followed by the scale dependence of angular and spatial LBG clustering.

\subsubsection{Depth and redshift dependence of the linear bias}
\begin{figure}[t]
    \centering
    \begin{subfigure}{0.5\textwidth}
    \includegraphics[width=\linewidth]{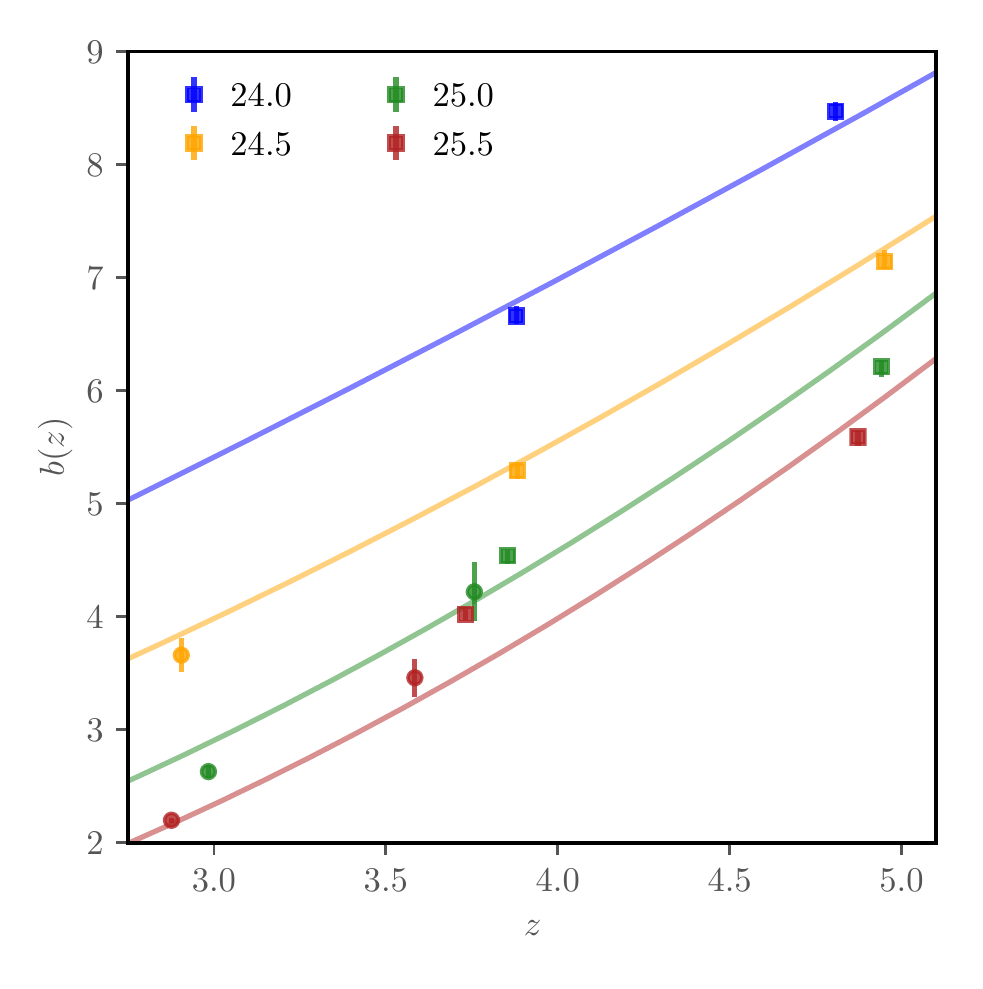}
    \end{subfigure}%
    \begin{subfigure}{0.5\textwidth}
    
    \includegraphics[width=\linewidth]{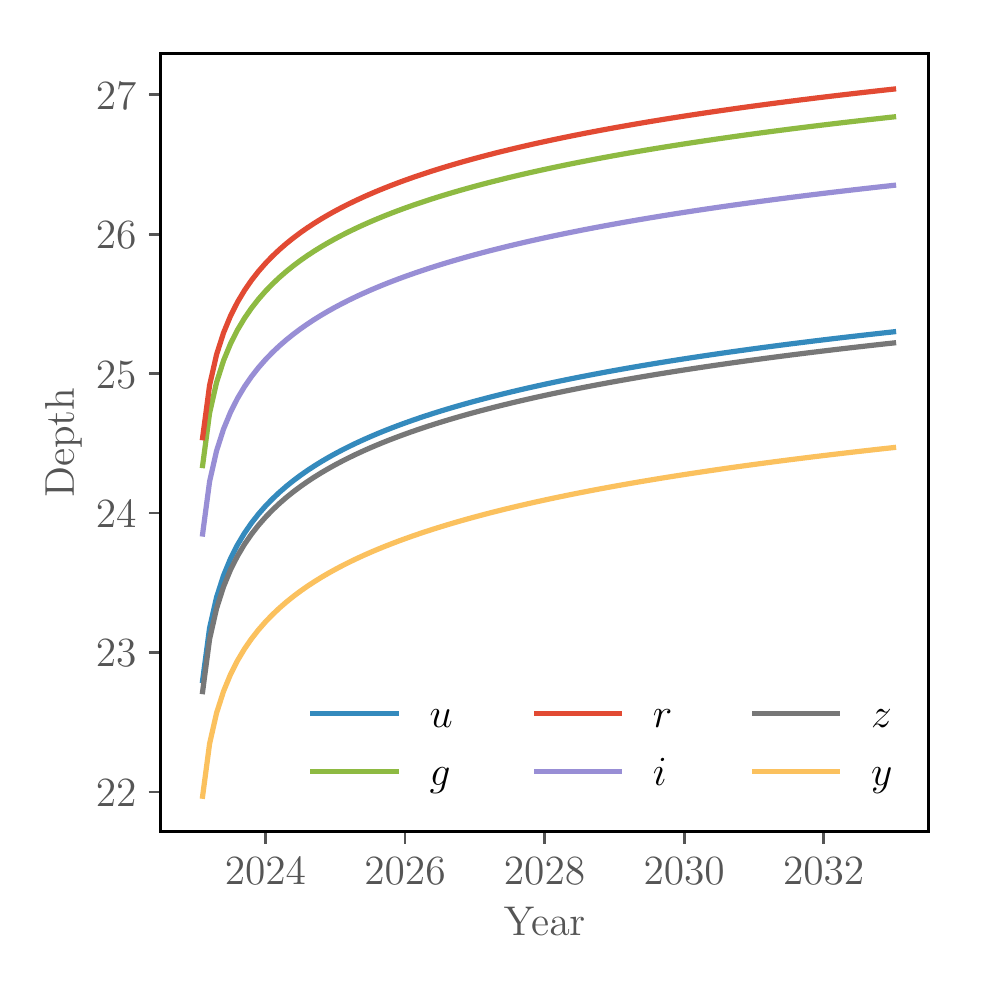}
    \end{subfigure}%
    \caption{\textbf{Left:} Redshift and apparent magnitude dependence of the linear galaxy bias for dropout selected galaxies from CARS (squares; Table 3 of ref. \cite{Hildebrandt09}) and GOLDRUSH (circles; Table 4 of ref. \cite{Ono18}).  Eqn.~\ref{eqn:bz} provides a simple compression necessary for interpolation (solid).  \textbf{Right:} Achieved depths after every year of the fiducial LSST program, assuming first light in 2023.   
    }
    \label{fig:bz}
\end{figure}

Fig.~\ref{fig:bz} shows a preliminary compilation of linear galaxy bias measurements \cite{Hildebrandt09, Harikane17}, to which we have fit a simple model that captures the basic evolutionary trends.  We assume a low-order polynomial in $(1+z)$ given that the growth factor $D_+ \propto a$ at these redshifts:
\begin{equation}
    b(z, m) = A(m)(1+z) + B(m)(1+z)^2,
\label{eqn:bz}
\end{equation}
finding $A(m) = -0.98 \ (m - 25) \ + \ 0.11$ and $B(m) = 0.12 \ (m - 25) \ + \ 0.17$.  The first term describes `stable' clustering \cite{Fry96, Jose17}, i.e.\ $b D_{+}$ constant, while the second coefficient captures the bias rising more steeply at high redshift -- almost certainly due to the apparent magnitude limits. This is certainly not a rigorous model, 
but does a satisfactory job for interpolation.  We show these fits as the solid lines in Fig.~\ref{fig:bz} and apply them to forecasts of the science return for our fiducial LBG samples in e.g.\ \S \ref{sec:ckg_like}.

\warn{For comparison, we show in Fig.~\ref{fig:bz} the expected depths achieved by LSST after each year of the survey, assuming first light in 2023.  For this, we assume the conservative depths quoted by DESC \cite{LSST18} and replicated in Table \ref{tab:depths}.  The raw imaging will likely be much deeper ($\simeq 27^{\rm{th}}$ magnitude in all bands), but stringent systematic requirements on e.g.\ inhomogeneity will impose something more modest.  We see that 2030 is sufficient for a $24^{\rm{th}}$ magnitude limited $u$-dropout sample, a $25.5^{\rm{th}}$ magnitude limited $g$-dropout sample and a $>25^{\rm{th}}$ magnitude limited $r$-dropout sample, assuming a year for data processing.  Given the relatively low sky background, the LSST $u$ band is read-noise limited and hence a linear gain with longer exposure times would be expected; this would greatly help limit the expected interlopers.  Unfortunately, the \rfree{repurposing} of visits from the $z$ and $y$ filters (of lesser importance for our purposes) would be difficult given these are acquired in twilight, during which the background is increased by a relatively blue, attenuated solar spectrum and hence a much reduced single-visit $u$ depth.}

\subsubsection{Expected scale-dependent bias}
\label{sec:nonlinear-bias}
\begin{figure}
    \centering
    \includegraphics[width=\linewidth]{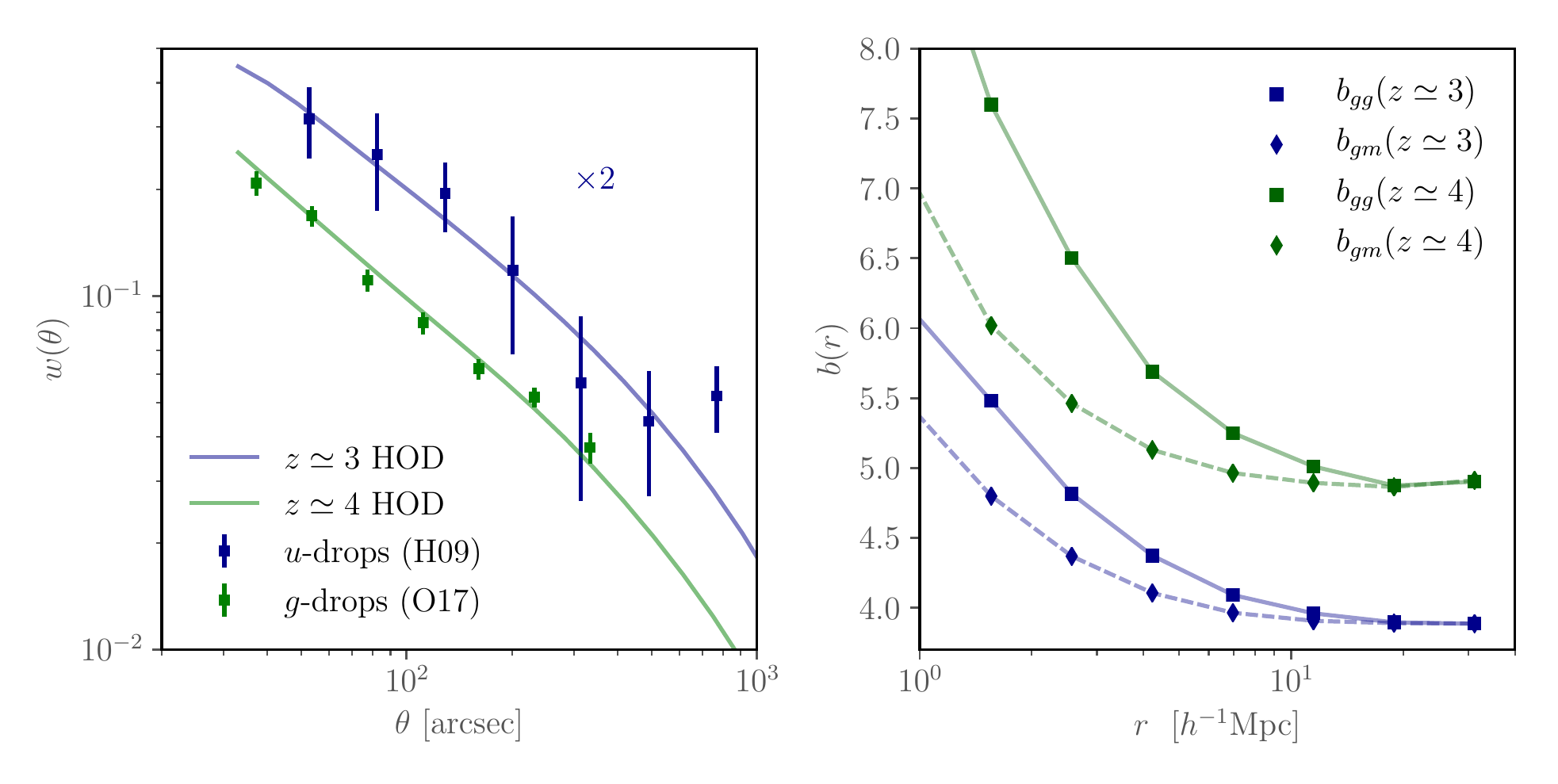}
    \caption{\textbf{Left:}  The observed angular correlation function, $w(\theta)$, of $m_{UV} < 24.5$, $u$ and $g$-dropout galaxies \cite{Hildebrandt09, Omori17} (points with errorbars, rescaled by $\times 2$ for $u$-dropouts).  \warn{Overplotted are model lines obtained by abundance matching halos found in a N-body simulation}.  At these redshifts, the data allow a comparison to be made for $\simeq 1-10\,h^{-1}$Mpc, i.e.~two-halo scales.  \textbf{Right:}  The corresponding bias of mock galaxies in configuration space.  Squares show estimates derived from the auto-spectra ($\sqrt{\xi_{gg}/\xi_{mm}}$), while diamonds show that from the cross-spectra ($\xi_{gm}/\xi_{mm}$).  Both are noticeably scale dependent below $10\,h^{-1}$Mpc.  The inconsistency of the two measures further indicates decorrelation between the galaxies and (non-linear) matter.
    }
    \label{fig:wt}
\end{figure}
\begin{figure}
    \centering
    \begin{subfigure}{0.5\textwidth}
    \includegraphics[width=\linewidth]{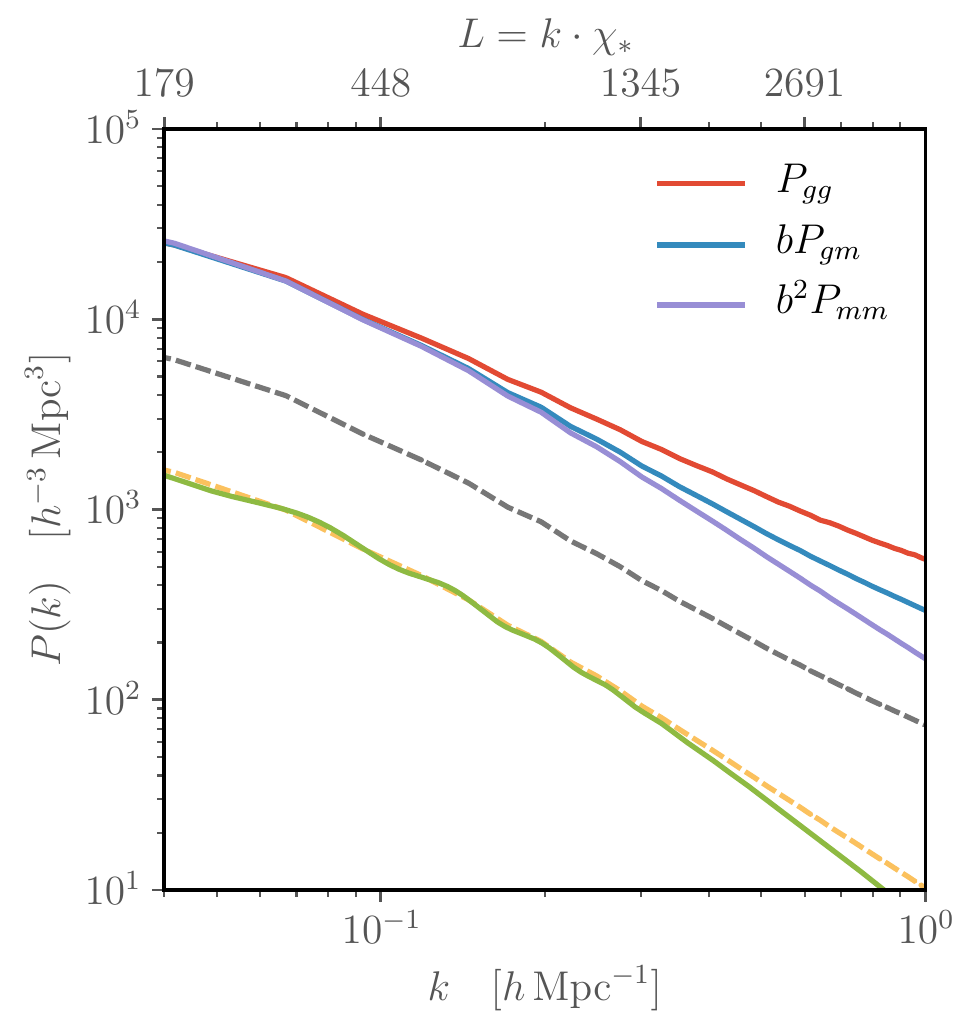}
    \end{subfigure}%
    \begin{subfigure}{0.5\textwidth}
    \includegraphics[width=\linewidth]{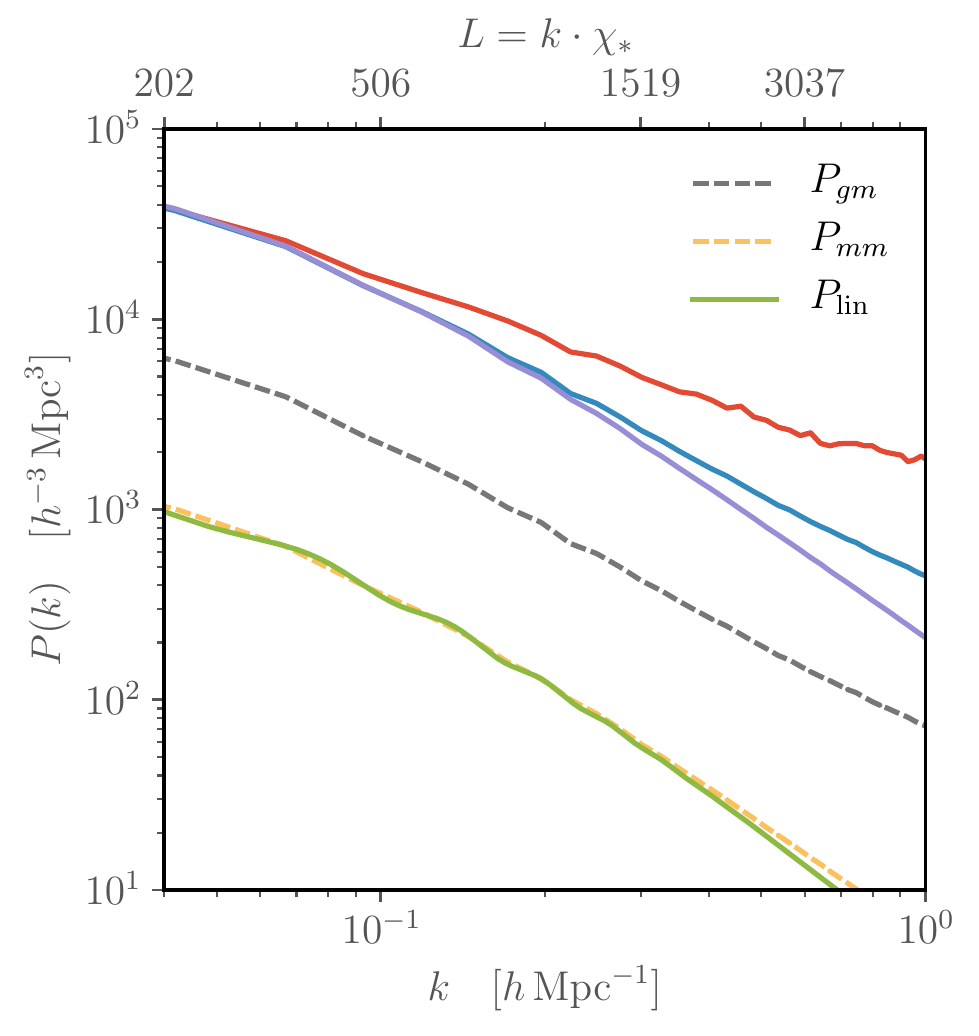}
    \end{subfigure}%
    \caption{A comparison of the real-space auto-power spectra for a simulation of abundance matched dropout galaxies with the same angular clustering as shown in Fig.~\ref{fig:wt}, with the dark matter auto-clustering and the cross-clustering between dark matter and galaxies.  The left panel corresponds to $u$-dropouts at $z\simeq 3$, while the right panel is for $g$-dropouts at $z\simeq 4$.
    }
    \label{fig:pk}
\end{figure}
Fig.~\ref{fig:wt} shows the angular clustering of $u$ and $g$-dropout galaxies with $m_{UV}<24.5$ from refs.~\cite{Hildebrandt09} and \cite{Omori17} respectively;  see also refs.\ \cite{Steidel03, Adelberger05, Ouchi05, Lee06, Bian13, Bielby13, Park16, Ishikawa17}.  For comparison, the $w(\theta)$ lines show predictions obtained by 
populating halos in N-body simulations with mock galaxies of the same abundance, measuring their $\xi(r)$ and converting to $w(\theta)$ with the Limber approximation and measured redshift distribution.  The angular scales shown emphasise the `two-halo' regime, i.e.~pairs comprised of two galaxies in separate halos.  The right hand panel shows bias estimates of the auto ($\sqrt{\xi_{gg}/\xi_{mm}}$) and cross-spectra ($\xi_{gm}/\xi_{mm}$) relative to the non-linear matter clustering.  This is noticeably scale dependent where the majority of constraining power lies, below $10\,h^{-1}$Mpc.  Note that this non-linear bias raises the signal significantly above that assumed in forecasts to date \cite{Schmittfull18, Yu18}.  The inconsistency between the auto and cross estimated biases provides further evidence that the galaxy and matter fields are significantly decorrelating on these scales.

In Fourier space, the bias also exhibits a strong scale dependence by $k\simeq 0.1\,h\,{\rm Mpc}^{-1}$, as can be seen for the real-space, galaxy-galaxy and galaxy-mass power spectra in Fig.~\ref{fig:pk}; a linear and non-linear matter power spectrum with a scale-independent bias is shown for comparison.
The difference in amplitude and shape of the galaxy clustering compared to the matter imply a high bias and significant scale dependence, in what is traditionally thought to be a linear regime.  Despite this scale dependence, these high-$z$ populations reside where the density field remains linear to smaller scales than in the local Universe (Figs.~\ref{fig:ang2spatial} and \ref{fig:pk}).  This opens up the possibility of modelling the observed field with some accuracy, given a suitable formalism \cite{Modi17a, Schmittfull18b, Castorina19}.

\subsection{Tailoring the color selection}
\label{sec:tailor}
To date, typical LBG studies have focused on comparatively small area -- by \rfree{large-scale structure} standards -- estimates of the UV luminosity function or HOD determination, the GOLDRUSH survey \cite{Ono18} serving as a recent counterexample.  The science case we consider is sufficiently different that one might tailor the color selection applied to large-scale structure studies, \rfree{i.e.} \ as the sensitivity to interloper contamination and completeness may be very different for CMB cross-correlation as compared to studies of e.g.\ the faint-end slope of the luminosity function.  Even without this optimisation, those previously deployed may not be suitable given that the imaging to which they were designed may be much deeper than that likely available to much larger area surveys and photometric scatter is a determining property.  Similarly, one may even consider defining cross-correlation only catalogues, e.g.\ with lower significance detection requirements and hence greater number density -- given that spurious detections will not correlate.  

Interloper contaminants should therefore be understood in the context of the known sources of systematic error to CMB lensing convergence maps:  low S/N point sources, residual fluctuations from the thermal and (frequency-independent) kinetic Sunyaev Zeldovich effect and contamination by the Cosmic Infrared Background (CIB).  \warn{CIB contamination may be challenging for other reasons, given it is largely sourced by the LBGs we consider}.  One is free to also take alternative steps, e.g.\ removing low-$z$ $\kappa$ fluctuations with already acquired galaxy samples, as we explore in \S\ref{sec:lowz_cleaning}, shear-dilation $\kappa$ estimators \cite{Schaan18b}, or simply applying strict $L$ cuts given the known contributions to $\kappa(L)$ from each redshift. 

With respect to color selection, our approach is to take small-area deep samples to first understand the selection applied, degrade these catalogues to the depths likely available to cosmology and then consider their effectiveness in this regime.  \warn{Clearly, this is preparatory work for much more in-depth studies to come}.  For degradation, we follow \S 3.2.4 of the Photo-$z$ Accuracy Testing studies (PHAT, \cite{Hildebrandt10}) in defining a magnitude error that behaves as a power law at bright magnitudes with an exponential cut-off beyond the effective depth.  As per Fig.~3 of this reference, we saturate the S/N at 100 for bright objects to replicate e.g.\ CCD saturation, blooming or a strongly non-linear detector response.  We choose effective depths, $m_\star$,  that are likely in the late 2020s, i.e.\ LSST-Y10 filters together with Euclid-like $Y,J,H = 24$, and otherwise retain the same fiducial values.  \rfree{We further explore $H$ dependent selection explicitly, assuming Euclid-like filters}.  As the majority of sources that we degrade are much fainter than these depths, we apply $5\sigma$ detection limits in an asinh (AB) magnitude system \cite{Lupton99} that is well defined for the negative and low signal-to-noise fluxes that result, quoting the respective colors; the asinh softening is $1.042\times$ the (PHAT) flux error.

For a well understood deep sample, we first degrade the 12 arcmin$^2$ Hubble UV-UDF survey (R15, \cite{Rafelski15}), 9960 galaxies with relatively secure EAZY photometric redshifts \cite{Brammer08} due to the eleven band photometry spanning both the UV and NIR ($0.2 \leq \lambda \leq 1.8 \mu$m) at depths $\gt 27.8$ magnitude; see Fig.~2 of R15 for the filter transmissions.  For the purposes of this exercise we treat these photometric redshifts as the truth as the coverage enables simultaneous detections of both the Lyman and Balmer breaks in the redshift interval $0.8 \leq z \leq 3.4$ and at least one break visible for the entire redshift range.  For this reason photometric redshift scatter is small ($\sigma_z = 0.035$)
based upon a spectroscopic sample of 169 galaxies, see \S 4.2.1 and Fig.~6 of R15, with an outlier fraction approaching 6\% -- largely due to blending close to the magnitude limit.  Photometric templates assumed by EAZY include emission lines based on estimated star formation rates and very dusty low-$z$ galaxies, together with a better accounting of the theoretical uncertainty associated to each template, particularly for $> 2 \mu$m \cite{Hildebrandt10, Wittman16}.  For these reasons, EAZY was one of the best performers in detailed comparisons \cite{Hildebrandt10, Dahlen13}; for $z \simeq 2$, there is generally agreement between alternative codes \cite{Dahlen13}.  \warn{However, our interest largely resides in the characteristics of low-$z$ interlopers with color selection, in which case a prediction of greater numbers of low-$z$ Balmer-break galaxies would be a concern -- as is the case for BPZ (v1.0, \cite{Benitez09}) -- and warrant further investigation}.  
\begin{figure}
  \begin{subfigure}{0.5\textwidth}
    \centering
    \includegraphics[width=\linewidth]{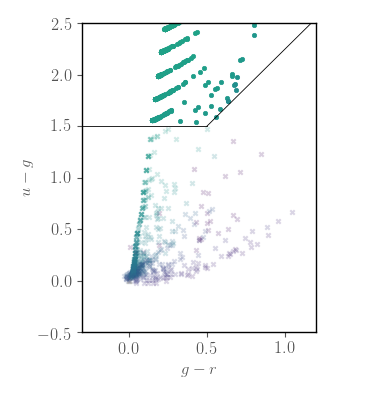}
  \end{subfigure}
  \begin{subfigure}{0.5\textwidth}
    \centering
    \includegraphics[width=\linewidth]{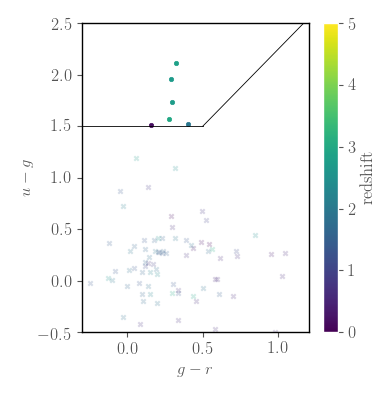}
  \end{subfigure}
  \begin{subfigure}{0.5\textwidth}
    \centering
    \includegraphics[width=\linewidth]{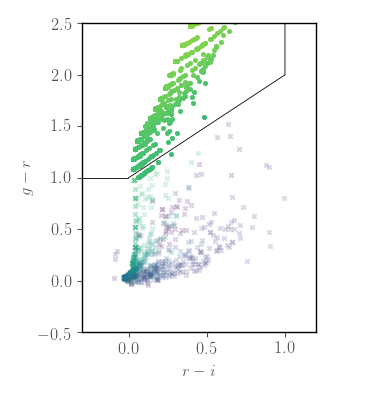}
  \end{subfigure}
  \begin{subfigure}{0.5\textwidth}
    \centering
    \includegraphics[width=\linewidth]{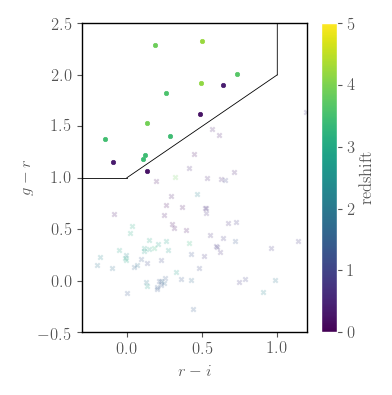}
  \end{subfigure}
  \caption{Distribution of UV-UDF \cite{Rafelski15, Skelton14, Momcheva16} photometric redshifts in dropout color-color space for the complete (left) and achieveable wide-area depths (right).  The top row shows $u$-dropouts \cite{Hildebrandt09} and the bottom row $g$-dropouts \cite{Ono18} with their selection boxes.  We require a 5$\sigma$ detection in the appropriate band for the degraded depths and further subsample the non-dropout objects for clarity.  Fig. \ref{fig:SchCounts} suggest of order 1 and 15 $u$ and $g$-dropouts at the degraded depths for this $\simeq 10 $ arcmin$^2$ sample, consistent with what we see here.
}
\label{fig:cat_ctrack}
\end{figure}
\begin{figure}[t]
  \begin{subfigure}{0.5\textwidth}
    \centering
    \includegraphics[width=\linewidth]{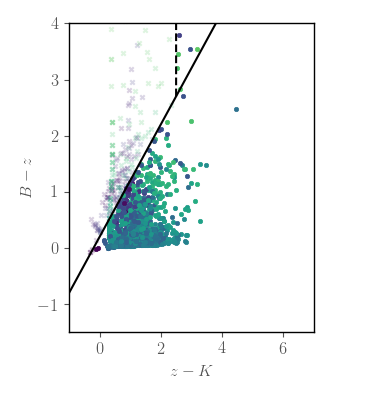}
  \end{subfigure}
  \begin{subfigure}{0.5\textwidth}
    \centering
    \includegraphics[width=\linewidth]{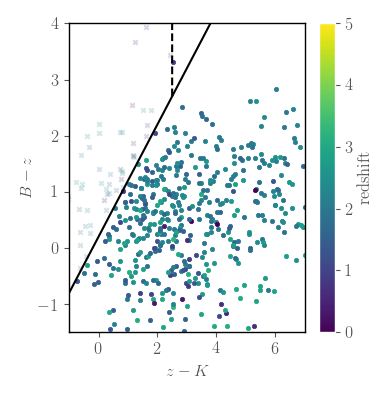}
  \end{subfigure}
  \begin{subfigure}{0.5\textwidth}
    \centering
    \includegraphics[width=\linewidth]{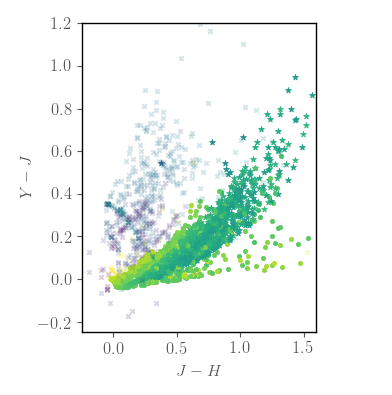}
  \end{subfigure}      
  \begin{subfigure}{0.5\textwidth}
    \centering
    \includegraphics[width=\linewidth]{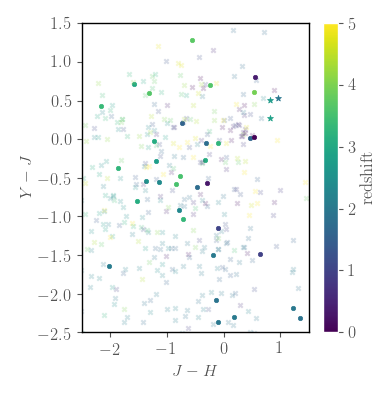}
  \end{subfigure} 
  \caption{Similar format to Fig. \ref{fig:cat_ctrack}, but for near-IR $BzK$ selection (top, \cite{Daddi04}) and the available Euclid colors (bottom).  For the latter, we mark objects passing $u$ or $g$ selection (stars and dots respectively).  These seem to be well isolated by an additional Euclid $(J-H)$ cut, despite the large photometric scatter of these bands.} 
\label{fig:cat_ctrackNIR}
\end{figure}

Fig.~\ref{fig:cat_ctrack} shows that $u$ and $g$ dropouts are well delineated from low-$z$ sources by traditional color selections at full depth, \warn{confirming our motivation for color rather than photometric redshift selection}.  However, clearly significant interloper contamination is to be expected for the depths relevant to wide-field cosmology surveys.  Where interlopers are present, their redshift is consistent with Balmer or $4000\,$\AA \ break confusion, namely $z \simeq 0.8$ and 1.0 for $u$ and $g$ dropouts respectively 
-- as $(1 + z_{\rm{int}}) = \left ( \lambda_{\rm{Lyman}} / \lambda_{\rm{Balmer}} \right )(1  + z_{\rm{LBG}} ) \simeq 0.2 \left ( 1 + z_{\rm{LBG}}\right )$.  At the degraded depths, there are very few genuine LBGs due to the small area of the UV-UDF survey, in line with Fig.~\ref{fig:SchCounts}.  For this reason, we appeal to the less secure redshifts of the CARS survey for estimating the resulting redshift distribution in \S \ref{sec:interlopers}.    

\warn{Fig.~\ref{fig:cat_ctrackNIR} shows that $BzK$ selection is also ineffective at isolating high-$z$ dropouts at relevant depths, perhaps unsurprisingly given the $z>1.4$ remit of this selection.  Given the deficit of planned wide-area, deep $K$ band imaging in the future this is perhaps not much of a concern}.  In contrast, the Euclid near-IR filters are likely to be a robust test for a Balmer break to $z=4$.  Unfortunately, as these targets are also faint, the photometric scatter significantly reduces the efficiency of a clean selection, but there is clearly some room for more efficient separation on the basis of Fig. \ref{fig:cat_ctrackNIR}. 

\subsection{Interloper redshift distribution}
\label{sec:interlopers}
\begin{figure}[t]
    \centering
    \begin{subfigure}{1.0\textwidth}
    \includegraphics[width=\linewidth]{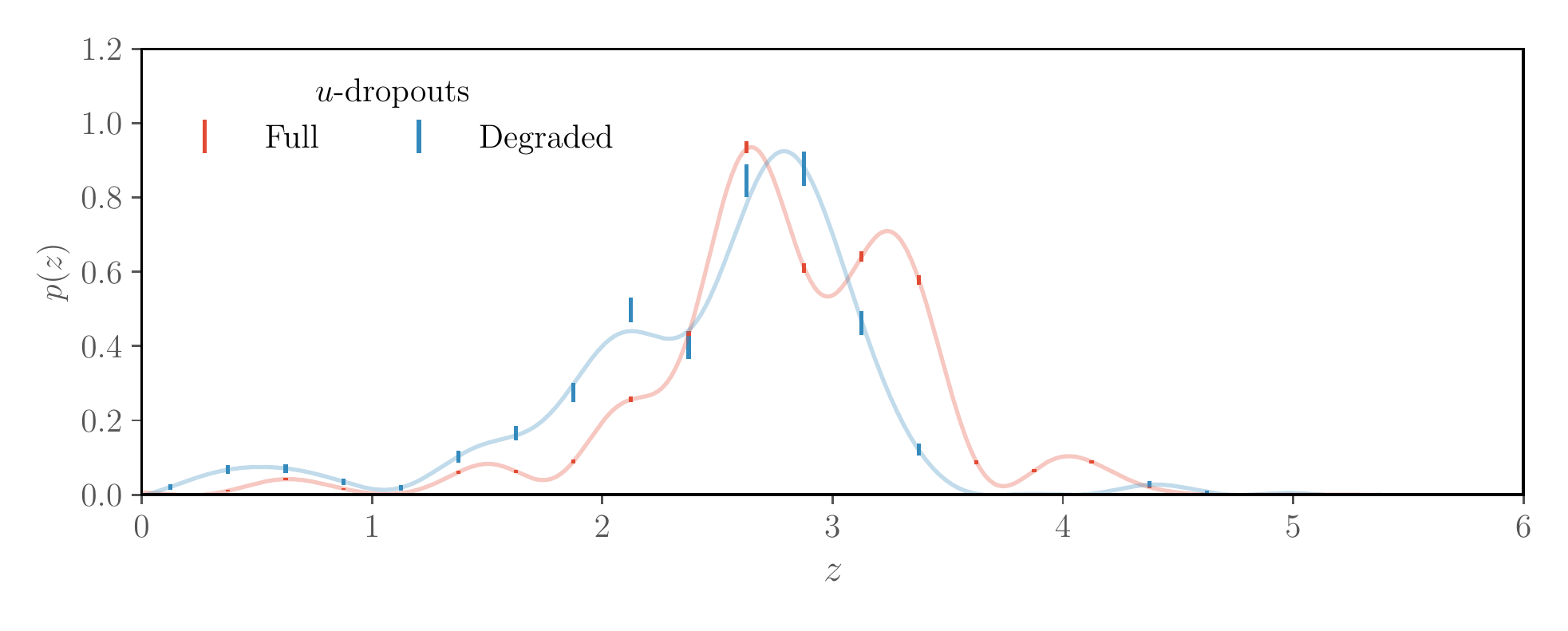}
    \end{subfigure}
    \begin{subfigure}{1.0\textwidth}
    \includegraphics[width=\linewidth]{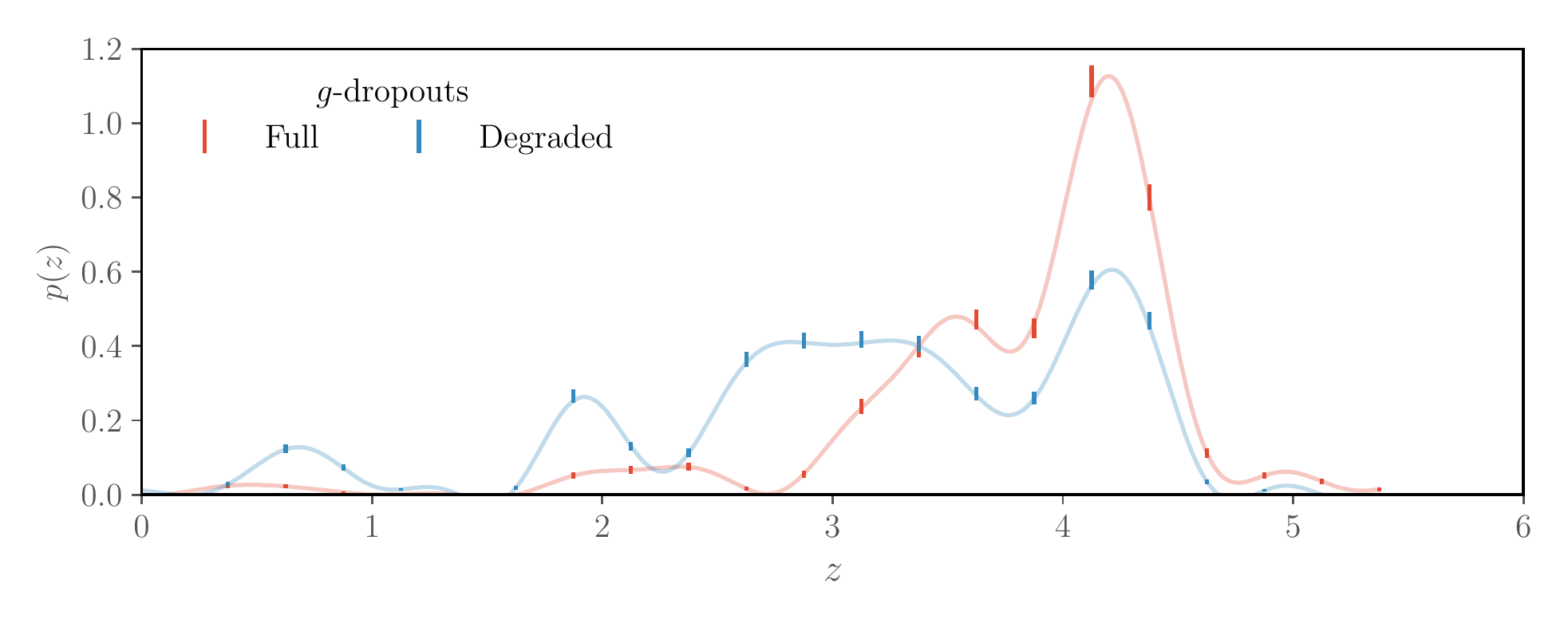}
    \end{subfigure}
    \caption{Estimates of the realised redshift distribution of $u$ (top) and $g$ (bottom) dropout samples using the CARS catalogue \cite{Hildebrandt09}.  Of those objects bright enough to be detected in the degraded catalogue, we explore their redshift distribution at both full-depth and at LSST DESC depths assuming a PHAT model for the magnitude error \cite{Hildebrandt10} and the Hyper-Z photometric redshifts to be the accurate.  Overlaid curves (translucent) indicate the best-fitting Gaussian process approximation.   
    }
    \label{fig:int_nz}
\end{figure}
Having considered the potential for low-$z$ contamination at the depths of LSST and Euclid, we now turn to quantitative estimates of the interloper redshift distributions;  this determines e.g.\ the expected CMB lensing cross-correlation signal.  

Application of color selection to the multi-modal distribution of galaxy SED types can lead to an involved form for the redshift distribution.  Estimates are seemingly rare and can differ in approach; e.g.\ ref.~\cite{Oesch07} proceeded with a template library and assumed $z \simeq 3$ luminosity function, but this determines only the LBG distribution, not interlopers; ref.~\cite{Vulcani17} used the photometric redshifts of the 3D-HST survey \cite{Skelton14} to determine the effect of color selection for $5<z<8$ LBGs; while ref.~\cite{Inami17} cross-referenced with a blind emission line search at similar redshifts using the MUSE HUDF survey \cite{Bacon17};  we also discuss the utility of narrow and medium band surveys in Appendix \ref{app:narrowband}.  

The approach we adopt mirrors that for the color-color plots above, but we degrade the CARS catalogue, rather than UVUDF, given problems with small-number statistics for small area samples.  This is not ideal, not least as the bands available for CARS are more limited and hence the photometric redshifts more questionable, but it suffices \warn{in lieu of a `Goldilocks' sample of deep, $i>25.6$ data with near-IR coverage and secure redshifts over a significant area}.  Moreover, our focus largely resides on the propagation of these trends to CMB lensing cross-correlation and these estimates may be refined in the future.

Fig.~\ref{fig:int_nz} shows our best estimate of the normalised redshift distribution of $u$ and $g$ dropouts, both at the full CARS depth (red), and after degradation to the LSST DESC depths (blue).  It can be seen this is considerably more involved than that of Fig.~\ref{fig:pz}, perhaps due to the field sample variance, means of estimate -- photometric redshift, image injection completeness estimates, etc.  Some expected trends are apparent after degradation to the depths expected in cosmology, namely an increased width due to the greater photometric scatter and a lower completeness for faint high-$z$ targets.  One must therefore be careful in utilising estimates derived from e.g.\ HST luminosity function studies blindly.  Overlaid curves (translucent) indicate the best-fitting Gaussian process approximation, which we apply in \S \ref{sec:cosmology}.  This serves for an easy communication of the populations present at reasonable accuracy.  Of course, the additional bands available for LSST and Euclid allows for multi-color, self-organising map \cite{Masters15}, extreme deconvolution \cite{Bovy11} or photometric redshift selection, all of which may facilitate a reduction in the number of interlopers.

\subsection{Propagation of interloper bias to cosmological parameters}
\begin{figure}[t]
  \begin{subfigure}{0.5\textwidth}
    \centering
    \includegraphics[width=\linewidth]{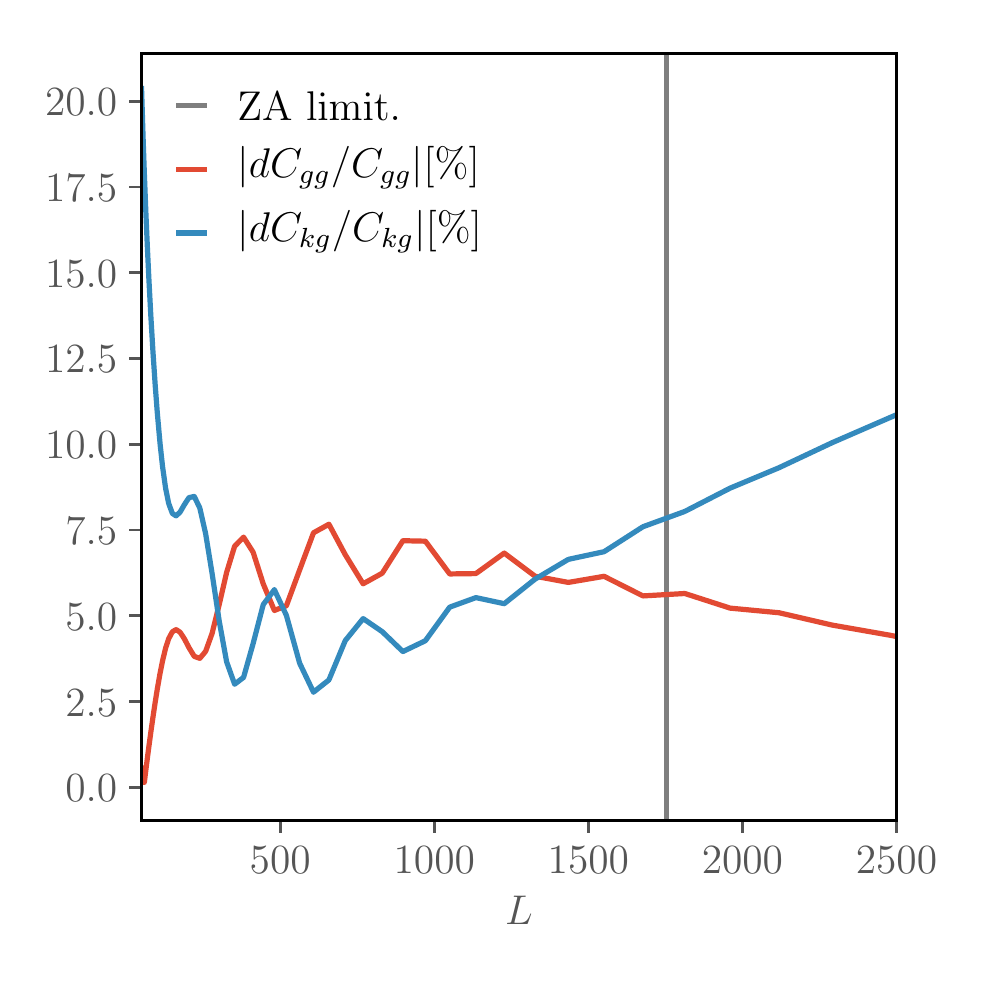}
  \end{subfigure}
  \begin{subfigure}{0.5\textwidth}
    \centering
    \includegraphics[width=\linewidth]{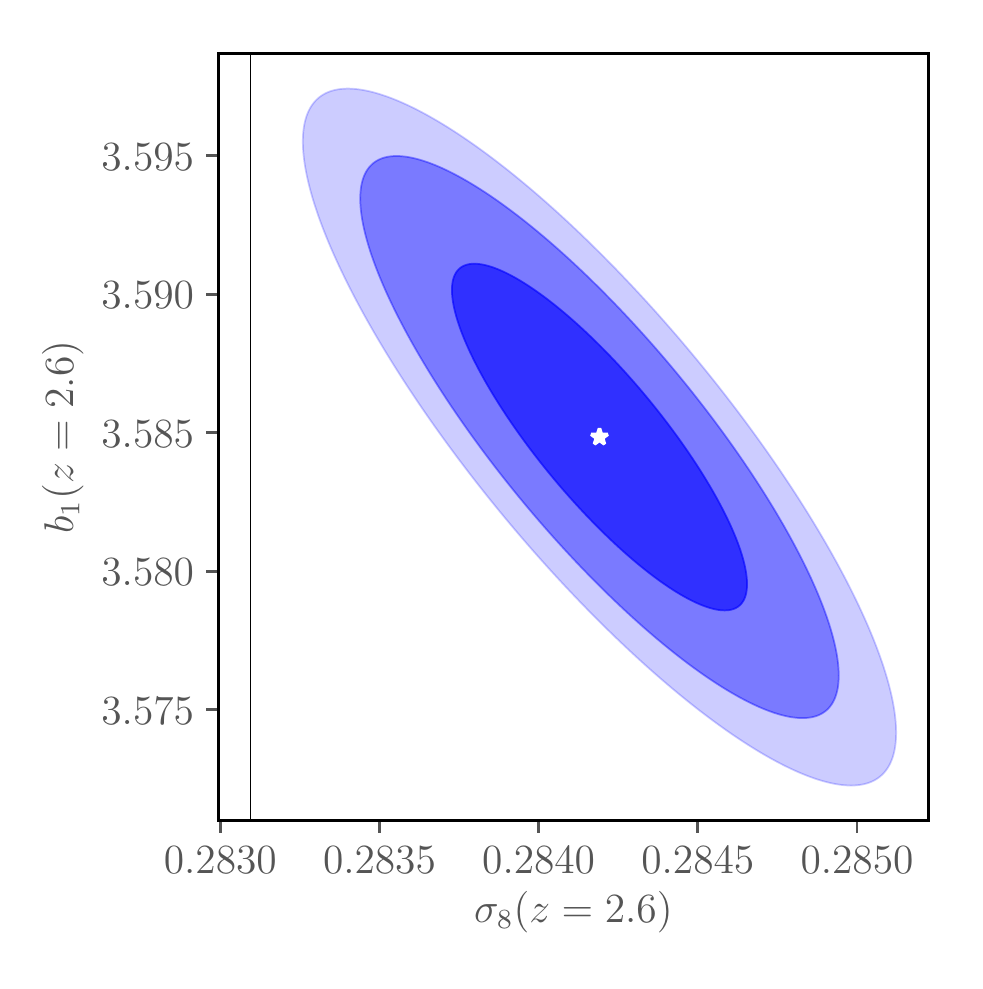}
  \end{subfigure}
  \caption{Linear effect of interlopers on a CMB lensing $\times$ dropout analysis if unaccounted for:  \textbf{Left:} fractional change in $C_{gg}$ and $C_{\kappa g}$ between the full and degraded cases of Fig.~\ref{fig:int_nz} when assuming a 6\% interloper fraction and a $(1+z)$ bias for $z<3$ interlopers.  \textbf{Right:} the linear propagation to a bias in cosmology.  The vertical line shows the fiducial $\sigma_8$ at the peak of the full depth $p(z)$, while the star and errors show the expected contours and $\simeq 3.5\sigma$ bias in $\sigma_8$ due to interlopers present in the degraded sample.  Note the expansion we consider is not strictly applicable to the large biases we find, but serves to confirm the bias is significant.}
\label{fig:int_bias}
\end{figure} 
Before embarking on detailed estimates of the feasibility of spectroscopic followup and the degree to which this interloper contamination can be mitigated, it makes sense to first estimate how biased the constraints would be if left uncorrected.  We propagate this $dN/dz$ to a bias in parameters following ref. \cite{Modi17a}, i.e.\ assuming a Gaussian covariance and the Fisher matrix to be restricted to the two parameters of primary interest,
$\boldsymbol \theta~=~\{\sigma_8(z), b_1(z)\}$, for which the bias is 
\begin{equation}
    \label{pbias}
    \Delta \boldsymbol{\theta}_\alpha = F_{\alpha \beta}^{-1} \ \frac{\partial \langle \boldsymbol{D}_L \rangle }{\partial \boldsymbol{\theta}_\beta} \ \rm{Cov}^{-1}_{L L'} \ \langle \Delta \boldsymbol{D}_{L'} \rangle, 
\end{equation}
where an Einstein summation over repeated indices is implied.  Note this formula is linearized in $\Delta \boldsymbol{D}_L$ and is invalid for significant interloper fractions that lead to greatly different spectra.  The Fisher matrix for each $\boldsymbol D_L \equiv (C_{\kappa \kappa}, C_{\kappa g}, C_{g g})$ is $F_{\alpha \beta} \equiv \  \partial_\alpha \boldsymbol D_L \  \rm{Cov}^{-1}_{L L'} \ \partial_\beta \boldsymbol D_{L'}$ \cite{Schmittfull18, Dodelson16}; the $C^{\kappa g}$ sub-matrix is illustrative, which is simply 
\begin{equation} 
    F_{\alpha \beta} = \sum_{L} \left ( \frac{S}{N} \right )^2 \left( {\begin{array}{cc}
   4/\sigma_8^2  & 2/(b_1\sigma_8)  \\
   2/(b_1 \sigma_8) & 1/b_1^2
  \end{array} } \right),
\end{equation}
and represents a rotation of the Fisher matrix for the bandpowers to the parameters.  Here $(S/N)^2 \equiv C_{\kappa g}^2 \ / \ \rm{var}(C_{\kappa g})$ with a diagonal covariance satisfying eqn.\ (\ref{eqn:var_ckg}).  

The interloper contribution has secondary impacts beyond the change in $p(z)$, given the $\simeq$ 6\% increase in areal density and change in the effective linear bias; the latter is well known to be highly variable with redshift given the color selection. Fig.~\ref{fig:int_bias} shows the resulting systematic bias to cosmological inference due to interlopers, if not actively corrected for.  We find a systematic shift of order $3.5 \sigma$ assuming a 6\% interloper fraction, the degraded $p(z)$ curve of Fig.~\ref{fig:int_nz} and a $(1+z)$ bias for (interloper) galaxies below the peak redshift.  We find this systematic to be even more significant for the $g$-dropout sample.
Thus, this interloper population must likely be corrected for explicitly.  Our chosen means to do so are foreground cleaning and `clustering redshifts', which we discuss in \S\ref{sec:lowz_cleaning} and \S\ref{sec:mcqw} respectively.  \warn{For the latter, dedicated followup of a small area is likely sufficient, although, in the absence of spectroscopic redshifts, any low-$z$ tail to the redshift distribution requires additional nuisance parameters to model the uncertain non-linearity of galaxy biasing and matter}.

\section{Feasibility of spectroscopic followup}
\label{sec:specz}

Having outlined the science case and means with which to tackle cosmology at $z \geq 2$ with dropout selection of LBGs, we now examine how this is facilitated by spectroscopy.  There are two cases to consider:  highly complete coverage of a significant area with the intention of achieving near-perfect redshift resolution, as is necessary for small-scale redshift space distortions or a large-scale \fnl \ analysis, and dedicated followup of a small, deep field to precisely estimate the redshift distribution required for a CMB lensing cross-correlation analysis.  Although we offer brief estimates of the first case in \S\ref{sec:rsd}, we shall focus on the latter and leave a detailed investigation of the implications of a broad-band clustering analysis to a followup work.  In \S \ref{sec:interlopers}, we have established the likely interloper distribution for our samples, which is typically dominated by stars and Balmer-break confusion with relatively bright $z<1$ red galaxies, and propagated this to the likely bias in cosmology.  Removal of this bias by unconstrained marginalisation over $dN/dz$ nuisance parameters would adversely degrade the cosmological constraints so it remains to estimate the priors that may be placed by clustering redshifts from spectroscopy \cite{Newman08, Menard13, McQWhi13}, as we estimate in \S\ref{sec:mcqw}.  

To do so requires knowledge of both the low-$z$ and high-$z$ spectroscopic samples that may be available with current and future instrumentation, which we survey in Table \ref{Table:Spectroscopy}.  Of these, we investigate DESI, PFS and M-DESI, DESI-like spectrographs on a 6.5m Magellan telescope, as three examples of forthcoming instrumentation.  \rfree{See refs.\ \cite{Schlegel19} and \cite{Bundy19} for concrete proposals}. 
\begin{table}
 \centering
 \begin{tabular}{|| c | c | c | c | c | c | c | c | c ||} 
 \hline
   Survey & Radius & Coverage & $R$ & Multiplex & FOV & Exp. & FOM
   & Ref. \\
          & [m]    
          & [$\mu$m] 
          & $[10^3]$
          & $[10^3]$   
          & [$\prime$]
          & [min.]
          & $f_{NL}$ (RSD) 
          & \\
 \hline\hline
 4MOST & 2.0 & 0.37 - 0.95 & 4.0 & 2.4 & 135 & 165 & 4.7 (1.55) 
 & \cite{deJong12} \\
 \hline
 DESI   
       & 2.0 & 0.36 - 0.98 & 5.0 & 5.0 & 180 & 132 & 12.2 (4.05)  & \cite{DESI16} \\
 \hline
 PFS  
     & 4.1 & 0.38 - 1.26 & 3.5 & 2.4 & 83 & 45 & 12.4 (5.72)
 & \cite{Tamura16} \\ 
 \hline
  M-DESI 
         & 3.3 & 0.36 - 0.98 & 5.0 & 20.0 & 90 & 50 & 13.0 (13.09)
  & \cite{Ferraro19} \\
 \hline
 MSE & 5.7 & 0.36 - 1.30 & 4.0 & 4.3 & 83 & 20.9 & 26.7 (22.2) & \cite{MSE18} \\
 \hline
 ESO/BOA 
     & 5.0 & 0.36 - 1.30 & 4.0 & 10/100 & 150 & 26.4 & 69 (40.5/69)
 & \cite{Pasquini16} \\ \hline
\end{tabular}
\caption{Instrumental properties for current, planned and proposed multi-object spectroscopic surveys. Here `Radius' is the primary mirror radius in meters, `$R$' is a point-estimate of the spectral resolution, `FOV' is the field-of-view diameter in arcmins and `Exp.' is the estimated exposure time necessary for the fiducial LBG \& LAE sample in \S \ref{sec:fiducial}.  Of these, we select DESI, PFS and M-DESI for detailed investigation.  We discuss the competitiveness of the alternative facilities in \S\ref{sec:surveys} and define an appropriate figure-of-merit for ranking such facilities in App. \ref{app:FOM}.  For this, we scale from the exposure time expected of M-DESI using the resolution and radius properties shown above.
}  
\label{Table:Spectroscopy}
\end{table}

\subsection{Exposure time calculator}
\begin{figure}
    \centering
    \includegraphics[width=\linewidth]{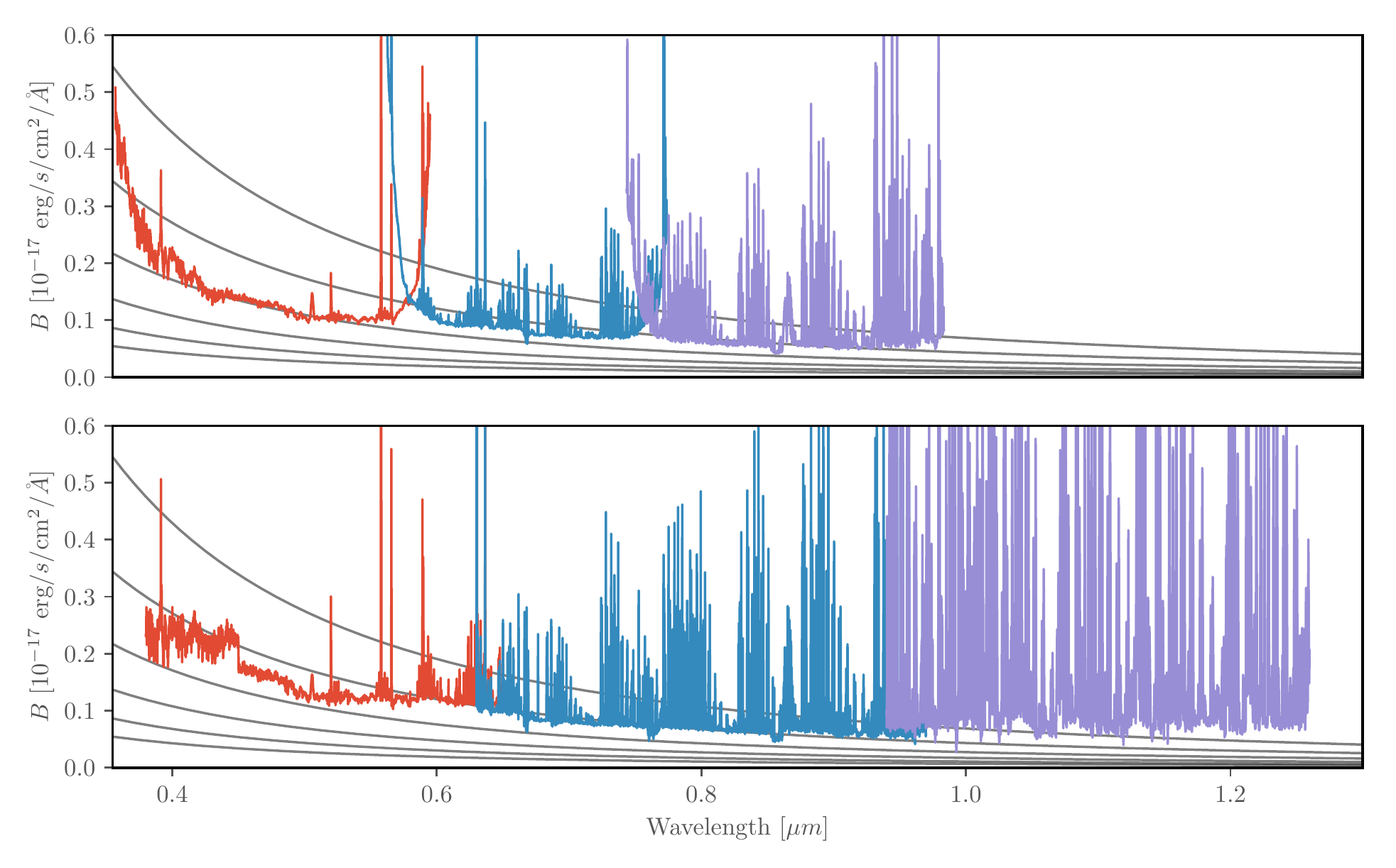}
    \caption{The effective background for 1\AA\ resolution elements in each of the three arms for M-DESI (top) and PFS (bottom) in $3000\,$s exposures, as derived from our model for the sky background and CCD specification.  Overlaid are flat $F_\nu$ spectra for 23 to 25.5 in half AB magnitude steps.  From this, we can appreciate that the continuum signal-to-noise is below unity for sources fainter than $24^{\rm th}$ magnitude.  For this reason, we consider either brighter LBGs or fainter Lyman-$\alpha$ emitters of significant rest-frame equivalent width ($>50$\AA ) as potential targets.  \rfree{Also apparent is the bluer coverage and lower background of DESI for <0.7$\mu$m, while PFS has greatly extended coverage albeit where the sky background is significantly larger}.  Note that a $7900\,$\AA\ limit is sufficient for Lyman-$\alpha$ detection to $z=5.5$.
    }
    \label{fig:background}
\end{figure}
We estimate the depths achievable for a range of exposure times when redshifting LBGs with DESI, PFS and M-DESI using the spectra simulator adopted by the DESI collaboration, \specsim\footnote{\url{https://specsim.readthedocs.io/en/stable/}}.  Note that ref.~\cite{Hirata12} provides the official PFS exposure time calculator\footnote{\url{https://github.com/Subaru-PFS/spt_ExposureTimeCalculator}}, but {\tt SpecSim} facilitates a simplified, direct comparison with DESI and M-DESI.  Where necessary, we assume publicly available characteristics for PFS\footnote{\url{http://pfs.ipmu.jp/research}} and include sufficient realism for our purposes -- e.g.\ spectral coverage, resolution, read noise and dark current for each of the three spectral arms.  For PFS, we further assume an atmospheric extinction curve appropriate to Mauna Kea \cite{Buton13} to replicate the better site conditions, and extend it to the near-infrared\footnote{\url{www.gemini.edu/sciops/telescopes-and-sites/observing-condition-constraints/}}.  For the infrared sky brightness model, we assume that of ref.~\cite{Sullivan12}.  This configuration is packaged with \specsim \ and publicly available\footnote{\url{github.com/michaelJwilson/specsim}}.  The resulting background noise curves for 3000s exposures are shown in Fig.~\ref{fig:background}.  The overlaid AB sources show the continuum signal-to-noise is below unity for sources fainter than $24^{\rm th}$ magnitude, \rfree{with DESI -- and similarly M-DESI, given the shared spectrographs -- providing the superior background in the blue spectroscopic arm and PFS having significantly extended coverage in the red}.  As such, we consider two samples for which secure redshifts are likely to be obtainable: bright $m_{UV} \simeq 24$ LBGs for which absorption line redshifts are potentially achievable and much fainter, $m_{UV} \simeq 25.5$, Lyman-$\alpha$ emitters (LAEs) with significant rest-frame equivalent width, e.g.\ $> 50\,$\AA \ (we assume a convention in which positive equivalent width corresponds to emission).  To represent the SEDs of LAEs for redshift efficiency estimates, we use the stacked composites of ref.~\cite{Shapley03}, split by quartiles according to Lyman-$\alpha$ equivalent width (-14.92, -1.10, 11.00 and 52.63\AA, labelled by Q0, Q2, Q3 and Q4 respectively).  We use these same templates for both forecasting the observed spectra and redshifting, as a first approximation, in addition to a range of QSO, low-$z$ galaxies and stars to assess degeneracies.

We use {\tt redrock}\footnote{\url{https://github.com/desihub/redrock}} for redshifting -- a state-of-the-art spectral classification and redshift fitting analysis developed for the DESI collaboration, which uses the complete spectral information available from `spectro-perfectionism' extractions \cite{Bolton10} and a new suite of PCA templates and archetypes based on stellar population synthesis modeling of $0<z<1.5$ galaxies, theoretical spectral models of stars and white dwarfs, together with a generative model of QSO spectra trained on spectroscopic observations for $2.2<z<3.5$.

\begin{figure}
\begin{subfigure}{0.45\textwidth}
\includegraphics[width=\textwidth]{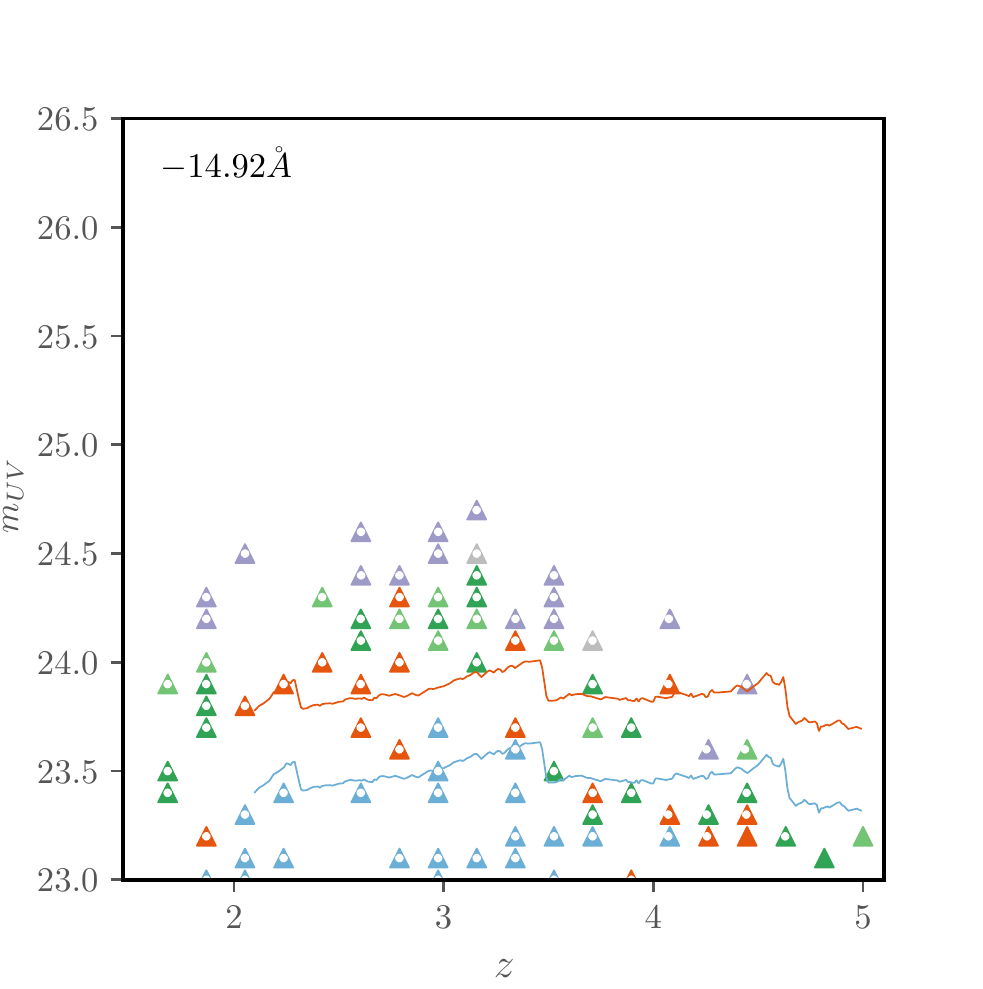}
\end{subfigure}%
\begin{subfigure}{0.45\textwidth}
\includegraphics[width=\textwidth]{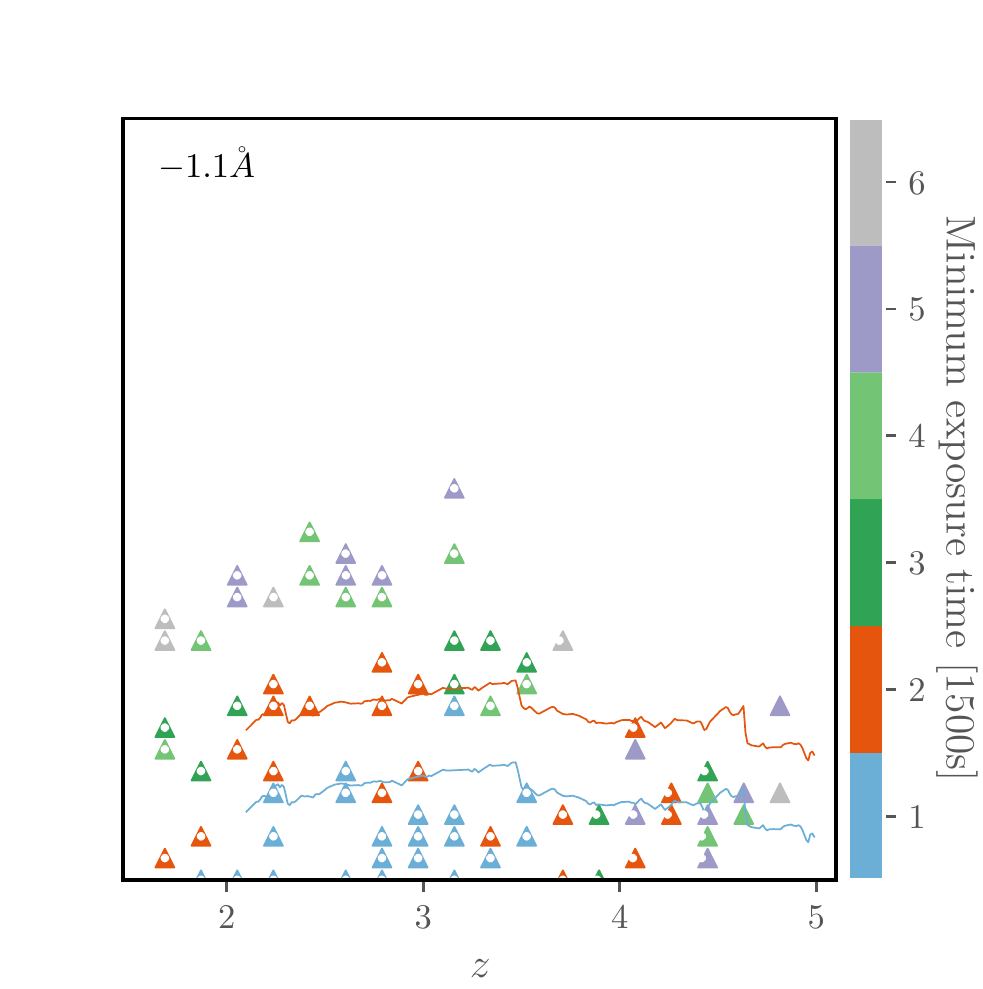}
\end{subfigure}
\begin{subfigure}{0.45\textwidth}
\includegraphics[width=\textwidth]{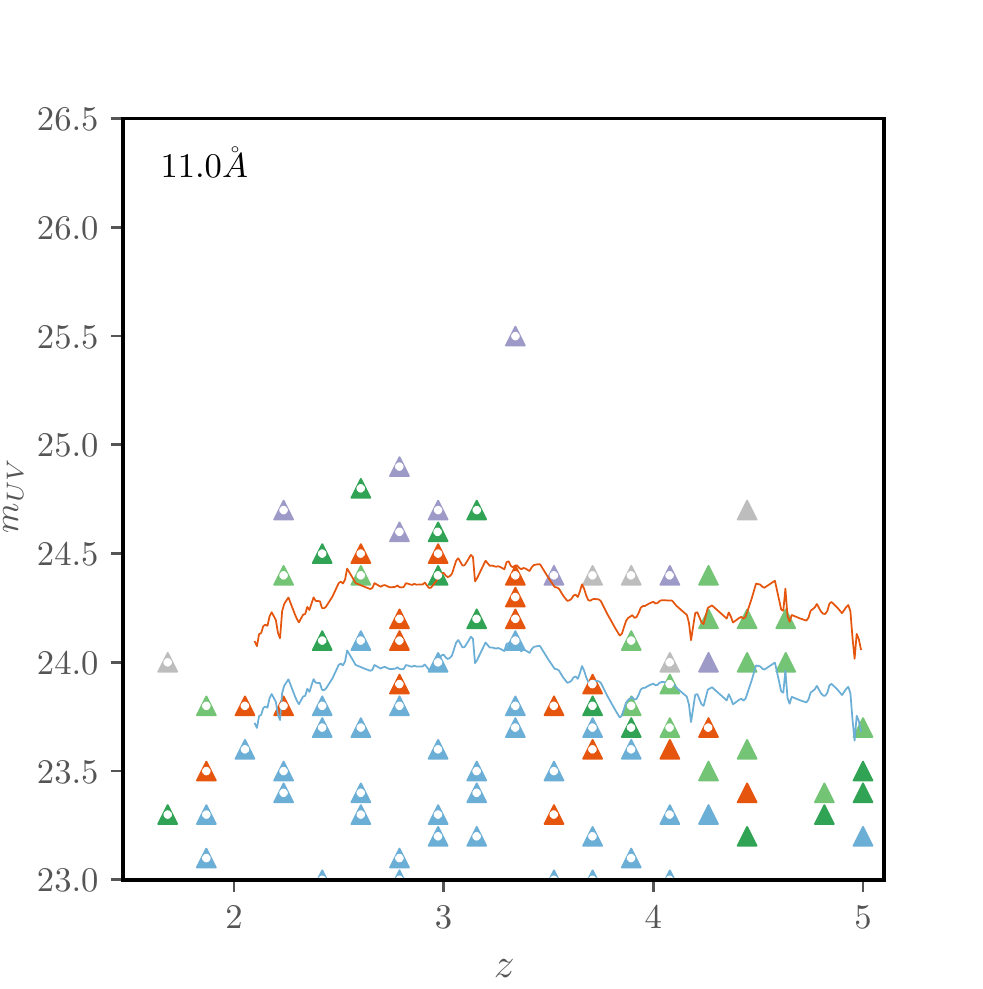}
\end{subfigure}%
\begin{subfigure}{0.45\textwidth}
\includegraphics[width=\textwidth]{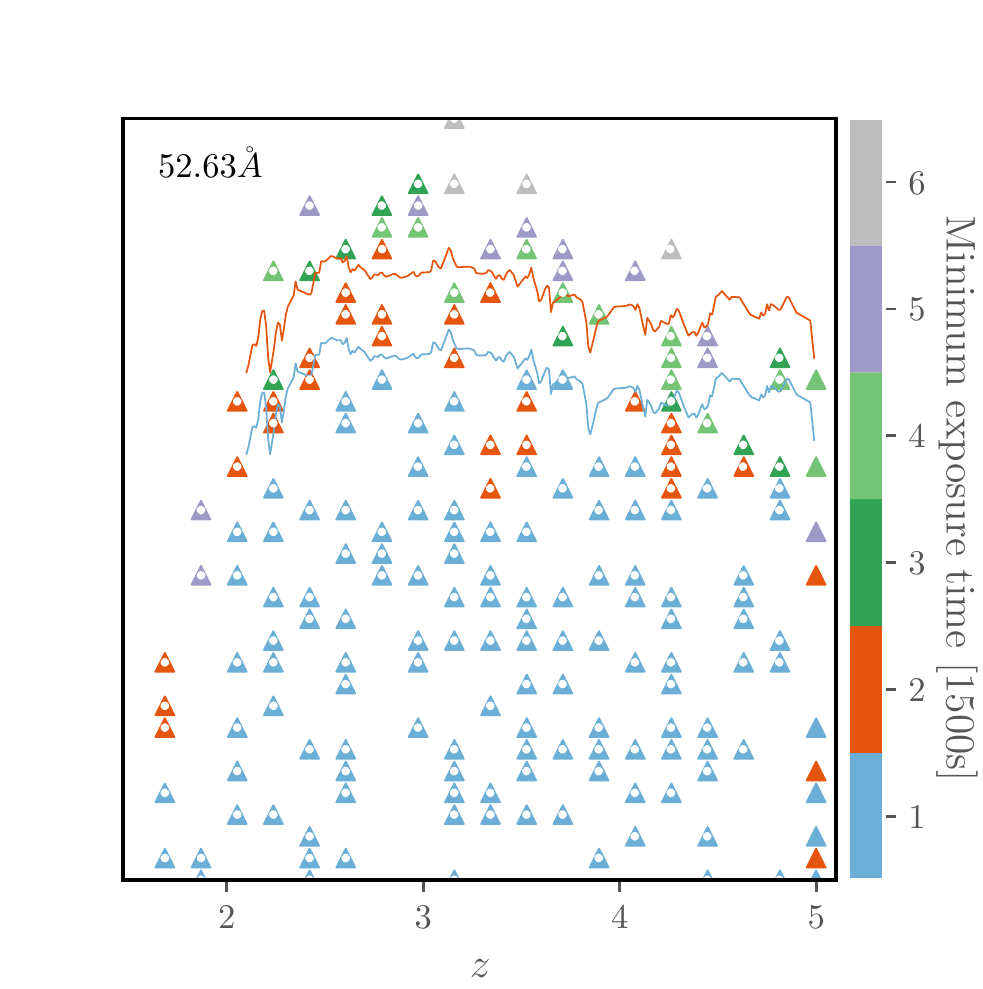}
\end{subfigure}
\caption{A detailed breakdown of redshift success with $m_{UV}$ and $z$ for M-DESI -- a 6.5m telescope equipped with DESI spectrographs.  Shown is the minimum exposure time required for {\tt redrock} to declare a confident redshift.  The four panels show composites with Lyman-$\alpha$ EW of -14.92, -1.10, 11.00 and 52.63\AA \ as labelled, which largely determines the depth possible.  In the absence of a white dot (marked at the true redshift), this is a false confidence and an erroneous redshift.  These primarily arise from confusion with [OII], H$\alpha$ and H$\beta$, e.g. beyond $z=4.1$ for 11\AA \ and for $z \simeq 5$ generally.  \warn{Assuming an accurate proxy for a secure redshift is the total signal-to-noise present in sharp emission and absorption features, see eqn. (\ref{eqn:rrs2n}), we may predict the achievable depth for a given redshift confidence, $\theta$.  Colored lines show this achieveable depth for each exposure time assuming $\theta = 5$ is appropriate, which can be seen to be approximately consistent with the {\tt redrock} results.}
}
\label{fig:beast-like-exposures}
\end{figure}

Fig.~\ref{fig:beast-like-exposures} shows the minimum exposure time necessary for {\tt redrock} to declare a successful redshift for each of the EW quantiles with the M-DESI instrument.  In the absence of a white dot, this is a false confidence and an erroneous redshift.  If we adopt fifty minute exposures as our fiducial, we see that depths (in the detection band, $m_{UV}$) of 25.5 mag are achieveable for a $\simeq 50$~\AA \ equivalent width,  which falls to $\simeq 24$ for smaller equivalent widths.  Typically, false confidence derives from line confusion with [OII], H$\alpha$ and H$\beta$, which is especially the case beyond $z=4.1$ for Q3, we later explore this in Fig.~\ref{fig:lines}.  Additional information, e.g.\ photometric colors or apparent size, or superior templates may help break this line degeneracy, but it represents a well-posed problem to be overcome.  We note that the declared redshift is typically a good one, but as we consider a range of science cases at both large and small-scales, we do not require a specific redshift precision at this stage. 
\begin{figure}[t]
  \centering
  \begin{subfigure}{0.49\textwidth}
    \centering
    \includegraphics[width=\linewidth]{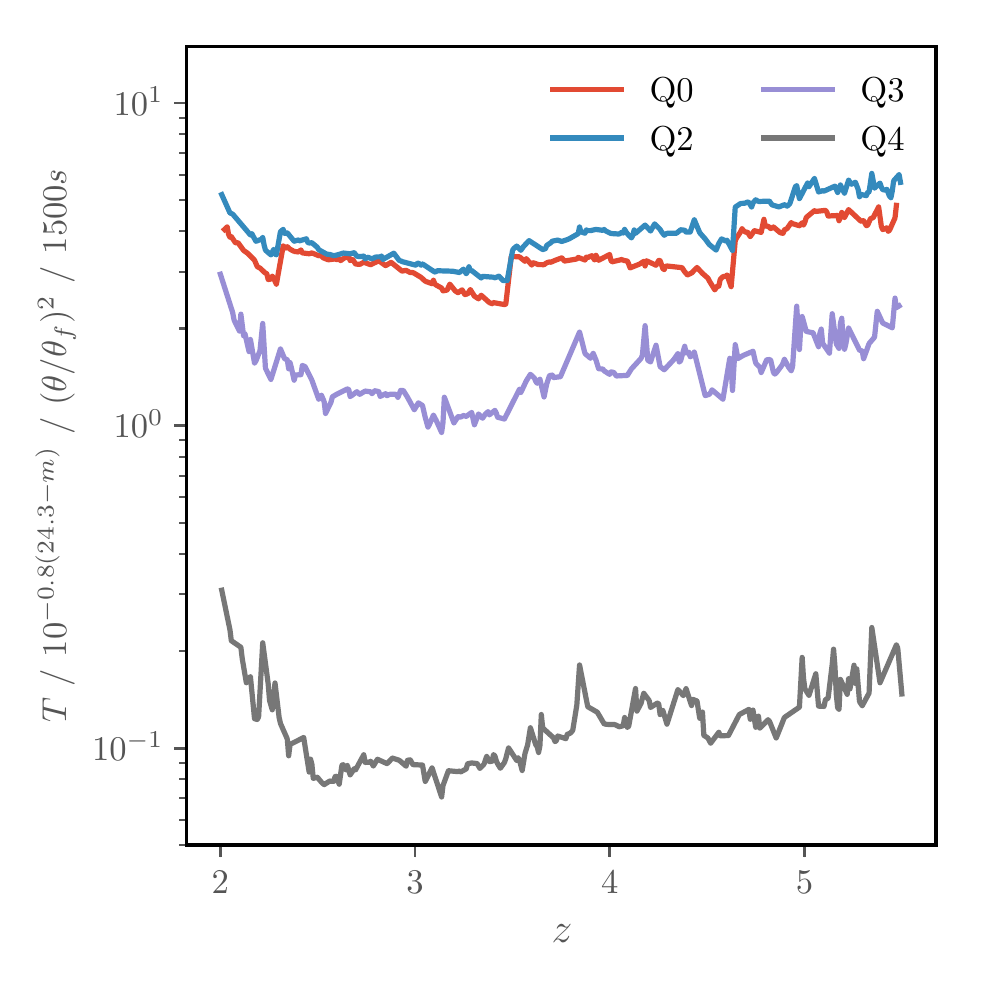}
  \end{subfigure} %
  \begin{subfigure}{.49\textwidth}
    \centering
    \includegraphics[width=\linewidth]{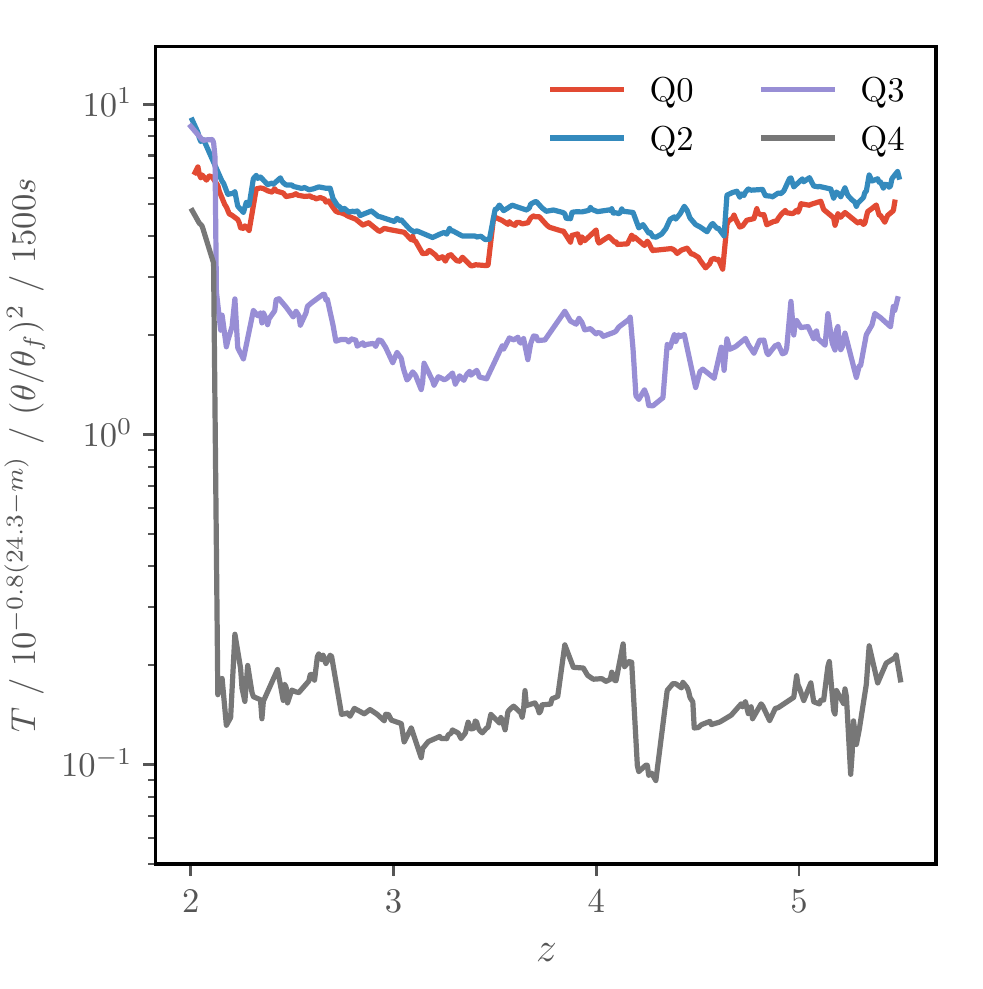}
  \end{subfigure}
  \caption{Our adopted model for the exposure time required, $T$, to obtain a successful redshift (of confidence $\theta$) for LBGs of given redshift, $m_{UV}$ and equivalent width for both M-DESI (left) and PFS (right).  Shown is that required (in units of 1500s) for each of the four quantiles, assuming a 24.3 magnitude source and a fiducial redshift confidence of 5.  The ordinate label shows the required scaling with $m_{UV}$ and redshift confidence, $\theta$.  To derive this result, we assume a S/N of sharp (<100 \AA \ FWHM, eqn. \ref{eqn:rrs2n}) spectral features greater than $\theta$ equates to a successful redshift.}
  \label{fig:Tz}
\end{figure}

To compress the {\tt redrock} results we adopt a  spectroscopic depth model of reasonable accuracy.  For this, we assume an accurate proxy for the redshift success is the S/N of sharp features i.e.\ those with \emph{narrow} emission or absorption lines ($<100\,$\AA\ FWHM, as tied to the redshift precision requirement), viz.
\begin{equation}
    \left(\frac{S}{N} \right )^2 = \sum_i \left ( \frac{F - \tilde F}{B} \right )^2 \geq \theta^2,
\label{eqn:rrs2n}
\end{equation}
where $F$ is the signal flux, $\tilde F$ is the filtered flux, $B^2$ is the expected background variance, i.e. that shown in Fig. \ref{fig:background}, the sum is over resolution elements and we equate a signal-to-noise greater than a confidence threshold $\theta$ to a secure redshift.  Given the trivial dependence of this $S/N$ on the apparent magnitude, exposure time and redshift confidence, $\theta$, one may straightforwardly predict the exposure time required for a secure redshift as a function of LBG redshift and equivalent width.  To do so requires calculating a S/N baseline for a fiducial magnitude and range of redshifts, the result of which is shown for each quantile in Fig.~\ref{fig:Tz}.  Conversely, this may predict the achievable $m_{UV}$ depth in a given exposure time, as shown by the limiting depths (colored lines) shown in Fig.~\ref{fig:beast-like-exposures}, which present a reasonable match to the {\tt redrock} results.  
\begin{figure}
\centering
\begin{subfigure}{0.45\textwidth}
\includegraphics[width=\textwidth]{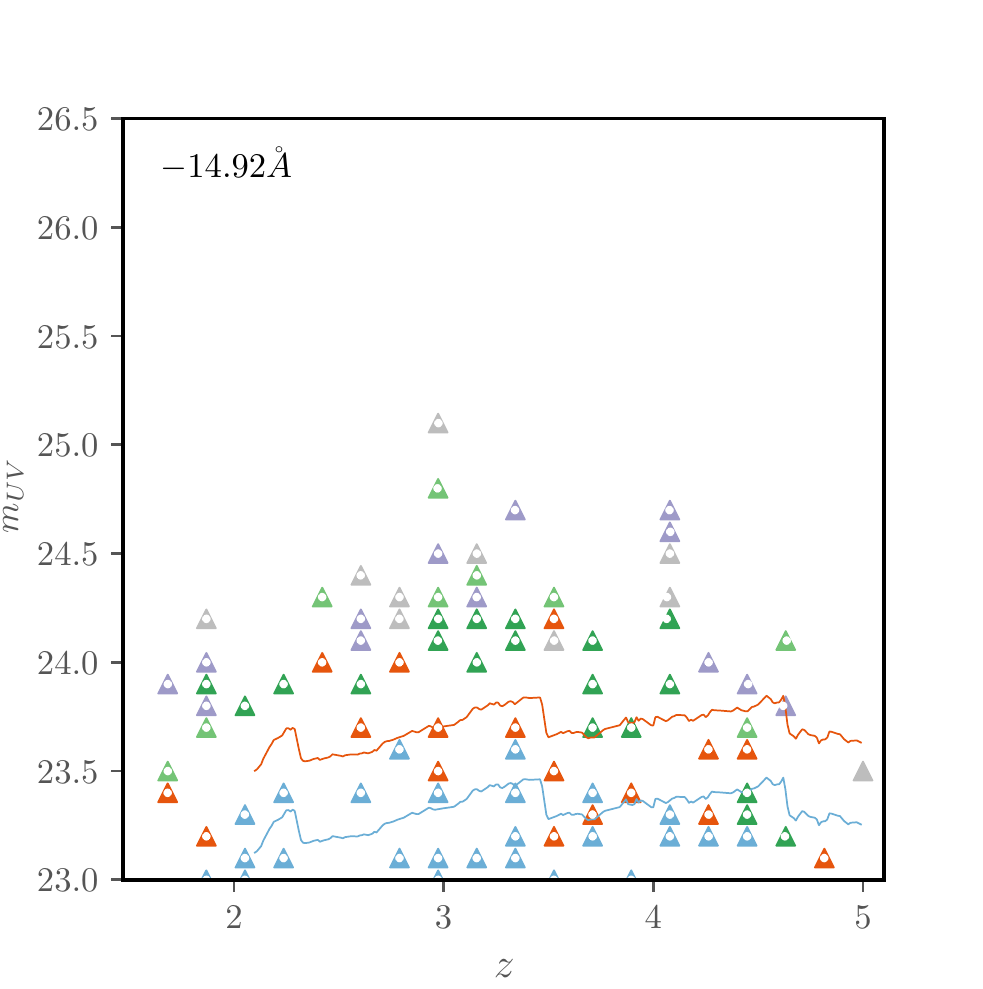}
\end{subfigure}%
\begin{subfigure}{0.45\textwidth}
\includegraphics[width=\textwidth]{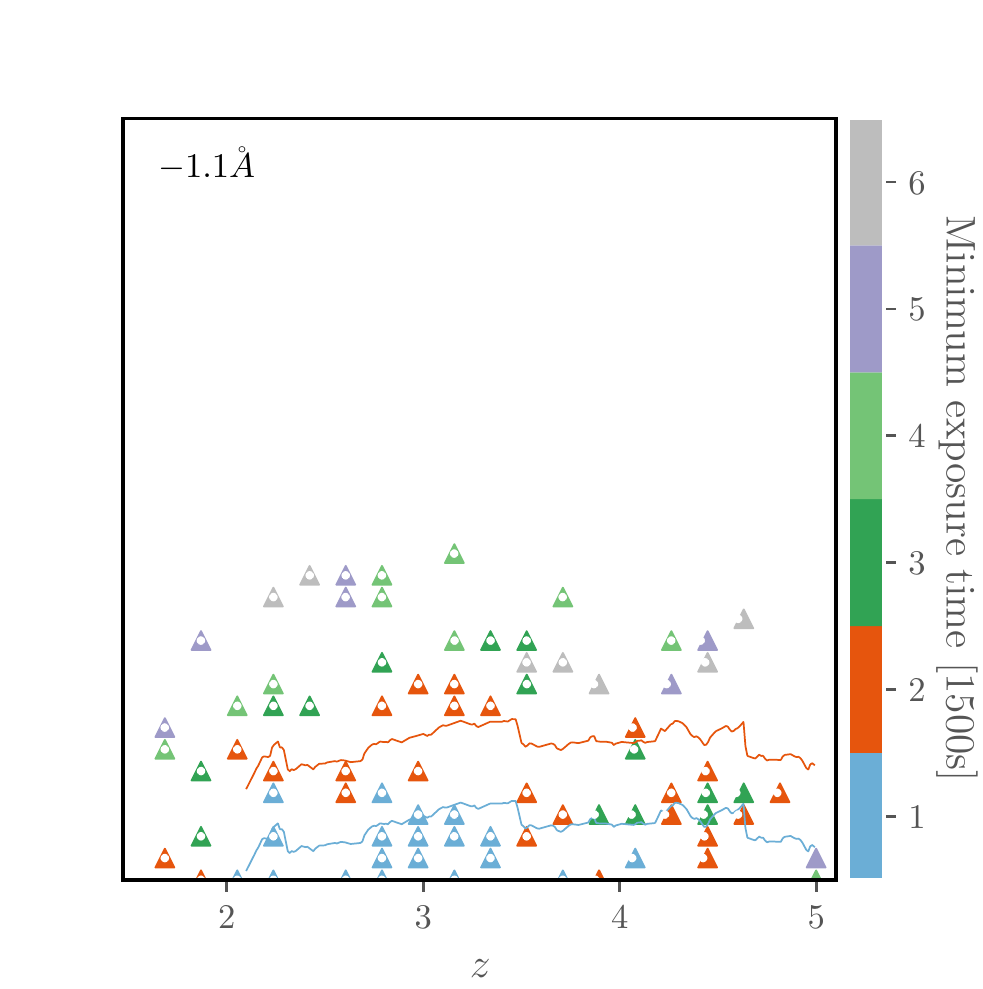}
\end{subfigure}
\begin{subfigure}{0.45\textwidth}
\includegraphics[width=\textwidth]{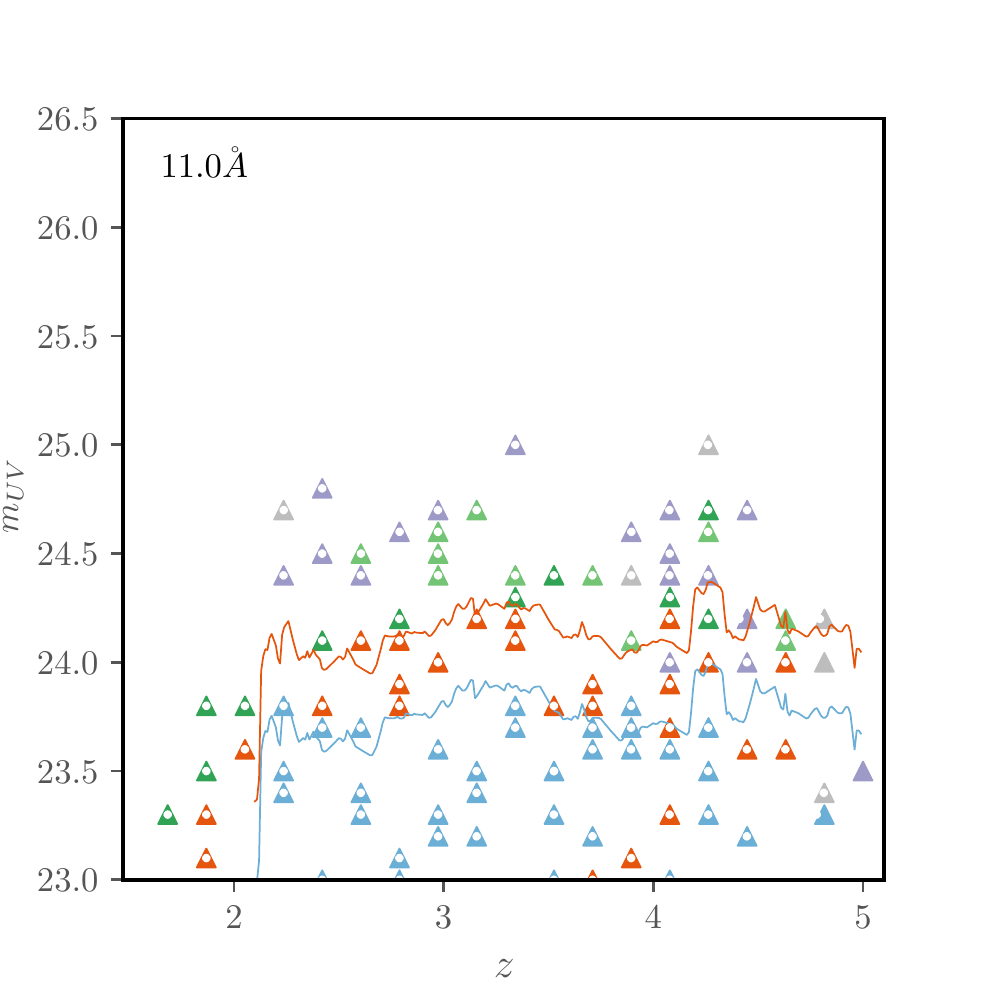}
\end{subfigure}%
\begin{subfigure}{0.45\textwidth}
\includegraphics[width=\textwidth]{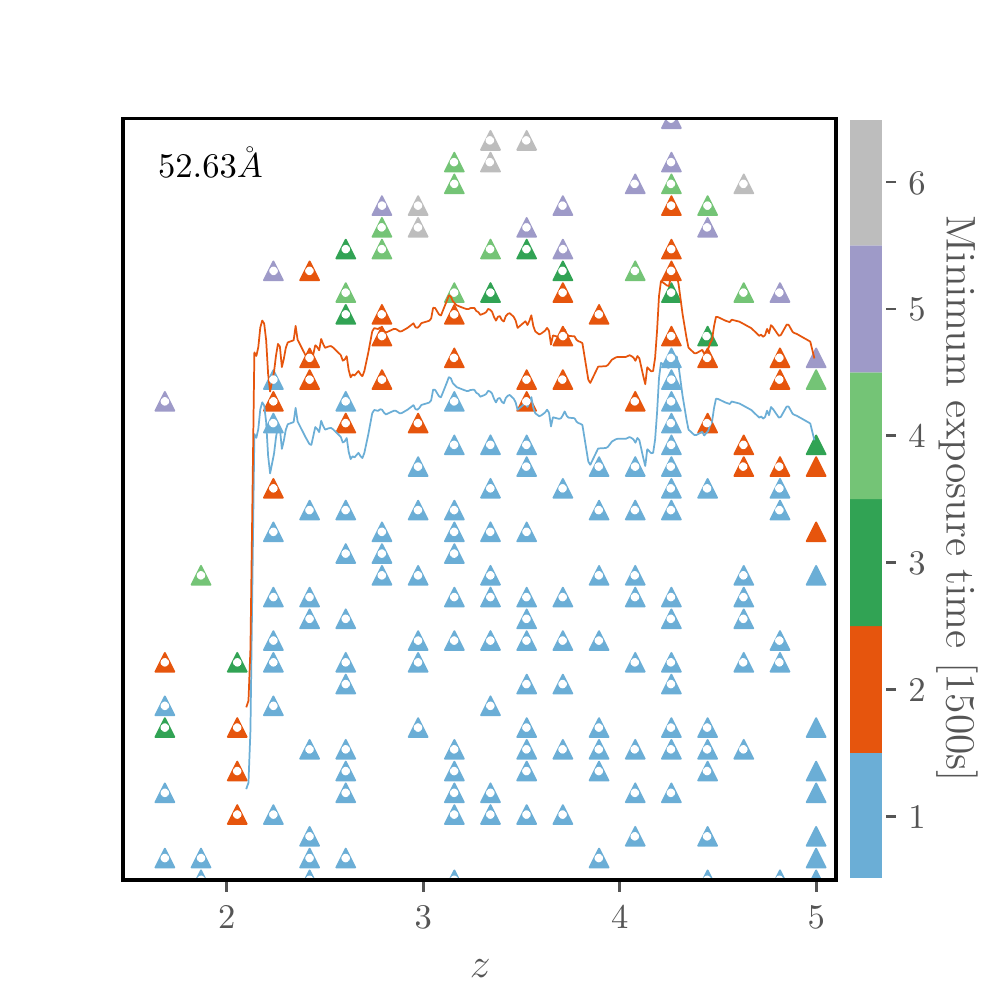}
\end{subfigure}
\caption{Same format to Fig.~\ref{fig:beast-like-exposures}, but for the PFS.  The achievable depths are largely consistent with M-DESI, perhaps reflecting that $1\mu$m coverage is sufficient to observe Lyman-$\alpha$ to $z=7.2$, together with the competitive transmission and obscuration of (M-)DESI.  
}
\label{fig:pfs-like-exposures}
\end{figure}  
  
Fig.~\ref{fig:pfs-like-exposures} shows results for the Prime Focus Spectrograph, which is largely able to access the same regimes as the extended coverage is largely insignificant for redshifting a Lyman-$\alpha$ line or break and the reduced transmission and obscuration acts to limit the 23\% added efficiency of an $8\,$m relative to a $6.5\,$m mirror.    In both cases, we see that these instruments are very efficient at acquiring LBG redshifts, which may be somewhat influenced by the presence of weaker emission lines in the composites.  Given this, and the number of additional assumptions involved -- template set, potential confusion with sky lines, redshift confidence required, etc. -- \rfree{in the following section} we seek to establish further grounding and confirm our estimates are reasonable based on \rfree{previous} observations.

\subsection{Comparison with available studies}
Firstly, ref.~\cite{Steidel99} obtained spectra for $I<25$ $G$-band dropouts in the range $3.8<z<4.5$ with LRIS on the $10\,$m Keck-II telescope.  With \rfree{$ \mathcal{R} \simeq$} 10-20\AA \ spectral resolution, typical two hour exposures achieved a 30-50\% redshift success rate.  This suggests five hour exposures are necessary for this depth given the relative area of M-DESI and Keck at LRIS resolution.  For a narrow peak or break, the wavelength range needed to integrate a fixed signal, and hence the integrated background, is \rfree{$\propto \mathcal{R}^{-1}$}.  Therefore we expect an exposure time of order sixty minutes for M-DESI on the basis of this Keck analysis.  By successfully targeting $z_{\rm{AB}}<25$ dropouts with 3.5hr exposures on DEIMOS, $\mathcal{R}<6000$, ref.~\cite{Mallery12} is roughly consistent with this revision; i.e. scaling this to $z_{\rm{AB}} \simeq 24$ and a 6.5m mirror suggests a 80 min exposure.  

More recently, the ongoing VANDELS survey \cite{McClure18} targets a photometric redshift selected sample with the $\mathcal{R}\simeq 600$ VIMOS on the 8.2m VLT.  With the restriction $25 < H < 27$ and $i<27.5$, achieving a continuum SNR $\approx 3$ for each 10\AA \ resolution element yields a Ly-$\alpha$ line flux limit of $2\times 10^{-18}$ ergs cm$^{-2}$ s$^{-1}$ .  To do so requires exposures totalling 20hrs for $i<25.5$, rising to 40hrs for $i<26$.  In comparison, ref.~\cite{Vanzella09} targeted $BVI$ GOODS dropouts with FORS2, using exposures totalling 5.5 to 22.2 hours at $\mathcal{R} \approx 660$.  Finally, ref.~\cite{Bielby16} is of interest as a narrow band selected search for $L_\alpha > 2 \times 10^{-17}{\rm ergs}\,{\rm cm}^{-2}{\rm s}^{-1}$ with VIMOS, achieving a 80\% redshift success rate for $\mathcal{R} \approx 2000$.  These examples are notable for their typically redder ($\approx 5500$ \AA) blue limit and much reduced resolution.

In terms of similar existing plans for future surveys, the galaxy formation program within PFS \cite{Takada15} includes estimates for a 100 night survey of 140K dropouts and 60K LAEs spanning $2 < z < 4$ over 16 deg$^2$ to $i < 24$.  Assuming CFHT archival $u>26$ imaging, this reference quotes six hour exposures as necessary to ensure continuum detection for $i<25$ galaxies, consistent with the depth model assumed above.  For further detail and a quantitative summary, see their Table 9.  Similarly, ref.~\cite{Percival19} also advocates for Inflationary science based on $30\,$min exposures of LBGs with the $11\,$m Mauna Kea Spectroscopic Explorer.  

\subsection{Broader considerations}
\begin{figure}[t]
  \centering
  \begin{subfigure}{0.49\textwidth}
    \centering
    \includegraphics[width=\linewidth]{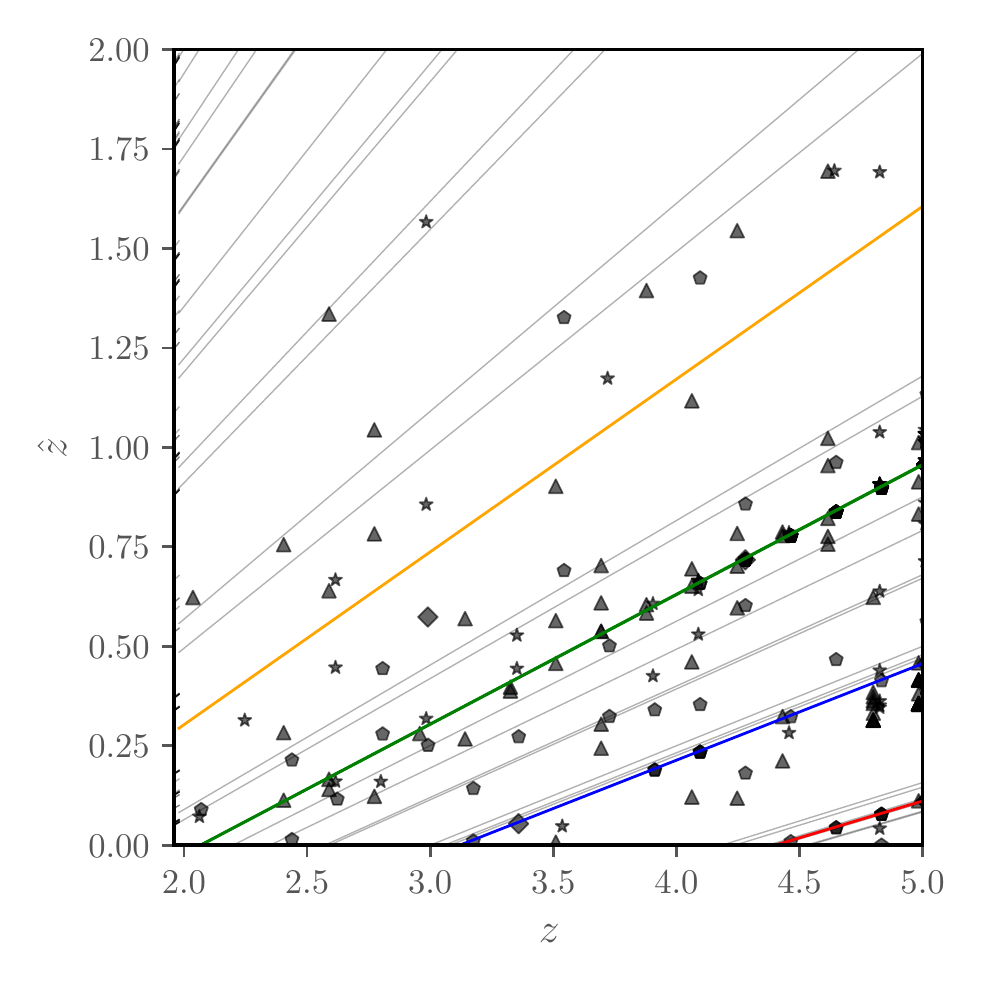}
  \end{subfigure} %
  \begin{subfigure}{.49\textwidth}
    \centering
    \includegraphics[width=\linewidth]{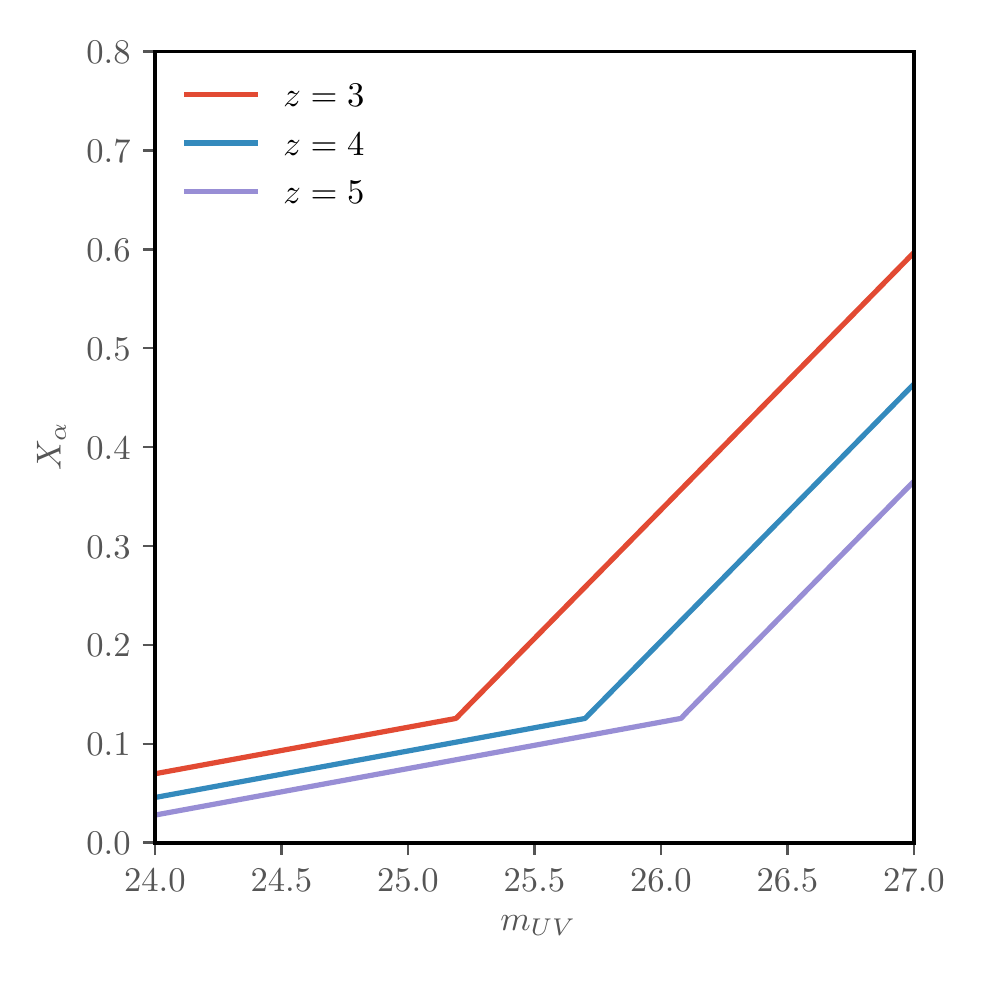}
  \end{subfigure}
  \caption{\textbf{Left}: The line confusion of estimated LBG redshifts, $\hat z$, that occurs for a given redshift, $z$.  Markers indicate the various equivalent widths, -14.92 (triangle), -1.10 (star), 11.00 (pentagon) and 52.63\AA \ (diamond).  Background lines show the confusion redshift for alternative emission lines.  In particular, significant confusion occurs for [OII] (green), [OIII] (blue) and H$\alpha$ (red).  \textbf{Right}:  Expected fraction of LAEs with $>50$\AA \ rest-frame equivalent width, as suggested by the relation of ref. \cite{Stark10}.  This allows for spectroscopic redshifts for significantly fainter targets at $z \simeq 4$ and 5, but we find no realistic sample that can reach the 40-50\% regime, see Table \ref{tab:selection}.}
  \label{fig:lines}
\end{figure}
In reality, the Lyman-$\alpha$ feature can have a multitude of forms, varying from absorption to strong one-sided or double peaked emission due to resonance scattering \cite{Steidel99, Dijkstra14, Verhamme06, Trainor15, Drake17}.  \rfree{While this potentially significant complication could revise the estimates above}, this can also help mitigate line confusion with e.g.\ the H$\alpha$ emission of low-$z$ galaxies.  Moreover, due to bulk flow of the emitting neutral hydrogen, the Lyman-$\alpha$ line may be significantly Doppler shifted (by $\approx 300-500\,{\rm km}\,{\rm s}^{-1}$) from the systemic stellar redshift -- most commonly, the Lyman-$\alpha$ redshift is the larger, \rfree{thereby} reflecting the increased escape fraction if the previous scattering redshifts the photon from resonance.  As these effects can have significant correlation with the dust content (geometry and large-scale environment), some care must certainly be taken in ensuring any cosmological analysis would be unbiased \cite{Hirata09}.
While the broadening of Lyman-$\alpha$ may in itself be a problem as an imprecise redshift indicator, it may be sufficient for certain studies, e.g.\ $f_{NL}$.  For highly precise redshifts, it may be the case that only absorption line redshifts will suffice, which would entail significantly slower survey speeds.  Similarly, the fraction of active quasars must be small given the imprecise redshifts resulting from line broadening, e.g. the sub-panel to Fig.~7 of ref.~\cite{Ono18} suggests a fraction of $\simeq 10$\% at $z\simeq 4$.

Fig.~\ref{fig:lines} shows an explicit test of the line confusion that occurs at the signal-to-noise expected for M-DESI in our fiducial exposure time.  Given the $\mathcal{R}=3000$ resolution assumed, one might hope that no confusion would occur between the Lyman-$\alpha$ line and the [OII] doublet.  Unfortunately, we find this not to be the case at this S/N.  This is in addition to the significant danger of confusion with sky lines, e.g. Fig. 4 of ref.~\cite{Oyarzun17} argues for a S/N of $\simeq 5.5$ to prevent false positives (for significantly fainter emitters).  Fortunately, the dominant confusion occurs in the volume that will be well mapped by planned surveys, which allows the degree of contamination to be estimated and mitigated \cite{Pullen16, Addison18, Grasshorn18}.  While the sky background is relatively dim for the LAEs of interest, percent-level subtraction of the sky for fibers and exposures approaching an hour is itself a technical challenge.  This may also impact the effective multiplex if a large fraction of fibers are required to account for the spatial variation \rfree{of the background}.  

Based on the material above, we suggest desirable straw-man surveys in the following section.  The properties of these samples are summarised in Table \ref{tab:selection}.   

\subsection{Fiducial sample}
\label{sec:fiducial}
\warn{Near $z\simeq 2$, BX selection to $\rfree{R} \simeq 25^{\rm th}$ magnitude yields approximately 13,000 galaxies per sq.\ deg.  A bright cut (e.g.~$\rfree{R} \simeq 24.5$) would help to limit interlopers, and further pre-selection on the basis of photometric redshifts reduces contamination from low-$z$ galaxies.  Such a sample would result in $\simeq 200$M galaxies for 20,000 sq.\ deg}.

The $u$-dropout sample at $z\simeq 3$ is severely limited by the available $u$-band depth, even with LSST-Y10.  Assuming a relatively shallow 0.7 magnitude drop, with associated larger contamination, $i<24.6$ would yield a sample of 2,200 galaxies per sq.\ deg.  
With respect to spectroscopic followup, this is seemingly too bright for there to be a significant Lyman-$\alpha$ fraction, see Fig.~\ref{fig:lines}.  This is unfortunate given the known means to efficiently pre-select large equivalent-width emitters based on the bluer slope of the continua in broad-band imaging \cite{Cooke09, Cooke13}.  However, it may be the case that the Lyman-$\alpha$ fraction is higher than expected for this less restrictive drop critera, given the effect on the colour due to the line itself.  To be conservative, we posit a bright $i \leq 24.0$ sample for which absorption line redshifts will be available based on the approximately eleven lines with significant flux decrements.  Such a sample would yield close to 500 good redshifts per sq.\ deg. given the magnitude limit, with a resulting higher bias, $b \simeq 5.2$.  Greater $u$-band depth would deliver significant returns in terms of this sample as this magnitude limit is brighter than $m_{UV}^{\star}$ -- as shown in Table \ref{Table:Schechter} -- and therefore the counts remain exponentially suppressed.   
\begin{table}
\centering
    \begin{tabular}{|c|c|c!{\vrule width 1.25pt} c|c|c|c!{\vrule width 1.25pt} c|c|c|c|}
      \hline
      \multicolumn{3}{|c}{} & \multicolumn{4}{c}{ \textbf{Photometric}}
      & \multicolumn{4}{c|}{ \textbf{Spectroscopic}}  
      \\
      \hline
      Sample & $m_{UV}$ & Ref. & $m_{\rm{lim}}$ & $\log_{10}|\bar n|$ & $n_{\theta}$ & $b$ & $m_{\rm{lim}}$ & $\log_{10}|\bar n|$ & $n_{\theta}$ & $b$ \\ 
      \hline
      BX & $R$ & \cite{Reddy08} & 25.5 & -2.06 & 26300 & -- & 24.0 & -3.22 & 2000 & 4.0 \\ 
      \hline
      $u$-dropouts & $i$ & \cite{Cooke13} & 24.6 & -3.15 & 2220 & 4.0 & 24.0 & -3.84 & 500 & 5.2 \\
      \hline
      $g$-dropouts & $i$ & \cite{Ono18} & 25.8 & -2.45 & 5250 & 3.2 & 25.5 & -3.67 & 330 & 3.8 \\
      \hline
      $r$-dropouts & $z$ & \cite{Ono18} & 25.8 & -3.00 & 1300 & 5.4 & 25.5 & -4.23 & 100 & 5.9 \\
      \hline
    \end{tabular}
    \caption{Selected properties for our fiducial photometric and spectroscopic dropout samples, including the comoving density, $\bar n$, in units of  $(h^{-1}\rm{Mpc})^{-3}$ and the projected density, $n_\theta$, in units of deg$^{-2}$.}
    \label{tab:selection}
\end{table}

The story is noticeably different at $z\simeq 4$, where $g$-dropouts can be efficiently selected to $i<25.8$ for a photometric sample.  This reflects the greater LSST depths in $g$ and $r$ due to the relative ease with which depth can be acquired in these bands.  The density would be significantly higher as a result, at 5250 galaxies per sq.\ deg.  At this depth, contamination is much reduced given that the interlopers are primarily bright and therefore much less dense; see Fig.~5 of ref.~\cite{Ono18}.  Of these galaxies, the expected fraction of Lyman-$\alpha$ emitters with >~50 \AA \ rest-frame equivalent width is shown in Fig.~\ref{fig:lines}.  On this basis, 10\% are likely strong emitters that may perhaps be pre-selected based on a $(i-z)<0$ cut \cite{Stark10}.  The bias we provide is tailored to LBGs and not revised for LAEs.  As such, it may be an overestimate given evidence that LAEs preferentially exist in lower mass haloes -- where the intrinsic dust content and therefore opacity is reduced. 

It is a similar situation at $z\simeq 5$, where the applied colour selection (e.g.\ that of ref.~\cite{Ono18}) results in a much reduced target density even for the photometric sample.  This perhaps reflects the greater possibility for confusion with stars and brown dwarves.  For $z<25.8$, there are 1,300 galaxies per sq.\ deg.\ with a similar 10\% LAE fraction and minimal interloper contamination, yielding perhaps 100 successful spectroscopic redshifts per sq.\ deg.  Note this is nominally fainter than the depths quoted by DESC, but within the magnitude 27 remit declared by LSST.  We leave further investigation of these samples to the future, but consider the facilities that may deliver them in the following section.  We forecast the likely returns for some potential science cases in \S\ref{sec:cosmology}.

\subsection{Potential spectroscopic facilities}
\label{sec:surveys}
Based on the above, we find a fifty minute exposure is reasonable for delivering our fiducial spectroscopic sample with an M-DESI experiment.  From this, we may estimate the relative exposure times based on the known scalings with area and resolution.  Beyond this, the relative significance of the instrument properties depends largely on the science case considered, as reflected by the figure-of-merit we define in Appendix \ref{app:FOM}.  For instance, the large field-of-view ensures DESI remains competitive for $f_{NL}$ despite the long exposure time required, when compared to those listed in Table \ref{Table:Spectroscopy}.  \rfree{While} a greatly increased multiplex pays dividends for a redshift-space distortions analysis due to the increased number density required to accurately sample small scales.  On this basis, M-DESI, MSE and BOA reflect logical steps for next generation facilities, given the factors of three to five improvement in the FOM between each while the BOA and SpecTel proposals represent more ambitious order-of-magnitude gains over current instrumentation.  There are further real-world restrictions that are not reflected by this FOM.  For instance, if the available (dark) time is dedicated to this science case, which increases the competitiveness of instruments such as DESI, M-DESI or BOA, while the imaging required must also be available on the relevant timeline.  This may limit the nearer-term potential e.g.\ of extensions to the DESI survey.  However, as an existing instrument, any proposed schedule is much more achieveable.  As such, perhaps the clearest opportunity is a modest proposal that determines the redshift distribution of the photometric samples and facilitates cross-correlation with CMB lensing. 
 
Of the properties in Table \ref{Table:Spectroscopy} under active consideration for future surveys, sufficiently cooled (77K) Germanium CCDs allow an extended coverage to $1.7\mu$m for MSE and ESO SpecTel\footnote{\url{http://www.iap.fr/pnc/pncg/Actions_files/eso_moswg_report.pdf}}.  This likely yields minimal returns as the coverage available is sufficient for LAEs well beyond the redshift range of interest, while the strong atmospheric cutoff at $1.3\,\mu$m prevents contiguous coverage even with a resolution high enough to mitigate the greatly enhanced sky background \cite{SEllis17, Dawson18}.  Although this does allow for the targeting of additional emission lines, e.g.\ [OII] or H$\alpha$, much of the returns of the $z < 2$ Universe will be well sampled by both Euclid and WFIRST.  One further consideration is the accuracy of sky subtraction required, which will approach percent-level requirements or even stricter.  This may affect any of the exposure time, effectively doubling that required if e.g.\ nod-and-shuffle \cite{Glazebrook01}; the fiber budget, if a large fraction of sky fibers is necessary due to spatial variation; or the coverage, depending on the mitigation strategy adopted.  This is less egregious for slit-based spectroscopy, but at the expense of greater source confusion, particularly at these depths. 

\section{Implications for cosmological studies}
\label{sec:cosmology}
\subsection{Likelihood of $C_{\kappa g}$ detection}
\label{sec:ckg_like}
Having previously considered the science case for CMB lensing cross-correlation and established a feasible dropout sample, we now forecast the detection significance for this science case.  We consider current and next generation CMB experiments, namely Planck, Advanced ACT, Simons Observatory and CMB-S4, as defined within the broader field in Table~\ref{Table:CMBxps}.

For a vanishing tensor-to-scalar ratio, the CMB has a purely $E$ polarisation and therefore there is no $EB$ correlation in the absence of lensing.  A measured $EB$ correlation is therefore a direct -- realisation dependent -- tracer of the lens distribution.  As such, the lensing S/N becomes dominated by $EB$ for a high-fidelity map -- an RMS of 3-5$\mu$K, as achieved by CMB-S4; see Fig.~47 of ref.~\cite{CMBS4}.  A further advantage is that polarisation is less sensitive to foregrounds than temperature typically, but perhaps not fundamentally so \cite{Schaan18b}.  We therefore forecast an internal maximum likelihood estimate of the CMB lensing potential by appealing to iterative delensing \cite{Smi12} and follow ref.~\cite{Schmittfull18} in reducing the  $EB$ reconstruction noise, and therefore total noise, by a factor of 2.5 for CMB-S4.  Otherwise, we assume that appropriate for the standard quadratic estimator \cite{Hu00}.  We neglect the contribution from lensing magnification throughout, as   this can be modelled and represents an $\mathcal{O}(10\%)$ effect at high redshift and will likely not change our S/N calculations significantly.

\begin{figure}[t!]
  \begin{subfigure}{0.5\textwidth}
    \centering
    \includegraphics[width=\linewidth]{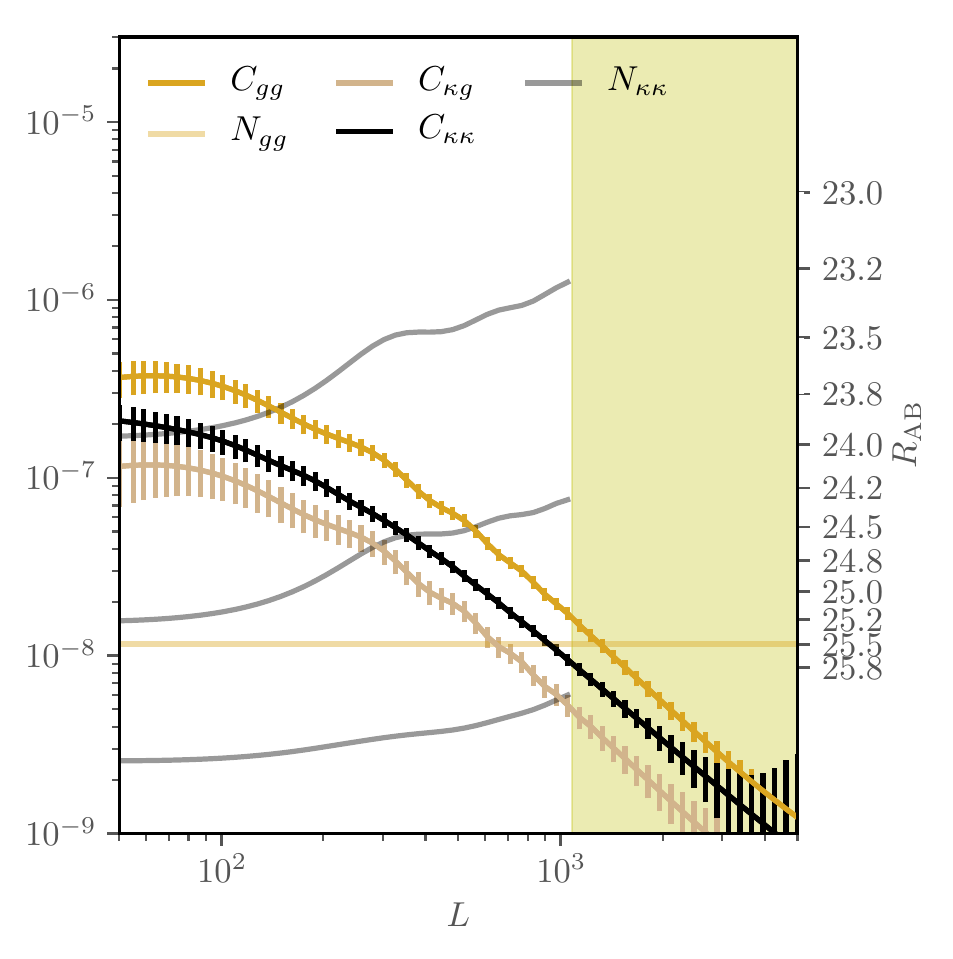}
  \end{subfigure}%
  \begin{subfigure}{0.5\textwidth}
    \centering
    \includegraphics[width=\linewidth]{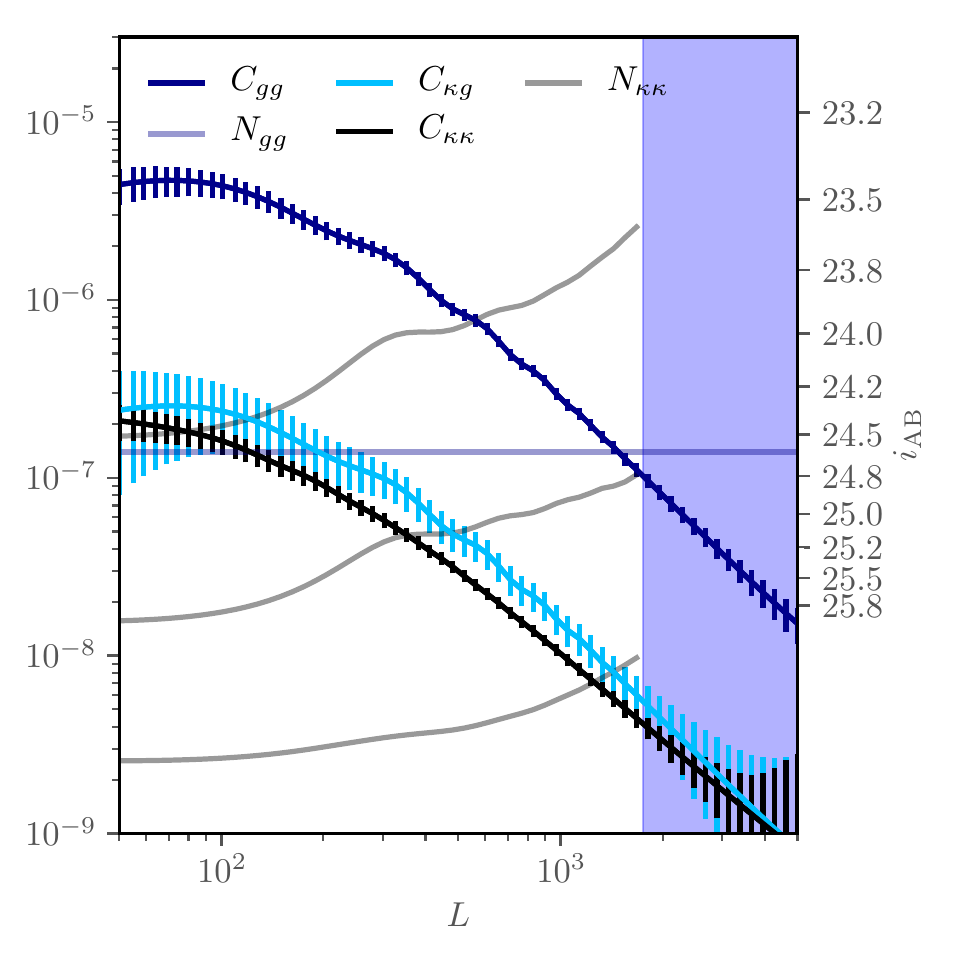}
  \end{subfigure}
  \begin{subfigure}{0.5\textwidth}
    \centering
    \includegraphics[width=\linewidth]{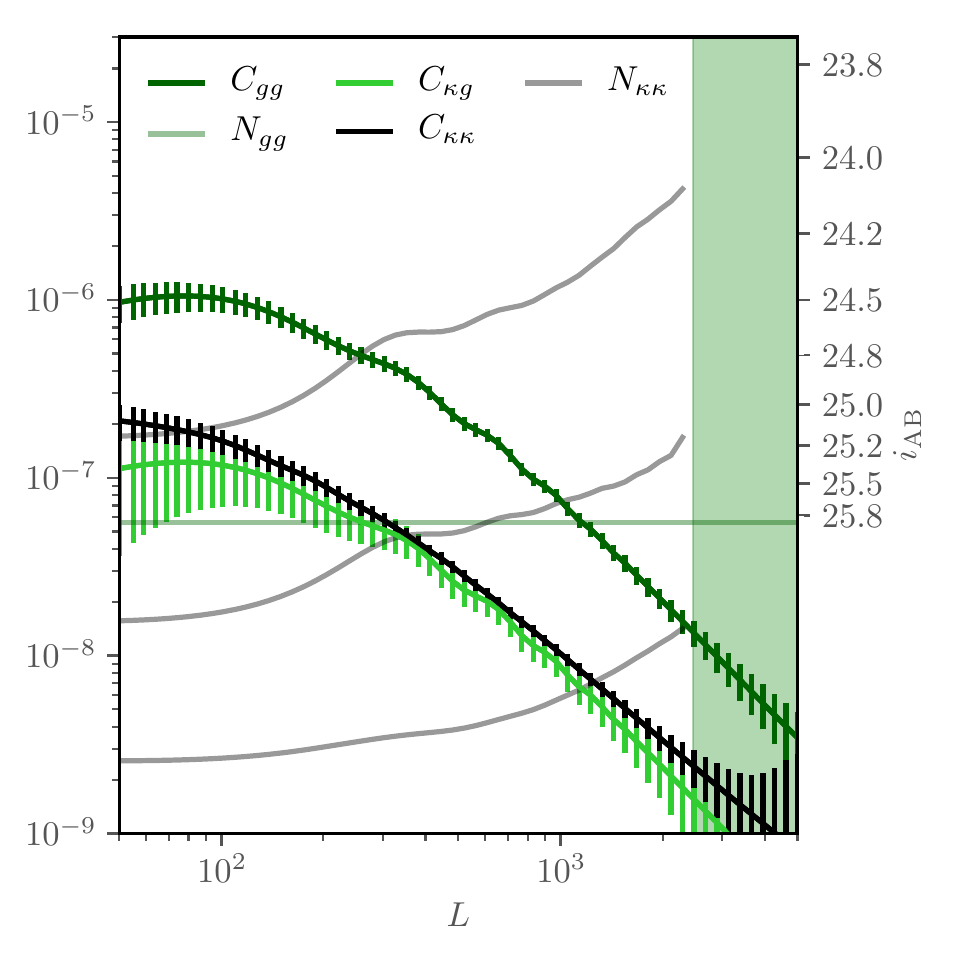}
  \end{subfigure}
  \begin{subfigure}{0.5\textwidth}
    \centering
    \includegraphics[width=\linewidth]{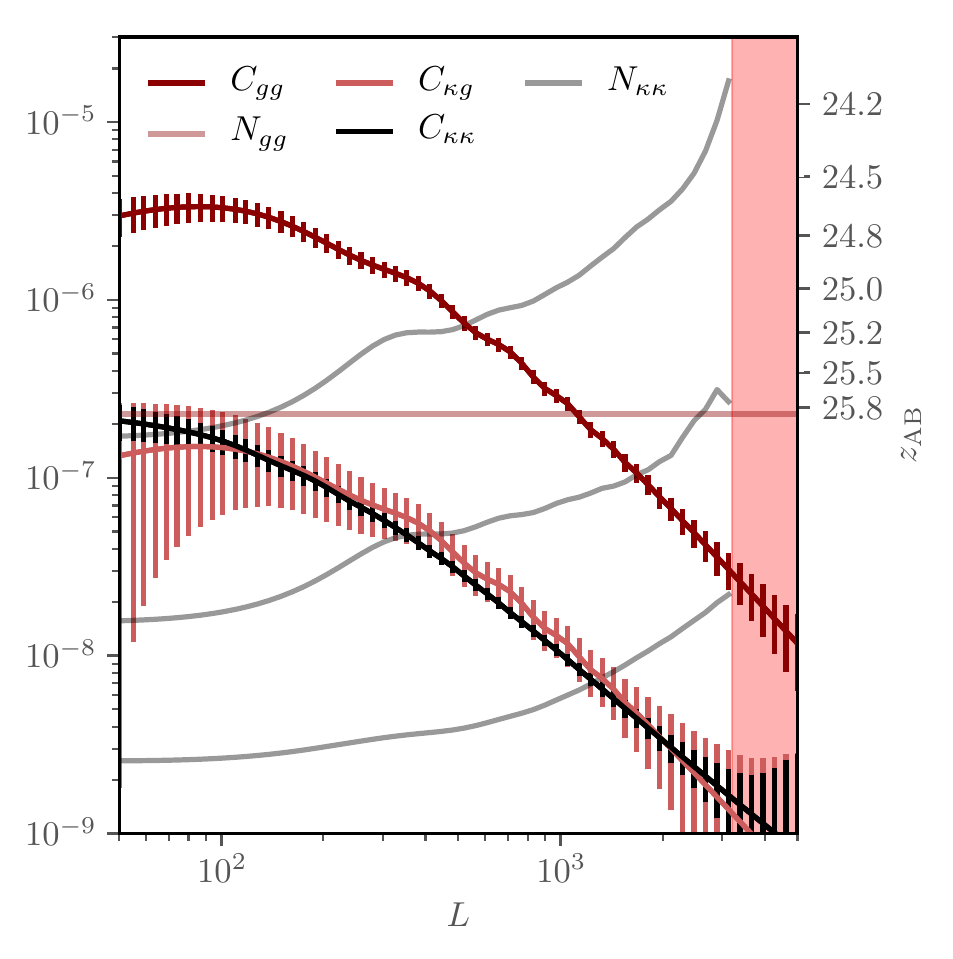}
  \end{subfigure}
\caption{Expected signal, $C_{gg}(L)$, statistical error and shot noise, $N_{gg}$, for the auto and cross-spectra of BX, $u$, $g$ and $r$ dropout samples (top left, right, bottom left, right) with a CMB-S4 (iterative delensing) estimate of the lensing potential.  For this, we assume the fiducial photometric samples shown in Table \ref{tab:selection}.  The right-hand axis shows the detection magnitude required to achieve the desired shotnoise.  The horizontal line shows the fiducial shotnoise in each case, while grey curves shows the lensing reconstruction noise, $N_{\kappa \kappa}$, for Adv-Act, SO and CMB-S4.  For comparison, the shaded band shows an estimate for the modelling limit based on the Zeldovich approximation. 
}
\label{fig:cgg}
\end{figure}

To calculate the detection significance, three input curves are necessary for each dropout sample, namely the $p(z), b(z)$ and $\bar n_{\theta}$ properties we have established previously.  
Fig.~\ref{fig:cgg} shows the potential limiting factors for a $C_{\kappa g}$ detection with dropouts -- reconstruction noise, shot noise and a proxy for the modelling limit.  The CMB-S4 $C_{\kappa \kappa}$ lensing reconstruction has a per mode S/N greater than unity beyond $L=1000$ and is not a limiting factor for any of the cases.  The $L_{\rm{max}}$ imposed by modelling limitations is likely constraining for $BX$ and $u$-dropouts, but negligible compared to the shotnoise for $g$-dropouts and unrestrictive for $r$-dropouts at our fiducial (photometric) depths.  Irrespective of the limiting factors, highly precise measurements are clearly possible in each instance.  Propagation of this detection significance to cosmological parameters is beyond the scope of this current exploration, but of interest for the future.  Both refs.~\cite{Schmittfull18} and \cite{Yu17} provide the necessary framework.  

Fig.~\ref{fig:snr} shows a  quantitative estimate of the detection significance, providing the expected cumulative signal-to-noise for a measurement of $C_{\kappa g}$ with $u$ and $g$ dropouts.  Note the y-axis corresponds to $f_{\rm{sky}}$ of unity for simplicity, with estimates scaling as $\sqrt{f_{\rm{sky}}}$.  We see that increasing returns are reached for $u$-dropouts with greater projected density up to $\simeq 10^4$ deg$^{-2}$, and similarly for $g$-dropouts.  CFIS may also be sufficiently interesting given the $f_{\rm{sky}}=0.25$ area and expected $90 \ \rm{deg}^{-2}$ $u$-dropouts and $300 \ \rm{deg}^{-2}$ $g$ dropouts given Fig.~\ref{fig:SchCounts}.  With a northern footprint, CFIS will overlap with Advanced ACT, but not SPT, in this respect.

\subsubsection{Foreground delensing}
\label{subsec:fgdelensing}
\rfree{The previous section estimated the detection significance for the correlation of a given dropout sample and the reconstructed CMB lensing map, $\hat{\kappa}$.  However, this is needlessly pessimistic because low-$z$ lenses that source fluctuations in $\hat{\kappa}$ are not sampled by our high-$z$ dropout galaxies, leading to a decorrelation between the two tracers and therefore a lower detection significance.  This significance can be raised simply by using already acquired low-$z$ galaxies to measure the exact realisation of extraneous fluctuations and cleanly subtract it, i.e.\ first delensing $\hat \kappa$.  In this way, we form a refined estimate of the $\kappa$ sourced by the high-$z$ Universe, $\hat \kappa '$, and exploit the low-$z$, high number density samples that have already been acquired.}

\begin{figure}[t]
    \centering
    \begin{subfigure}{.5\textwidth}
    \includegraphics[width=\linewidth]{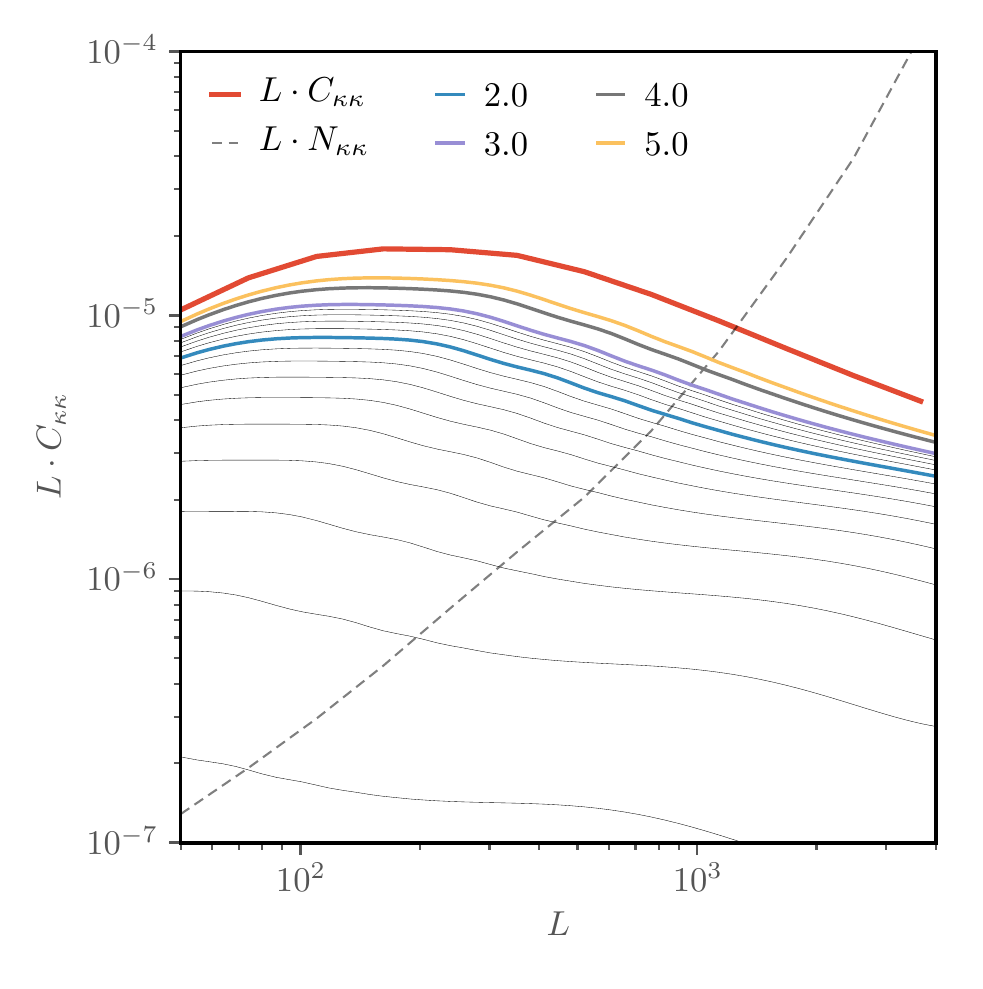}
    \end{subfigure}%
    \begin{subfigure}{.5\textwidth}
    \includegraphics[width=\linewidth]{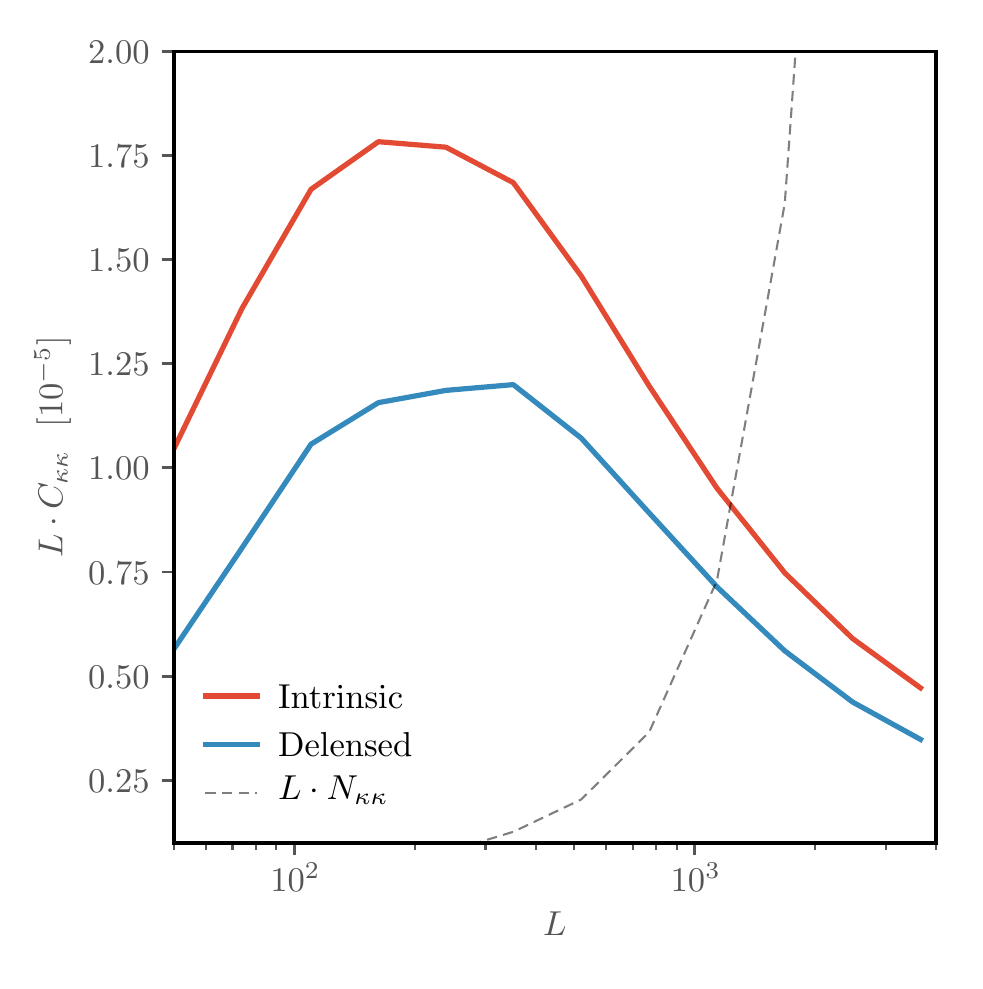}
    \end{subfigure}
\caption{\textbf{Left}: Cumulative form for $L \cdot C_{\kappa\kappa}(L)$ as a function of redshift.  The dashed line shows the CMB-S4 noise curve, while the solid red line shows the full signal.  Black lines starting in the lower left show the contribution from $z<0.2$, $0.4$, ..., up to $z=2$ (blue line), and additional cumulative contributions to the redshifts shown.  \textbf{Right}:  The efficiency of delensing of $\hat \kappa$ with low-$z$ spectroscopic may be quantified by the reduction in $C_{\kappa \kappa}$.  We find that BOSS and DESI can reduce this by 50\% and even greater for CMB-S4.
}
\label{fig:yu}
\end{figure} 
To do so requires assuming a fiducial lensing kernel in order to form an estimate of the low-$z$ lensing potential.  To date, this delensing has seen most applications for the suppression of cosmic variance on the $B$-mode power spectrum when determining the tensor-to-scalar ratio, which is primarily derived from $L<100$.  Conversely, our focus is on the Limber regime, $L>60$, and science cases dominated by small scales, e.g.~a measurement of neutrino mass with $\sigma_8(z)$ \cite{Yu18}.  

The cross-correlation coefficient for the optimally combined tracer \cite{Smith12, Sherwin15, Yu17},
\begin{equation}
    \rho^2(L) = \rho_{i \kappa} \left ( \boldsymbol \rho^{-1} \right )_{ij} \rho_{j \kappa}, 
\end{equation}
quantifies the effectiveness of this approach.  Here $\rho_{i\kappa}$ is the cross-correlation coefficient between tracer $i$ and $\kappa$, and the elements of $\boldsymbol \rho$ are these same coefficients but between the galaxy samples.  As the variance, and hence significance, of the dropout cross-correlation is\footnote{Further background may be found in e.g. ref. \cite{Modi17a}.}
\begin{equation}
  {\rm Var}\left[C_{\kappa g}(L)\right] =
  \frac{1}{(2L+1)f_{\rm sky}} \left\{
  \left(C_{\kappa\kappa}+N_{\kappa\kappa}\right)
  \left(C_{gg}+N_{gg}\right) +
  \left(C_{\kappa g}\right)^2\right\},
  \label{eqn:var_ckg}
\end{equation}
assuming the fields are Gaussian, the S/N on a detection of $C_{\kappa' g}(L)$ can be higher than that of $C_{\kappa g}(L)$ by $C_{\kappa' \kappa'}(L) = \left [ 1 - \rho^2(L) \right ] \ C_{\kappa \kappa}(L)$ if the  reconstruction noise is small. 

Fig.~\ref{fig:yu} shows how each redshift shell contributes to $C_{\kappa\kappa}(L)$ and therefore the extent to which this variance can be suppressed given sufficiently dense tracers residing in each shell.  Adapting Fig.~1 of ref.~\cite{Sherwin15} and Fig.~5 of ref.~\cite{Schmittfull18}, we estimate the actual delensing efficiency given known spectroscopic samples (BOSS and DESI, including QSOs) in the right panel of the same figure.  The potential for systematics derived from catastrophic redshift failures is much mitigated by this restriction, at the  cost of a more inefficient delensing.  The Legacy Survey \cite{Dey18} and recently approved SPHEREx space mission\footnote{\url{http://spherex.caltech.edu}} of low-$z$ galaxies over the full sky has much potential, and risk, in this respect.  We see that even with spectroscopy, a significant delensing fraction is possible -- 50\% and greater.

\begin{figure}[t!]
  \begin{subfigure}{0.5\textwidth}
    \centering
    \includegraphics[width=\linewidth]{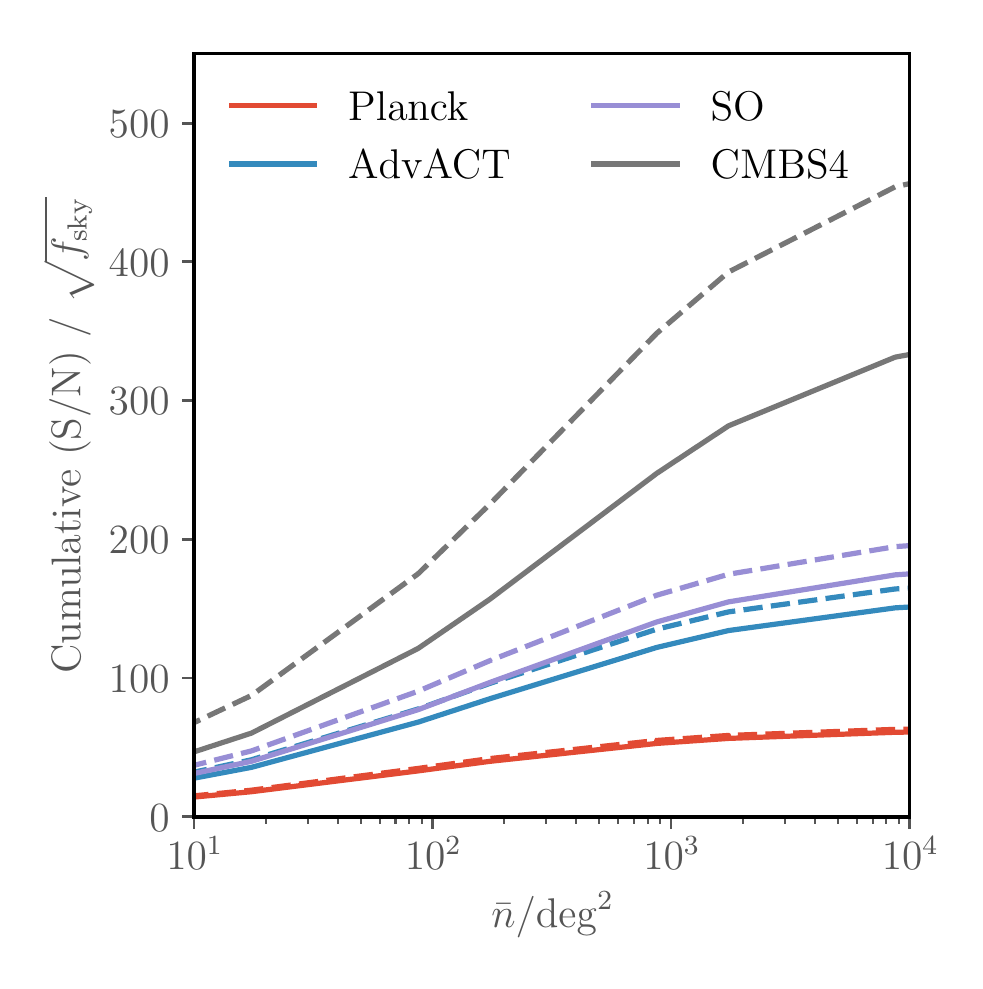}
  \end{subfigure}%
  \begin{subfigure}{0.5\textwidth}
    \centering
    \includegraphics[width=\linewidth]{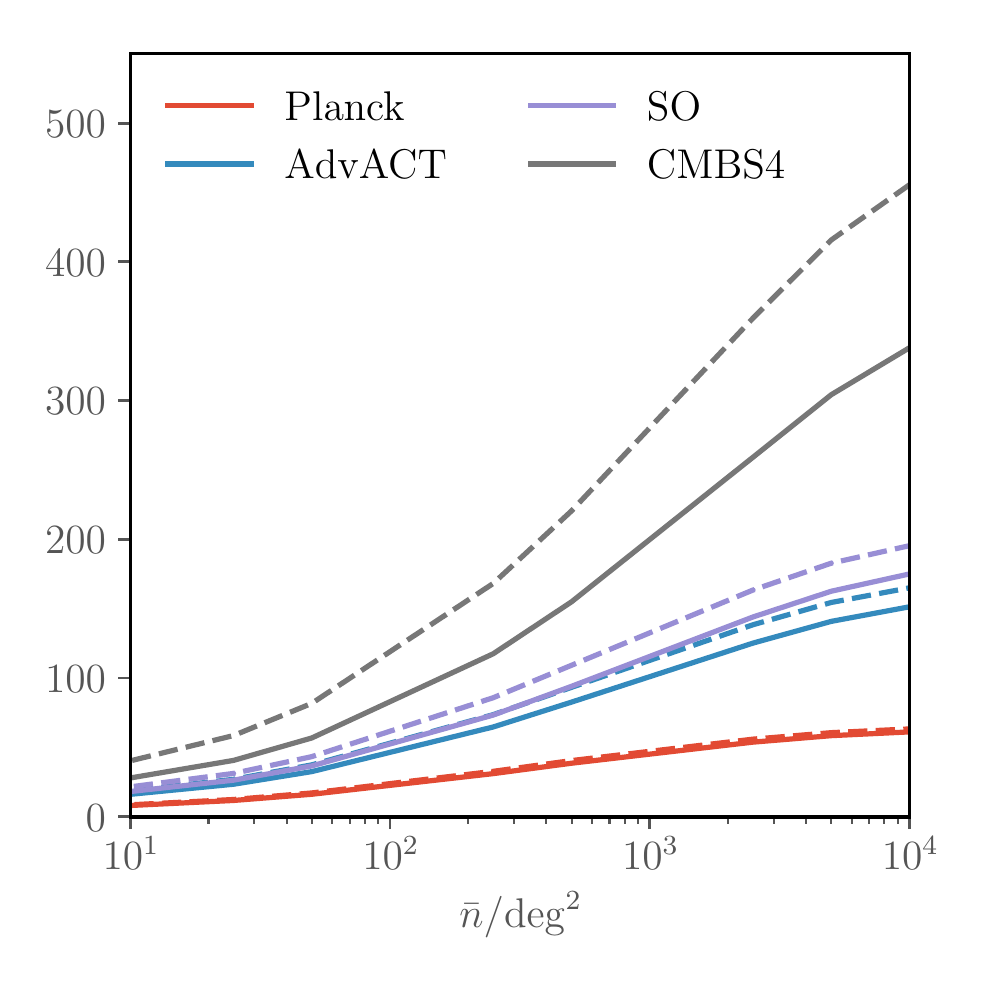}
  \end{subfigure}
  \caption{Cumulative S/N curves for the cross-correlation of the given CMB lensing experiments with our fiducial $u$ (left) and $g$ (right) dropout samples; note we assume an $f_{\rm{sky}}$ of unity for simplicity.  A hard $L_{\rm{max}}$ modelling limit has been applied, as further described in the text, but the limit is effectively set by the reconstruction noise on $\kappa$ for all but CMB-S4.  Note that the non-linear halo bias shown in \S\ref{sec:clustering} would raise the expected signal-to-noise above these values.  Dashed lines indicate the same curves when employing delensing with planned low-$z$ spectroscopic surveys (BOSS \& DESI), for which catastrophic outliers are better controlled.}
\label{fig:snr}
\end{figure}

One can then reconsider our estimates from \S\ref{sec:ckg_like} for this high-$z$ only $\hat \kappa'$ map, which should provide the relevant signal-to-noise for $\sigma_8(z)$ at $z>2$.  As we consider $L>50$, we ignore the internal Planck $\kappa$ estimator that is noise dominated beyond this $L$.  The dashed lines of Fig.~\ref{fig:snr} show these revised signal-to-noise estimates when including this spectroscopic low-$z$ delensing.  We can see that the greatest returns are for CMB-S4 given the much reduced reconstruction noise, achieving a $\simeq 30$\% increase in the significance.  With an increased map RMS, and therefore reconstruction noise, this boost is suppressed for both SO and Advanced ACT; a fact compounded by the inability of AdvACT and SO to reach a map RMS where a maximum likelihood lensing estimate outperforms the simple quadratic estimator.

One complication is that, at sufficiently low-$z$, delensing in this manner is unfeasible even at moderate $L$, as non-linearity and complex bias cause decorrelation of the galaxy distribution from the matter field as shown in Fig. \ref{fig:wt}; but low-$z$ contributes a few percent to $C_{\kappa \kappa}(L)$ and may also be removed by LSST shear measurements.  The overlap in redshift and the signal-to-noise ratio are lower in this case, but likely sufficient.

Finally, we note that Figs.~3 and 6 of ref.~\cite{Smith12} also suggests an infinitely dense tracer to $z \simeq 4$ is as effective at delensing the $L<100$ $B$-mode power as an $EB$ measurement of a ($2'$, 2$\mu$K-arcmin) CMB experiment, which would nicely complement and provide a valuable cross-check on a LiteBIRD, COrE or similar experiment.  In this respect, dropout samples would play a role competitive with Cosmic Infrared Background delensing.  In particular, the finer redshift resolution provided by dropouts would allow for an optimal redshift weighting.

\subsubsection{Cleaning of dropout interlopers}
\label{sec:lowz_cleaning}
We have previously seen that selection of high-$z$ galaxies from optical imaging inevitably leads to undesirable interlopers, e.g.~Fig.~\ref{fig:int_nz}.  In many cases, the contaminating redshift range is well-known in advance -- on the basis of the relatively small number of SED features, e.g. confusion of the Lyman and Balmer breaks, or with the aid of deeper fields -- and typically resides at low-$z$ where we have already mapped, or will soon map, the large-scale structure with high fidelity.  Whereas, in the previous section, we assumed the dropouts had zero contamination and effectively treated the low-$z$ $\hat \kappa$ as a contaminant to be removed, we may instead use the same formalism and low-$z$ spectroscopic redshifts to instead clean the dropout sampled projected density map.  The optimal method for removing the signal depends upon the degree to which the interloper $b \cdot dN/dz$ and properties are known and the fidelity of the map being used in the cleaning.  If the redshift range is known, but the overall amplitude of the signal is not, then a method based upon cross-correlations may be used.  As more information is gained, a joint fit involving priors on the nuisance parameters should perform better.

As an example, we imagine removing the $z<1$ contribution from a $u$-dropout sample with e.g. BOSS or DESI and a minimum variance method.  Specifically, the dropout sample, $g$, is a mix of low and high redshift populations.  We assume these are uncorrelated since they are widely separated in redshift and we neglect the small contribution from lensing as a first approximation.  If we have an additional low-$z$ spectroscopic sample, $t$, and good knowledge of $b \cdot dN/dz$ of the interlopers, then we can match the tracer to the interloper sample and remove the bias.  One approach is to minimize the variance of a weighted difference between the fractional overdensities, $\delta_g$ and $\delta_t$, to isolate the high redshift piece.  The result is $\delta'_{\ell m}=\delta^g_{\ell m}-w_\ell\delta^t_{\ell m}$ with $w_\ell=C_\ell^{gt}/C_\ell^{tt}$ (where $C_\ell^{tt}$ includes shotnoise) as per the standard Wiener filter.  
This can be implemented directly at the two-point function level, viz.
\begin{equation}
    C^{\kappa g'} = C^{\kappa g} - \frac{C^{gt}C^{\kappa t}}{C^{tt}},
    \qquad \qquad
    C^{g'g'} = C^{gg} - \frac{(C^{gt})^2}{C^{tt}}.
\end{equation}
For high S/N, we expect excellent cleaning up to the scale of decorrelation \cite{Modi17a}.  Further gains would require a more sophisticated forward model of both the low and high-$z$ populations.  It is straightforward to also include lensing magnification in this case, while self-consistently inferring $dN/dz$ from the large-scale cross-correlation with spectroscopic tracers, as we discuss next.

\subsubsection{Clustering redshifts}
\label{sec:mcqw}
\begin{figure}[t]
  \centering
  \begin{subfigure}{\textwidth}
  \includegraphics[width=\linewidth]{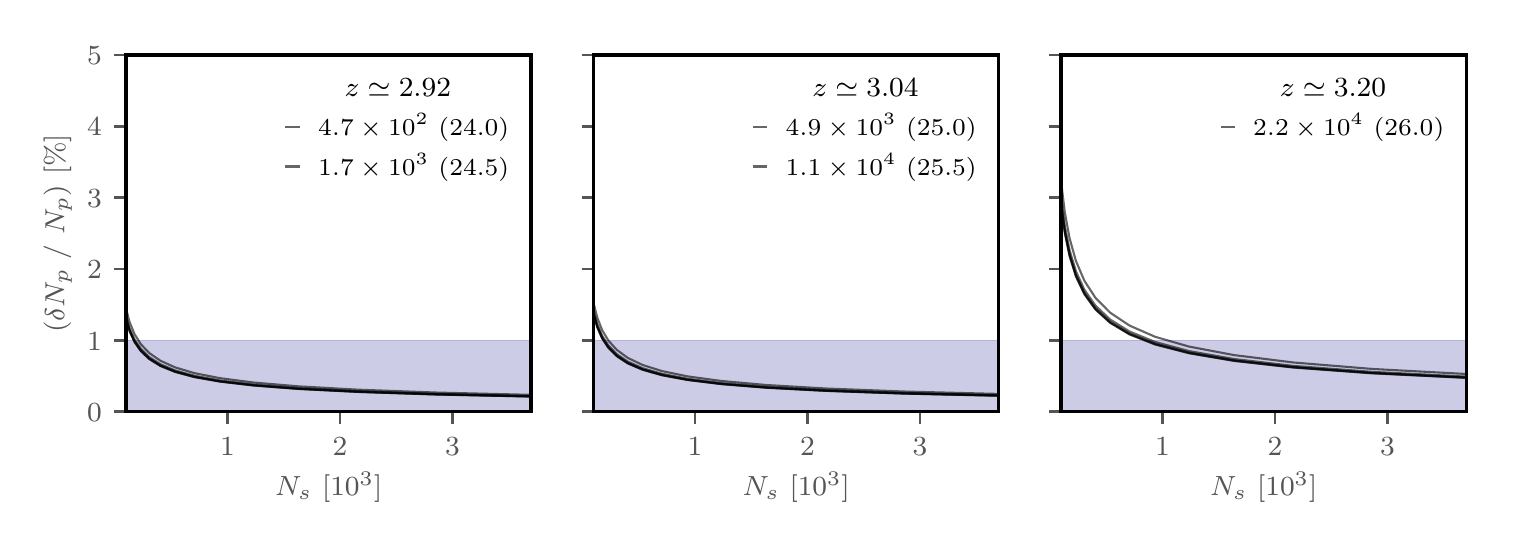}
  \end{subfigure}
  \begin{subfigure}{\textwidth}
  \includegraphics[width=\linewidth]{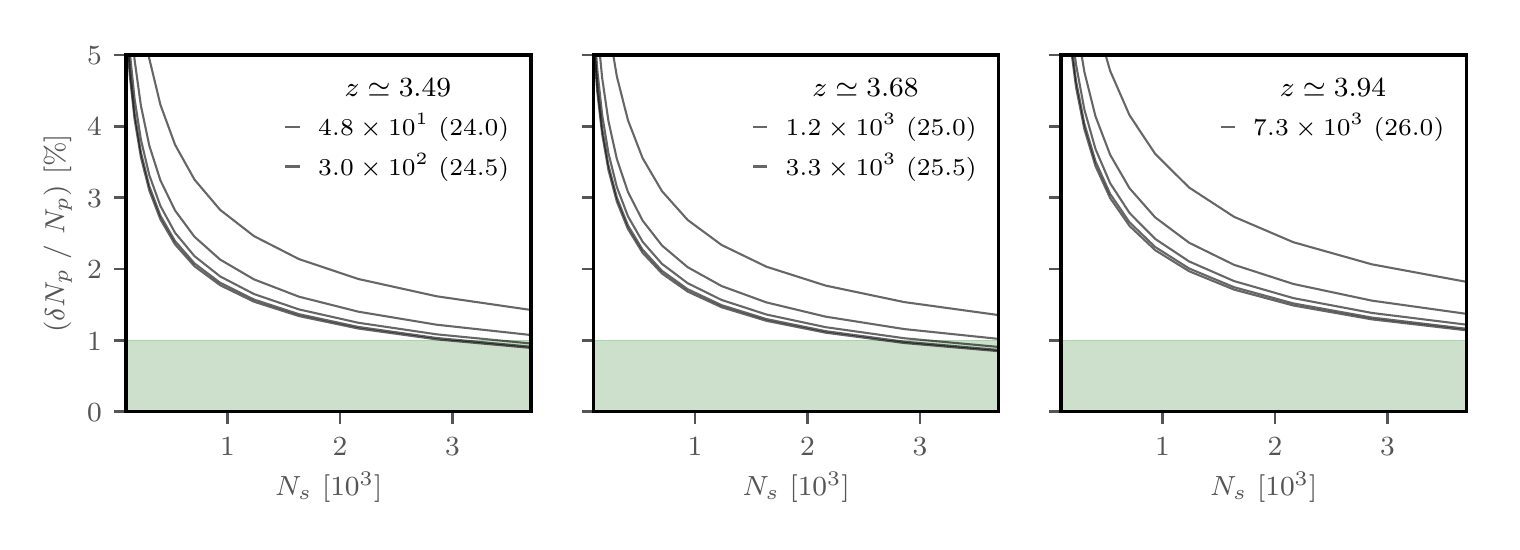}
  \end{subfigure}
  \caption{\textbf{Top}:  Error on $dN_p/dz$ in $dz=0.1$ redshift shells vs.~number of dedicated spectroscopic redshifts required per deg${}^2$ for $u$-dropouts.  We show the constraints at the quartile redshifts of $p(z)$, for photometric number densities and magnitude limits as listed in the legend.  For simplicity, we assume an overlap area of $f_{\rm{sky}}=10^{-3}$, with the errors scaling as $\sqrt{f_{\rm{sky}}}$ (for surveys with areas $\gg$ 1 deg$^2$).  \textbf{Bottom}:  Similarly for $g$-dropouts.  Note that the errors increase in the tails of $p(z)$ where the shotnoise is large in both samples, but that percentage-level constraints are feasible by a few thousand spectroscopic gals.  deg$^{-2}$.  For this, we assume low-$z$ surveys will mitigate the impact of low-$z$ interlopers and hence assume a strong prior on the redshift limits.} 
\label{fig:mqw}
\end{figure}
Having established a feasible spectroscopic sample in \S \ref{sec:fiducial}, we now consider the constraints that can be placed on the priors of a (discretized) photometric $dN/dz$ with \emph{dedicated followup spectroscopy} over 0.1\% of the sky.  Specifically, we apply a Fisher formalism to forecast the gains achievable with a large-scale, minimum-variance, quadratic estimator of the angular cross-correlation between the dropout sample, $p=N_p \ \cdot \ \delta_p(\ell, m)$, and overlapping objects in a spectroscopic sample, $s$, residing in the same, relatively fine, shells in spectroscopic redshift, $s_i$.  This provides sufficient information to reconstruct a proxy for the photometric redshift distribution, i.e.\ the number of photometric galaxies in this same spectroscopic bin, $N_i$.  A real-world complication is that is it $b_i \cdot N_i$ that may be recovered \cite{Phillipps87, Padmanabhan07, Ho08, Newman08, McQWhi13, Bond98}.  Nevertheless, the primary advantage of this approach over photometric training samples is that the spectroscopic sample need not replicate the distribution of the photometric in physical properties, but must only fall within its sky coverage, as is the case for our fiducial survey.  Therefore a significantly faster spectroscopic survey speed is achieved as a result.  

Quantitaively, the error with which $b^p_i \cdot N_i$ can be recovered is \cite{McQWhi13}
\begin{equation}
    \sigma_i^{2} = F^{-1}_{ii} \quad \mathrm{where} \ \  F_{ii} = \frac{S(1+2S\,r_i^2)}{\langle p^2 \rangle\ \langle s_i^2\rangle} \left(\partial_i \langle p s_i \rangle\right)^2, 
\end{equation}
and the Schur parameter, $S=(1 - \sum r_i^2)^{-1} \geq 1$, quantifies the loss of correlation between the spectroscopic tracer in shell $i$ and the photometric sample, $p$, due to a misoverlap in redshift or the decorrelation of galaxies and matter.  Here $r_i$ is the cross-correlation coefficient of $p$ and $s_i$ and the expression (and correlators, e.g.\ eqns.~(6) and (8) of ref.~\cite{McQWhi13}) are those in the Limber limit -- in which the Fisher matrix reduces to these diagonal components; see eqn.~(30) of the same reference.  
We assume a satellite fraction of zero and take the overlap fraction, $f_{\rm{over}}$, to be unity, such that the rarest min($N_s, N_p$) sources are the same in both samples.  The former would be appropriate, if e.g. emission line galaxies formed a subset of the LBGs and were underrepresented among satellites.  Finally, this is a prediction for the error on the number of galaxies in each redshift, whereas the relevant quantity is often the mean redshift for each the shell, which would be significantly smaller.

In the case of complete redshift overlap of $s$ and $p$ and in the absence of shotnoise, $S\to \infty$ and the products $b_p \cdot N_p(z)$ are perfectly constrained.  If the unknown sample is limited by shotnoise, or if there is imperfect redshift overlap, then $S\to 1^{+}$ and many modes are required to compensate for the decorrelation.  In the derivation, $\delta_p(\ell, m)$ is further assumed to be Gaussian so the estimator is valid only on large, linear scales.  This limits the number of modes available, and thus the precision, in the absence of shotnoise.

Fig.~\ref{fig:mqw} shows the error on $dN/dz$ of the photometric sample as a function of its depth (lines) and the number of spectroscopic objects (abscissa) for $u$ and $g$-dropout samples.  The three panels in each row show the fractional error on the number of photometric galaxies in three redshift slices of width 0.1 chosen at the $p(z)$ quartiles.  With a few thousand spectroscopic galaxies per sq.\ deg.\ over $f_{\rm sky}=10^{-3}$, percent-level constraints per redshift bin are obtainable for surveys deeper than $24.5$ magnitude.  Since the bins are largely independent \cite{McQWhi13}, this leads to tight limits on $dN/dz$ under the assumption of a smooth redshift evolution.

Despite being encouraging in terms of enabling a $C_{\kappa g}$ science case, in practice this has a number of further complications.  Typically the linear bias is a smooth, slowly varying function of redshift.  This is not the case for the sharp color boxes shown in Fig.~\ref{fig:ctrack}, as the distribution in galaxy type can \rfree{increase considerably} in the tails of $p(z)$.  A secondary complication is the correlation of low-$z$ samples with magnification in the high-$z$ shells, which does have the silver-lining that it may break the degeneracy with $b(z)$.  Dust in our own galaxy may also correlate different redshift shells in a similar manner \cite{Monaco18}.  Ref.~\cite{Sanchez19} suggests one means to jointly solve for many of these effects -- the color-redshift degeneracy, bias evolution, etc. -- by including both the flux measurements and density fluctuations in the likelihood. 

\subsection{Redshift-space distortions}
\label{sec:rsd}
\begin{figure}[t]
    \centering
    \includegraphics[width=\linewidth]{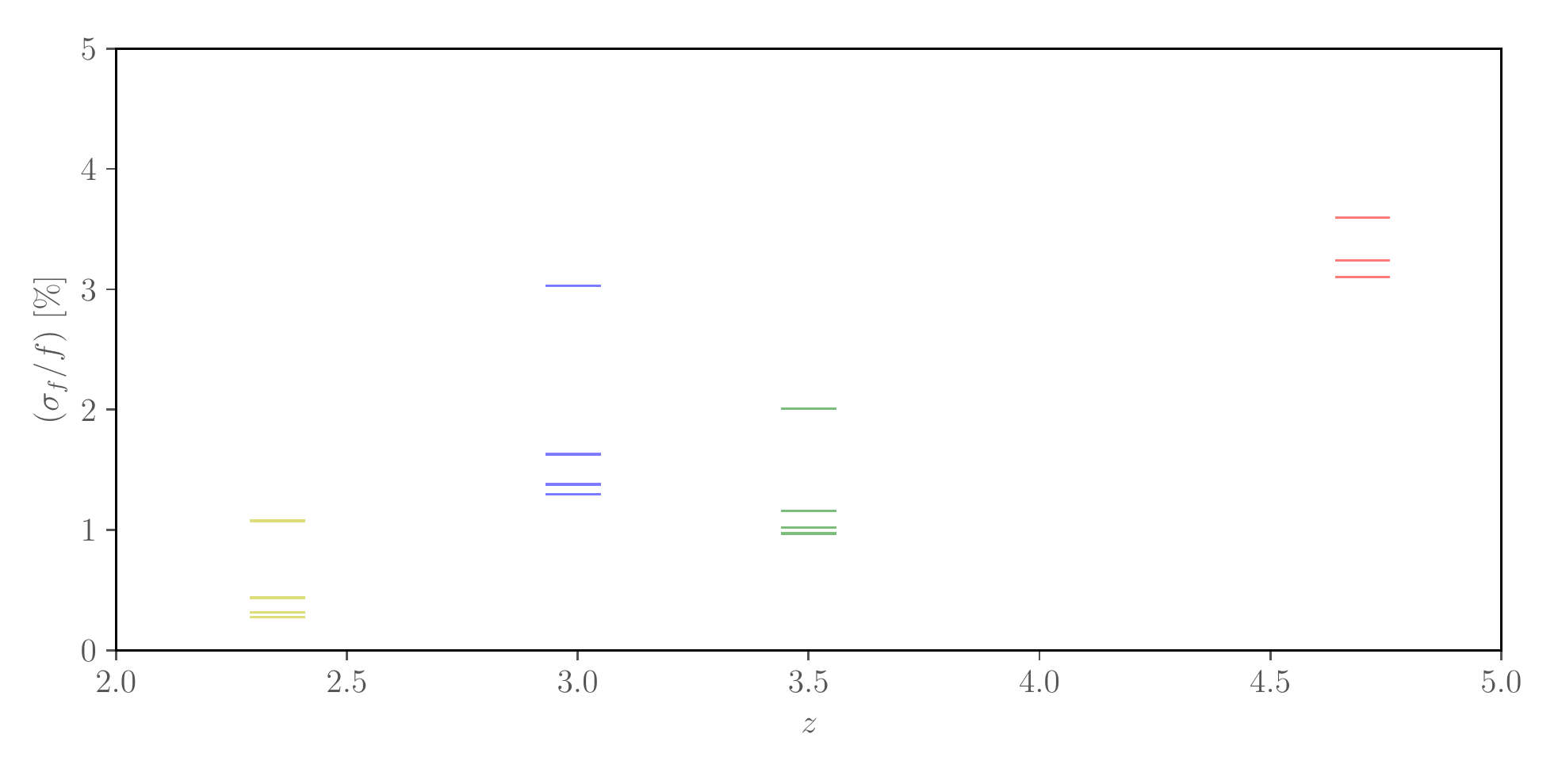} %
\caption{Fractional error on the linear growth rate from the anisotropic redshift-space power spectrum of our fiducial spectroscopic sample, as defined in \S \ref{sec:fiducial}.  The markers shown indicate the dropout samples at each redshift, with an ordering of $k_{\rm{max}} = 0.1$ to 0.4 \rfree{$\hompc$} (top to bottom).  The gains \rfree{rapidly saturate for larger wavenumber limits}, as can be seen from the convergence of the points.
}
\label{fig:RSD}
\end{figure}
Having shown cross-correlation of LBGs and CMB lensing to be a compelling science case for $z>2$, we now make preliminary estimates for the returns of a redshift-space distortions analysis with the fiducial spectroscopic sample described in \S \ref{sec:fiducial} and the linear bias model of eqn.~\ref{eqn:bz}.  For this science case, complete spectroscopic followup would be required \rfree{-- perhaps over a smaller area --} such that small-scale information is not lost to imprecise photometric redshifts.  

With the common assumption of a Kaiser model, and in the FKP approximation \cite{Feldman94}, we perform forecasts for $\sigma_f$ following ref.~\cite{White09}, see also ref. \cite{Tegmark97}.
We assume 15,000 deg$^2$ as our fiducial configuration.  There are two main restrictions on the critical wavenumber limit, $k_{\rm max}$.  The first is that it should represent modes which may be modelled, which we enforce by assuming $k_{\rm{max}} \cdot \Sigma$ is less than about unity; here $\Sigma$ is the (1D) Zeldovich displacement.  At these scales, perturbative models are valid \cite{Carlson13,Vlah16,Foreman16,Modi17a} and constraints are not overly impacted by marginalisation over a number of nuisance parameters describing non-linear structure formation and biasing.  We find $\Sigma^{-1}=0.41$, 0.54, 0.67 and $0.81\,h\,{\rm Mpc}^{-1}$ for $z\simeq 2, 3, 4$ and 5.

A  further limitation arises due to the difficulty of determining an accurate redshift from broad lines or systemic outflows.  An intrinsic redshift uncertainty, $\delta z$, corresponds to an uncertainty in radial distance of $\sigma_\chi=[c(1+z)/H(z)]\ [\delta z/(1+z)]$.  This reduces $P(\mathbf{k})$ by $\exp[-k^2 \mu^2 \sigma_\chi^2]$ and begins to reduce the achievable S/N whenever $\sigma_\chi$ is comparable to $\Sigma$.  This is the case for $\delta v=c\,\delta z/(1+z)=250\,{\rm km}\,{\rm s}^{-1}$ at $z=2$ and $170\,{\rm km}\,{\rm s}^{-1}$ at $z=5$.  A detailed investigation of the consequences of systemic redshift offsets due to the outflow of the emitting neutral hydrogen and resonance scattering is beyond our scope. As such, we crudely model this with \rfree{an effective} $\simeq 400\,{\rm km}\,{\rm s}^{-1}$ dispersion and plot the resulting constraints in Fig.~\ref{fig:RSD}.  

From this figure, we see that compelling constraints on the growth rate are possible, achieving percent level by $k=0.3\,h\,\rm{Mpc}^{-1}$ for our fiducial area, at which point the gains saturate due to the suppression of the signal by the neutral hydrogen outflow.  Here, we marginalise over the linear bias but not the dispersion, which might have a tight prior based on simulations.  On this basis, it is clear that the number density and intrinsic outflows are more limiting than \rfree{our estimate of the non-linear scale}, $\Sigma$, would suggest.  This would not be the case for the absorption line redshifts \rfree{we advocate at $z \simeq 2$ and 3}, 
\rfree{but at the cost of a much slower survey speed}. 

\section{Summary and conclusions}
\label{sec:discussion}  
Forthcoming generations of cosmological imaging surveys promise to map close to a celestial hemisphere at depths approaching twenty-seventh magnitude in six optical bands, together with twenty-fourth magnitude depth in the near-infrared \rfree{($Y$, $J$ and $H$)}.  Simultaneously, the next generations of CMB experiments will cover $\simeq20,000$ deg$^2$ at a few $\mu K$-arcmin sensitivity.  We therefore review and investigate the science enabled by a combination of CMB lensing and `dropout' selected $2 \leq z \leq 5$ Lyman-break galaxies.

Together with known local tracers, we find that \rfree{the physically motivated (Fig.~\ref{fig:lbgsed}}) and well-established color selections shown in Fig.~\ref{fig:ctrack} can achieve an efficient tomographic decomposition of the CMB lensing kernel to $z=5$, as shown in Fig.~\ref{fig:pz}, and \rfree{provide} a proxy for (the time evolution of) matter density fluctuations that \rfree{yields} compelling tests of General Relativity, inflation and neutrino masses via simultaneous measurements of the Bardeen potentials, curvature and local primordial non-Gaussianity.  These color selected samples \rfree{would} represent a subset of the photometric redshifts that are likely to be available, but could well comprise robust `Gold' samples that are both well-understood and require minimal imaging requirements (e.g.\ a dropout and one to two detection bands) -- resulting in the largest possible area and homogeneous selection, as favoured by our chosen science cases.

A literature search largely provided the modern versions of classic color selections necessary for LSST \cite{Cooke13, Ono18} and we provide an approximate conversion where necessary, e.g.\ for BX selection (App.~\ref{app:filter_conversion}, \cite{Adelberger04}) -- based on a linear regression between the model magnitudes of a large Bruzal and Charlot template set \cite{BC03} with a range of intrinsic extinction \cite{Calzetti00}.  We establish \rfree{useful characteristics, including} the projected number density with depth in Fig.~\ref{fig:SchCounts}, together with the completeness and contamination rates, and consider tailoring the color selection to CMB cross-correlation based on the dissimilar propagation of e.g.\ interloper bias \rfree{for} luminosity function and \rfree{CMB} cross-correlation analyses.  In particular, by degrading the ultra-deep Hubble UV-UDF \cite{Rafelski15} and CARS \cite{Hildebrandt09} studies to depths appropriate for LSST \& Euclid and examining the expected interloper characteristics and redshift distribution; see Figs.~\ref{fig:cat_ctrack}, \ref{fig:cat_ctrackNIR} and \ref{fig:int_nz}.  This represents the first steps to establishing this as a viable science case.   

Moving to the clustering of these populations, we consider how the apparent magnitude limit and redshift typically leads to highly biased tracers ($b\simeq 4-8$, Fig.~\ref{fig:bz}) with an \rfree{associated strong non-linear biasing}, see Fig.~\ref{fig:pk}.  Eqn.~\ref{eqn:bz} provides a straightforward relation for \rfree{the dependence of linear biasing on redshift}, which approximately interpolates between much of the available data.  This is a necessary input for the later cosmology forecasts we consider.  Using the legacy CARS \cite{Hildebrandt09} and GOLDRUSH \cite{Harikane18} angular correlation function measurements shown in Fig.~\ref{fig:wt}, we then provide a fresh study of the halo occupation using high-$z$ N-body simulations and consider how this might affect our CMB lensing science.

Having assembled the necessary framework for the dropout samples, viz.\ $b(z)$, $p(z)$ and $\bar n_{\theta}(m_{UV})$, we propagate this to the signal-to-noise on CMB lensing cross-correlation and examine the potential limiting factors in Fig.~\ref{fig:cgg}.  In Fig.~\ref{fig:snr}, we \warn{find configurations that are able to deliver $>100\,\sigma$ detection of the cross-correlation at $z \simeq 3$ and $4$ given sufficient overlap area};  \rfree{the precise degree of overlap remains uncertain given the continuing optimisation of the footprint for the various surveys}.  Refined estimates (dashed) that include delensing by already acquired low-$z$ spectroscopic redshifts deliver higher significance by a CMB-S4 map RMS, but may be limited by the \rfree{larger statistical error} of nearer-term experiments and difficulty in modelling halo biasing.    

A limiting factor to any $C_{\kappa g}$ science will likely be how well the redshift distributions of the dropout samples can be known, which will necessitate additional spectroscopic followup -- perhaps of a small number of fields -- to avoid biases or degraded constraints due to nuisance parameters.  \warn{We forecast the potential for this to be constrained by `clustering redshifts', see e.g.\ ref.\ \cite{McQWhi13}, and find in Fig.~\ref{fig:mqw} that a few thousand secure redshifts per sq.~deg.~are necessary to achieve the required percent-level constraints given strong, but reasonable, priors on the redshift range.  The achieveable near-term spectroscopic suveys we define in Table \ref{tab:selection} suggest this number may be challenging to achieve at $z=3$ and 4 with a large-area (14K deg$^2$) cosmological survey, but dedicated follow-up of absorption line redshifts may well be warranted - likely on an 8m class telescope.}   

These spectroscopic samples facilitate both RSD and \fnl \ science by targeting bright LBGs at $z\simeq 2$, $3$ and a high Lyman-$\alpha$ equivalent width (10\%, Fig.~\ref{fig:lines}) fraction at $z=4$ and 5.  Although extraordinarily large samples, these would have increased shot-noise with respect to the (greater than a magnitude) deeper photometric samples, but much more secure radial positions.  \rfree{We assume broad-band colors are effective at pre-selecting sources with strong Lyman-$\alpha$ emission}, but previous studies are optimistic based upon the expected bluer continuum of bright LAEs \cite{Cooke09, Stark10} -- \rfree{as opposed to any perceptible change in color due to the line emission directly}.  The nature of these large-scale surveys requires relatively low signal-to-noise spectra.  As a result, Fig.~\ref{fig:lines} suggests that significant line confusion is possible, including of the Lyman-$\alpha$ line with the [OII] doublet.  This, together with sky line confusion, would have to be very well controlled for any potential \fnl \ survey.     

The spectroscopic sample would be highly biased, leading to high signal-to-noise clustering measurements, an amplified \fnl \ signal and a reduced RSD anisotropy.  Nevertheless, Fig.~\ref{fig:RSD} illustrates percent level constraints on the linear growth rate are achievable for a $\simeq 15,000$ sq.\ deg.\ redshift survey.  The difference between the Lyman-$\alpha$ and systemic redshift, rather than shot-noise or modelling uncertainties, will likely be the limiting factor. In principle, a wavenumber limit of $0.67\,h\,{\rm Mpc}^{-1}$ is feasible at $z\simeq 4$ if sufficiently precise redshift tracers can be found, e.g.\ absorption line or emission lines in the near-IR.  The resulting increase of $\simeq (0.67/0.1)^3 = 300$ in the number of available modes relative to low-$z$ studies would represent an unprecedented discovery space \rfree{and include a significant legacy for the fields of galaxy formation and radiative transfer}.  

We show in Table \ref{Table:Spectroscopy},  Fig.~\ref{fig:beast-like-exposures} and Fig.~\ref{fig:pfs-like-exposures} that redshifts for this fiducial spectroscopic sample can be obtained in $\simeq 130$, $45$ and 50 mins.\ for DESI, PFS and M-DESI (DESI-like spectrographs on a 6.5m Magellan) respectively.
Fig.~\ref{fig:Tz} shows the necessary scaling of these exposure times to fainter objects, more secure redshifts or a more restrictive redshift range. 
In Appendix \ref{app:FOM}, we provide a simple figure-of-merit that ranks these competing facilities according to their ability to deliver for the science cases we consider.  On this basis, M-DESI, MSE and BOA reflect logical steps for next generation analyses given the factors of three to five improvement.  The BOA and SpecTel proposals represent \rfree{much more ambitious, order-of-magnitude, gains}.

Despite providing strong motivation and outlining the worthwhile steps to be taken by the community, there are clearly numerous avenues for investigation given the scope and exploratory nature of this work.  Firstly, there remain open questions as to the optimal target selection, which may (in)directly affect the area, homogeneity and potential systematic biases for our CMB lensing cross-correlation science.  In particular, residual CIB will no doubt strongly correlate with the samples we define.  At this point, propagation of the $C_{\kappa g}$ signal-to-noise curves to parameter forecasts would be an obvious step, together with refinements that include e.g the effect of foreground removal on the lensing reconstruction noise.  Delensing with highly secure redshifts is seemingly also beneficial for raising the detection significance by CMB-S4, which motivates \rfree{inclusion} of additional \rfree{lower-precision} redshifts, e.g.\ `redmapper' \cite{Rykoff14}.  Further investigation of clustering redshifts is warranted, e.g.\ with respect to priors on the bias evolution and redshift range of colour-selected galaxies.

There is a clear need to design a means to isolate large equivalent width LAEs with broad-band imaging, perhaps using already acquired narrow band or spectroscopic data.  This is a critical aspect of the proposal we have defined.  Further work is necessary to confirm the \rfree{realism} of our spectroscopic redshift forecasts -- with open questions remaining on the impact of sky lines, \rfree{line misidentification and its mitigation}.  A greater understanding of the impact of radiative transfer on the selection of Lyman-$\alpha$ emission must be developed, including templates that properly sample the range of one-sided, double-peaked and absorption features that may be present.  There should be clear cross-community interest in determining whether observed trends of Lyman-$\alpha$ equivalent width with apparent magnitude are physical or selection based \cite{Stark10, Du18, Caruana18}.  The modelling of RSD must evolve to incorporate such scenarios, e.g. systematic biases in redshift due to intrinsic outflows, and our forecasts updated to better model this and non-linear biasing.  Further simulation work will no doubt be required to facilitate this.  The relevant limits of potential RSD surveys: a fast, wide area survey of Lyman-$\alpha$ emitters, but with relatively limited $k_{\rm{max}}$ due to systemic redshift differences, or a much slower absorption line survey that better samples the modes available, represents an interesting optimisation problem.  Sparse sampling strategies \cite{Kaiser86, Chiang13} should be considered in this respect, in particular for those primarily focused on large-scale \fnl \ analyses. 

\rfree{Exploiting fore-runner surveys will form a large part} of the necessary work and coordination, e.g. modest proposals for DESI or other instruments, together with extensions of the Clauds $u$-band survey to a larger, perhaps discontiguous, area.  The next release of HSC data will be important in this respect, due to the added area, depth and first Clauds release.  Given the relative ease with which the three critical inputs curves, $p(z), b(z)$ and $\bar n_\theta$, needed to enable an accurate $C_{\kappa g}$ forecast could be obtained, it should be a priority to determine these in time for next-generation CMB experiments, e.g.\ Simons Observatory, being finalised.  Finally, the brighter galaxies present in the samples we define enable measurements of the Lyman-$\alpha$ forest, a fact we have entirely neglected. 

In short, we have seen that it is entirely within our means to deliver significant advancements of our knowledge of gravity and Dark Energy, inflation and the neutrino mass hierachy based on well-tested and efficient selection of high-redshift Lyman-break galaxies \rfree{of varying Lyman-$\alpha$ equivalent width}.  A particularly exciting frontier is the combination with CMB lensing, due to the redshift overlap between these two tracers.  

\acknowledgments
We benefited from discussions with S. Ferraro, Z. Cai, A. Dey, M. Schmittfull, K. Dawson, J. Cooke, S. Bailey, D. Kirkby, J. Moustakas, D. Schlegel, R. Bowler, J. Guy, M. Jarvis, H. Hildebrandt and K. G. Lee.  
We also thank F. Bian, H. Hildebrandt, M. Rafelski and Y. Ono for providing their data in electronic form.
MJW gratefully acknowledges the hospitality of the University of Oxford during completion of this work.  M.W.~is supported by the U.S.~Department of Energy and by NSF grant number 1713791.  This research used resources of the National Energy Research Scientific Computing Center (NERSC), a U.S.~Department of Energy Office of Science User Facility operated under Contract No.\ DE-AC02-05CH11231.  This work also made extensive use of the NASA Astrophysics Data System and of the {\tt astro-ph} preprint archive at {\tt arXiv.org}.

\appendix

\section{BX selection with LSST}
\label{app:filter_conversion}
To facilitate comparison of previous BX selected samples with modern LSST filters, we quote the conversion to Johnson and Steidel filters utilising the extincted BC03 templates described in \S\ref{sec:sed}.  Based on linear regression, we find
\begin{equation*}
\begin{array}{llr}
  (U-G) &= 0.97\ (u-g) \ + \ 1.27                                    & (0.41), \\
  (G-R) &= 0.32\ (r-i) \ + \ 1.11                                    & (0.13),
\end{array}
\end{equation*}
where the final parentheses show the RMS residual.

\section{Medium and narrow band surveys}
\label{app:narrowband}
Narrowband selection -- equivalent width limited samples found by excess flux in the narrow band over an encompassing broad band -- has traditionally served to isolate Lyman-$\alpha$ emitters, e.g.\ refs.\ \cite{Ouchi03, Sobral18}, or provide increased precision for photometric samples (given a large number of filters).  The former suffers some severe drawbacks for our purposes: only thin slices in redshift can be obtained, which have limited volume and access to large scales; the limiting flux in a filter is $\propto (1/\sqrt{\Delta \lambda})$ \cite{Benitez09} where $\Delta \lambda$ is the filter width.  Hence the limiting flux is approx.\ $\sqrt{5}$ times shallower for narrowband ($R>50$) with respect to broadband ($R<10$) at fixed exposure time.  This is particularly difficult approaching one micron, as the limited resolution with respect to spectroscopy yields greater sensitivity to the sky background and hence a further reduced depth.  For instance, the ALHAMBRA survey \cite{Molino14} presents a 2.4 deg$^2$ study (\cite{Viironen15}, V15) of $2.2 < z < 5.0$ LBGs with 20 contiguous filters (3500 - 9700\AA) on the $3.5\,$m Calar Alto telescope.  They find the effective depth to drop from $\geq 24$ in the blue to $\simeq 21.5$ by the red.  This suggests a combination of broad-band imaging and spectroscopy \rfree{would} be more competitive. 

However, such surveys fulfil many purposes for informing our science case.  Firstly, they characterise the effectiveness and contamination rates of small dropout samples at a greater completeness than \rfree{current} spectroscopy.  For instance, at a limiting magnitude of $\simeq 24$, \S4.4 of V15 confirms color-color selection to be highly efficient -- typically >95\%, but slightly lower with for BX, 84\%-- at selecting the same redshift range as with twenty-filter photometric redshifts. The SHARDS survey \cite{ArrabalHaro18} reaches the same conclusion for $3.35 < z < 6.8$ LBGs at greater depths ($\simeq 26.75$), with 25 medium filters spanning 5000 to 9410\AA\ over $\simeq 130$ arcmin$^2$ on the $10.4\,$m Gran Telescopio Canarias.
An unexplored opportunity in this respect is the design of broad-band color selection of large Lyman-$\alpha$ equivalent-width samples using such narrowband samples, providing this is not done overzealously with respect to any redshift evolution.

\section{Figure-of-merit for spectroscopic surveys}
\label{app:FOM}
To assess the relative utility of a given facility for our science cases, we argue as follows.  As an example, to deliver competitive $f_{NL}$ constraints, one requires a number density of approx.\ $10^{-4} \ [(h^{-1}\rm{Mpc})^{-3}]$.  For an RSD measurement, this number is higher by a factor of $\simeq 3$ depending on the linear bias, with a similar number required for cross-correlation estimates of $dN/dz$.  Assuming an airmass-limited 14K sq.\ deg. survey, this suggests a 14M galaxy survey at a minimum. To satisfy both the area and multiplex requirements, the minimum number of required pointings is
\begin{equation}
    N_{\rm{point}} = \rm{max} \ \left \{ \frac{14M}{\rm{Multiplex}}, \frac{14K}{\rm{FOV}} \right \},
\end{equation}
which sets one \rfree{appropriate} metric.  \rfree{For a clustering redshift case, the minimum number of pointings must also be equal to or greater than that required to beat the mode-sampling variance}. 

\rfree{The most realistic figure-of-merit} is (1 / survey time [decades]), where the survey time is the product of the exposure time and this minimum number of pointings.  This better accounts for the dependence on multiplex, field-of-view and science than the traditional `$A\Omega$'.  

For the exposure time, eqn. \ref{eqn:rrs2n} provides a good proxy for the redshift success.  Given our fiducial M-DESI {\tt redrock} runs and the known exposure time scalings (see e.g.\ ref.\ \cite{Chromey16}):
\begin{equation}
\left ( \frac{S}{N} \right ) = \frac{N_{*}}{\sqrt{N_{*} + \langle N_s \rangle + n_p (T \dot D + B^2)}},    
\end{equation}
we may estimate this for any given telescope and spectrograph.  Note any obscuration must be included in the effective mirror radius.  An important scaling is \rfree{spectral} resolution, for which we assume \rfree{$T \propto (1/\mathcal{R})$} (for fixed signal) given the sky background variance scales as the width of the resolution element in wavelength.  Here $N_* = \pi R^2 T \times \lambda E_\lambda F_\lambda d \lambda / (hc)$, \rfree{$R$ is the mirror radius}, $T$ is the exposure time, $E_\lambda$ is the (atmospheric and end-to-end optical) transmission, \rfree{$F_{\lambda}$ is the flux density}, $\dot D$ is the dark current, $B$ is the read noise and $\langle N_s \rangle$ is the sky brightness in photons across the aperture -- the product of the fiber area projected on the celestial sphere and the angular surface brightness of the night sky.  Specifically, a fiber projects to an angular scale of $\phi_s = (d / f)$
, where $d$ is the physical fiber size ($\simeq 100\,\mu$m) on the focal plane -- with plate scale $2.1\times 10^{5}/D \ [''/ $mm] and \rfree{effective focal length, $f$}.  Typically, this brightness is of order 18.3 AB mag~/~arcsec$^2$, e.g.\ at $\approx 9134$~\AA \ of the SDSS $z'$ band\footnote{\url{https://subarutelescope.org/Observing/Instruments/SCam/exptime.html}}.  If this is less than the local seeing, significant fiber loss can be expected.  For comparison, ref.~\cite{Bian13} suggests a angular scale of $0.3^{\prime\prime}$ for a $z\simeq 3$ LBG with $R_{AB} \simeq 22.3$, which corresponds to $2\,$kpc in physical distance.    

In the absence of, and as a sanity check on, {\tt redrock}, this may be further estimated by hand.  To do so requires a few additional characteristics, e.g. the required resolution sets the projected fiber shadow on the CCD as the intuitive \cite{DESIb}: 
\begin{equation}
\frac{1}{\mathcal{R}(\lambda)} = \frac{\sqrt{3} md (\lambda_{\rm max} - \lambda_{\min})}{2 L \lambda}, 
\end{equation}
which yields $n_{\rm p}$ given typical characteristics of the camera, e.g. a physical detector size $L$ and pixel size, $\approx 15 \mu$m for a 4K$\times$4K CCD.  In the absence of anamorphic magnification by the spectrograph, the $md$ factor is simply $d (f_{\rm{CAM}}/f_{\rm{COL}}) \simeq (d/2)$ \cite{Chromey16} \rfree{for a typical ratio of the camera and collimator focal lengths}.  Typical detectors of interest today are designed to resolve the 2\AA\ separation of the [OII] doublet across the coverage.

\bibliographystyle{JHEP}
\bibliography{lbgcmb}
\end{document}